\documentclass[sigconf]{acmart}
\usepackage{soul}
\usepackage{hyperref}
\usepackage{enumitem}
\usepackage{listings}
\usepackage{cleveref}
\usepackage{booktabs} 
\usepackage{xspace}
\usepackage{xcolor}
\usepackage{microtype}
\usepackage{subcaption}
\usepackage{balance}
\usepackage{mathtools}
\usepackage{scalerel}

\usepackage{tcolorbox}
\usepackage{microtype}
\usepackage[section]{placeins}
\usepackage{graphicx}
\usepackage[T1]{fontenc}
\usepackage{svg}
\usepackage{amsmath}
\usepackage{amsmath}
\usepackage[mathletters]{ucs}
\usepackage[T1]{fontenc}
\usepackage[utf8x]{inputenc}
\usepackage{pict2e}
\usepackage{wasysym}
\usepackage[english]{babel}
\usepackage{tikz}
\usepackage{array}

\DeclareMathSymbol{:}{\mathord}{operators}{"3A}

\def\mathbi#1{\textbf{\em #1}}
\def\HS{\hspace{\fontdimen2\font}}
\definecolor{darkgreen}{rgb}{0.15,0.55,0.15}
\definecolor{darkblue}{rgb}{0.1,0.1,0.5}
\definecolor{blue}{rgb}{0.01,0.40,.8}
\definecolor{darkgreen}{rgb}{0.15,0.55,0.15}
\definecolor{mred}{rgb}{.80,.12,.30}
\definecolor{grey}{rgb}{0.5,0.5,0.5}
\definecolor{Purple}{rgb}{.75,0,.85}
\definecolor{light-gray}{gray}{0.95}
\definecolor{mid-gray}{gray}{0.85}
\definecolor{darkred}{rgb}{0.7,0.25,0.25}
\definecolor{rose}{rgb}{1.0, 0.01, 0.24}
\newcommand{\red}[1]{\textcolor{red}{#1}}

\newcommand{\blue}[1]{\textcolor{blue}{#1}}
\usepackage[ruled,vlined]{algorithm2e}

\SetCommentSty{mycommfont}
\newtcbox{\redbox}{on line,
  colframe=white,colback=red!10!white,
  height=1em,valign=bottom,
  boxrule=0.5pt,arc=2pt,boxsep=0pt,left=2pt,right=2pt,top=1pt,bottom=1pt}
\newtcbox{\bluebox}{on line,
  colframe=white,colback=blue!10!white,
  height=1em,valign=bottom,
  boxrule=0.5pt,arc=2pt,boxsep=0pt,left=2pt,right=2pt,top=1pt,bottom=1pt}

\newcommand{\eat}[1]{}

\newcommand{\stitle}[1]{\smallskip\noindent\textbf{#1}}

\newtheorem{ex}{Example}
\newtheorem{pr}{Problem}
\newtheorem{example}[ex]{Example}

\newtheorem{problem}[pr]{Problem}

\newlength{\listingindent}                
\usepackage[LGRgreek]{mathastext}

\DeclareMathOperator*{\argmin}{arg\,min}

\setlist{leftmargin=*}

\graphicspath{ {images/} }
\begin{document}

\newcommand{\total}[0]{\emph{TOTAL}}
\newcommand{\cnt}[0]{\emph{COUNT}}
\newcommand{\cof}[0]{\emph{COF}}
\newcommand{\ewu}[1]{\red{ewu: #1\xspace}}
\newcommand{\sys}[0]{\texttt{Reptile}\xspace}
\newcommand{\matlab}[0]{\texttt{Matlab}\xspace}
\newcommand{\revise}[2]{\blue{#1\xspace}}

\author{Zezhou Huang}
\email{zh2408@columbia.edu}
\affiliation{
  \institution{Columbia University}
}
\author{Eugene Wu}
\email{ewu@cs.columbia.edu}
\affiliation{
  \institution{DSI, Columbia University}
}

\renewcommand{\shortauthors}{Z. Huang, E. Wu}

\begin{abstract}
Recent query explanation systems help users understand anomalies in aggregation results by proposing predicates that describe input records that, if deleted, would resolve the anomalies.  However, it can be difficult for users to understand how a predicate was chosen, and these approaches are limited to errors that can be resolved through deletion.  In contrast, data errors may be due to group-wise errors, such as missing records or systematic value errors.   

This paper presents \sys, an explanation system for hierarchical data.  Given an anomalous aggregate query result, \sys recommends the next drill-down attribute, and ranks the drill-down groups based on the extent repairing the group's statistics to its expected values resolves the anomaly. \sys efficiently trains a multi-level model that leverages the data's hierarchy to estimate the expected values, and uses a factorised representation of the feature matrix to remove redundancies due to the data's hierarchical structure.  We further extend model training to support factorised data, and develop a suite of optimizations that leverage the data's hierarchical structure.   
\sys reduces end-to-end runtimes by ${>}6\times$ compared to a Matlab-based implementation, correctly identifies 21/30 data errors in John Hopkin's COVID-19 data, and correctly resolves 20/22 complaints in a user study using data and researchers from Columbia University's Financial Instruments Sector Team.

\end{abstract}

\title{Reptile: Aggregation-level Explanations for Hierarchical Data}
\maketitle

\section{Introduction}

Data exploration tools follow the ``overview, zoom, then details'' analysis pattern~\cite{North2000SnaptogetherVA} to help users analyze their datasets at a high level before diving into the individual records.  However, modern datasets are often hierarchical and multi-dimensional.  Thus, when users identify an anomalous aggregate value that is too high or too low (in an overview), it can be difficult to know which attributes to drill-down (zoom in) and which of the drill-down results to focus on.   This is particularly relevant when anomalous results are due to systematic data errors (e.g., missing or duplicate records, measurement errors) that the user wants to find and address.

Recent query explanation systems~\cite{wu2013scorpion,roy2014formal,abuzaid2020diff} have been proposed to identify predicates over the query input ({\it explanations}) that, if the records are deleted, will most repair the anomalous result ({\it complaint}).  However, they are limited to deletion-based repairs, and not errors due to missing records, value errors, or anomalous statistics of subsets of the input data.  Further, instead of being presented with a (potentially complex) predicate, users often want assistance while incrementally drilling down, so they can verify data at each step.  We illustrate with an example based on Columbia University's Financial Instruments Sector Team (FIST):

\begin{figure}
     \centering
     \begin{subfigure}[b]{0.3\textwidth}
         \centering
         \includegraphics[width=\textwidth]{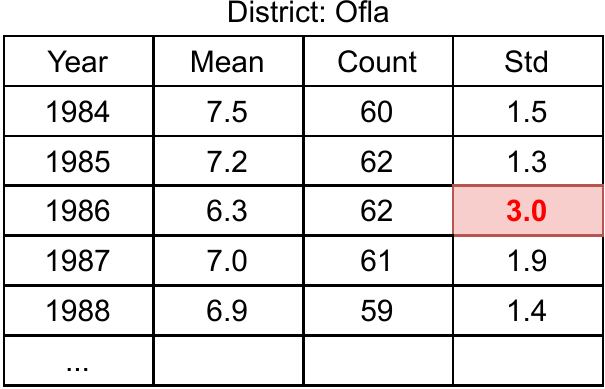}
         \caption{Drought statistics per year in Ofla.}
         \label{fig:exampleini}
     \end{subfigure}
     \hfill
     \begin{subfigure}[t]{0.3\textwidth}
         \centering
         \includegraphics[width=\textwidth]{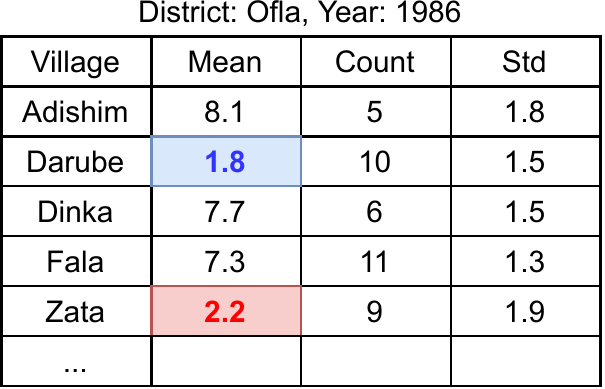}
         \caption{After drill-down by geography to village stastistics. 
         }
         \label{fig:exampledrill}
     \end{subfigure}
     \hfill
    \begin{subfigure}[t]{0.16\textwidth}
         \centering
         \includegraphics[width=\textwidth]{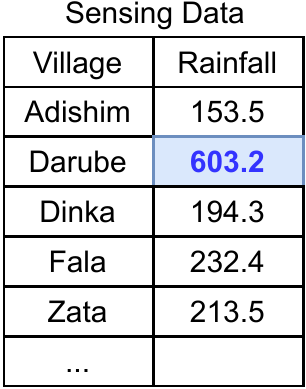}
         \caption{Auxiliary satellite sensing dataset. 
         }
         \label{fig:exampleext}
     \end{subfigure}
     \hfill
     \caption{Example FIST use case.  (a) The researcher thinks that Ofla's 1986 standard deviation of severity is suspiciously high (the complaint).  (b) After drilling down to villages, Darube and Zata have low means and are potential explanations.  (c) Darube is explained away by the high rainfall in the sensing data.}
    \label{fig:examplecase}
\end{figure}

\begin{example}
\label{exp:firstexample}
FIST surveys and collects drought severity data from farmers in villages throughout African countries (e.g., Ethiopia) to develop sustainable drought insurance plans for the countries.  
As a toy example, \Cref{fig:exampleini} shows drought severity (from 1 (not severe) to 10 (severe)) statistics between 1984 to 1988 collected from the Ofla District 
in Ethiopia. The FIST researcher {\it complains} that the standard deviation in 1986 is higher than expected and suspects bias in the reporting.   Rather than read the raw records, the researcher wants to drill-down one level before looking at individual records. 

Although most villages in Ofla reported high severity due to low 1986 rainfall, Darube and Zata have abnormally low means that may contribute to the complaint (\Cref{fig:exampledrill}).  However, Darube's high rainfall in the auxiliary satellite sensing data (\Cref{fig:exampleext}) explains the low severity.  Zata's low severity is unexplained and thus highlighted when the researcher drills down to villages.  
\end{example}

This process is laborious to FIST for many reasons.  When there are multiple possible drill-down hierarchies, the choice depends on which of the many resulting groups (e.g., villages) most contributed to the complaint. A group's contribution depends on how much its statistics diverge from its expectation, and the extent that repairing the statistics would address the complaint.  However, it is hard to manually estimate a group's expected statistics.  Instead, the FIST researchers wish for the ability to submit complaints about anomalous aggregate statistics during data exploration, and be recommended drill-down results to investigate.  By repeating this process, they can incrementally zoom into their data and arrive at the data errors.

Existing explanation algorithms fail because each makes assumptions about data errors that do not hold in this context.  
Density-based approaches such as Smart Drill-Down~\cite{joglekar2015smart} are designed for count-based complaints, and identify high-cardinality groups;  in \Cref{exp:firstexample}, they would return Fala, despite its normal drought severity.
Sensitivity-based methods~\cite{wu2013scorpion,roy2014formal,abuzaid2020diff} such as Scorpion support complaints over general aggregation functions, but are limited to deletion-based interventions inappropriate for FIST's needs.
For instance, deleting Darube village's records would most reduce the standard deviation, but is an incorrect recommendation. 
Counterbalancing~\cite{miao2019going} looks for sibling aggregates that offset (counterbalance) the deviation specified in the user's complaint. However, a village's drought is not offset by higher rainfall elsewhere.

The paper presents \sys which tackles the above problem.
Given a hierarchical dataset and complaint (e.g., an aggregation query result value is too high or low),
\sys recommends the next drill-down attribute and highlights the most relevant groups to examine so the user can understand and verify the recommendations in each step.
\sys ranks groups by the extent that repairing their statistics (count, mean, sum, or a user-provided aggregate) to its expected value would resolve the complaint.  \sys estimates the expected statistics for each drill-down attribute by fitting a multi-level model\footnote{\sys supports custom models, but is optimized for multi-level models.} to the drill-down's results, and uses the model to exploit the data's hierarchical structure. Users can improve the model by providing additional predictive signals such as custom features (e.g., nearby village severities) or joining with auxiliary datasets (e.g., satellite rainfall estimates).

We address two main challenges.   The first is the scarcity of training examples when ranking the immediate groups in a candidate drill-down operation. For example, Olfa only contains a handful of villages, which is likely insufficient to train an accurate model.  \sys addresses this by using parallel groups (e.g., villages in other districts), and the multi-level model accounts for systematic variation between parent groups (e.g., different districts).  

The second challenge is scalable model training. The number of possible models is exponential in the number of attributes, and unrealistic to fully precompute.  
\sys efficiently trains models online by exploiting functional dependencies inside hierarchies and independence between hierarchies.  Instead of materializing a feature matrix exponential in the number of hierarchies, \sys computes a succinct factorised matrix representation~\cite{olteanu2015size} that reduces the matrix representation by orders of magnitude. We extend prior work~\cite{schleich2020lmfao,schleich2016learning}, which developed model training procedures over factorised matrices derived from join queries, to matrices based on join-aggregation queries that exhibit fewer redundancies.  We further design factorised matrix multiplication operators, and develop a suite of novel work-sharing and caching-based optimizations. 

\smallskip\noindent In summary, our contributions are as follows:
\begin{itemize}
  \item We propose and solve the {\it complaint-based drill-down problem}.

  \item We adapt factorized representations to compactly represent the feature matrix, and extend matrix operations to support factorized representations.  We develop precomputation, work sharing, and caching optimizations to further accelerate successive drill-down operations.

  \item On synthetic data, our factorized matrix operations accelerate matrix materialization and gram matrix computations by orders of magnitude, and work-sharing reduces runtimes by $4\times$ over LMFAO~\cite{schleich2020lmfao}.  On real-world data, \sys reduces end-to-end performance by $6\times$ as compared to a Lapack-based Matlab implementation.

  \item We show that \sys robustly identifies multiple classes of errors (missing and duplicate data, systematic value errors) with 70\%-100\% accuracy, and leverages auxiliary data when it has predictive power; baseline complaint- and outlier-based approaches have 0\% to ${<}60\%$ accuracy.
  
  \item We evaluated \sys with FIST team members, who explored their drought survey data and submitted 22 complaints.  \sys was able able to correctly identify data errors for 20 of them.  One failure was inherently ambiguous (team members disagreed about the cause), and \sys partially explained the other.
  
\end{itemize}


\section{Background}

\subsection{Usage Walkthrough and Architecture}

We will use \Cref{exp:firstexample} to illustrate how the Financial Instruments Sector Team (FIST) uses \sys to identify errors in their farmer-reported drought severity dataset.  \sys is initialized with the database as well as metadata about the attribute hierarchies (e.g, geographical and temporal for this example).  
A FIST researcher studies the annual severities in the Ofla district.  She suspects that the standard deviation in 1986 is too high and submits it as a complaint.  She also provides village-level rainfall as an auxiliary joined dataset because she feels it can help indicate droughts.

\begin{figure}
  \centering
      \includegraphics [scale=0.7] {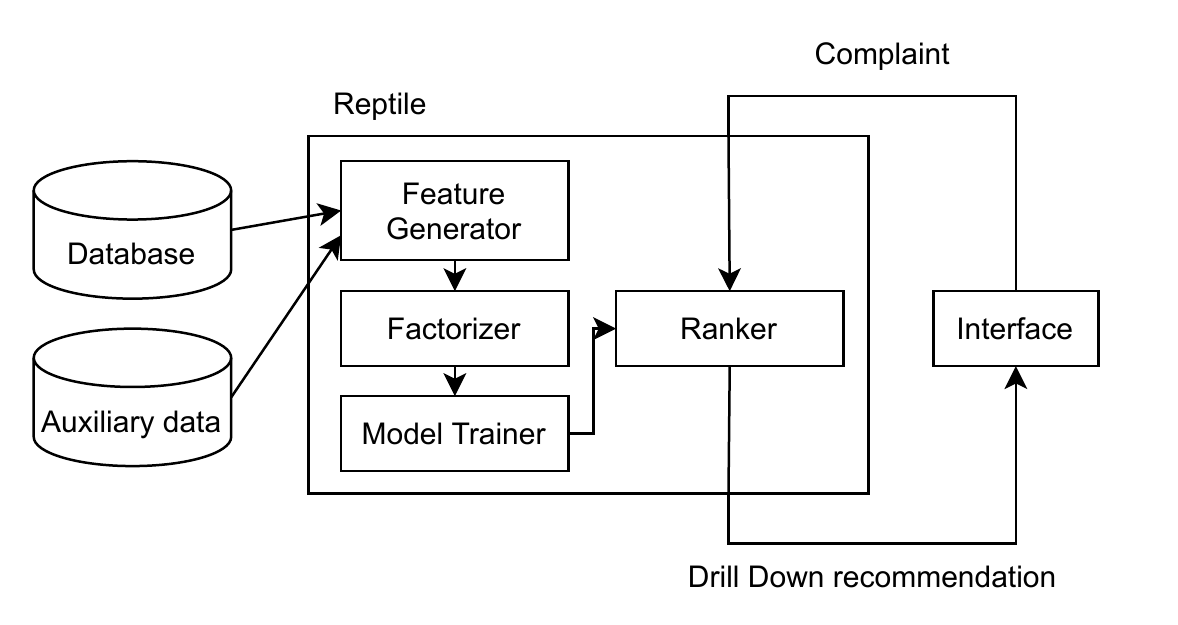}
  \caption{\sys architecture}
  \label{fig:arch}
\end{figure}



At this point, \sys follows the architecture in \Cref{fig:arch}.  It first combines the queried tables with the auxiliary sensing dataset, and uses them to extract model features. The {\it Factorizer} stores the features in an efficient factorised representation (described below), and the 
{\it Model Trainer} fits a predictive model to estimate the statistics for each group in the next candidate drill-down.  For instance, if \sys drills down along geography, the model estimates village level statistics in Ofla 1986.  \sys uses multi-level models to account for hierarchical relationships, and introduces optimized matrix operations over factorised representations.

The {\it Ranker} first evaluates each group (e.g., village) based on the extent that repairing the group's statistics to its expected value would address the complaint; it assigns a score to the groups for every possible drill-down (geography and time in this example), and recommends the top-K.
The researcher can then submit a village-level complaint to continue drilling down.

\subsection{Overview of Factorised Representations}
\label{backfrep}


Joins and hierarchical data exhibit redundancy when encoded in a tabular format, and factorised representations~\cite{olteanu2015size} (f-representations) remove this redundancy.  Assuming a fully normalized database (e.g., in BCNF), f-representations encode query results as an algebraic expression composed of unions and cartesian products. In a join query, for instance, the set of left and right records that have the same join key will emit the logical cartesian product and not materialize it.  Matrix operations in train models reduce to batches of aggregations~\cite{schleich2020lmfao}, which can be efficiently executed over f-representations by pushing them through joins.

Given a relational table with schema S, 
The following notations are used for f-representations: 

\begin{itemize}
\item $\{ (v): i \}$: a unary relation with tuple (v) whose count is i .
\item $(E_1 \cup ... \cup E_n)$: union of relations $E_1,...,E_n$ with the same schema.
\item $(E_1 \times ... \times \, E_n)$: cartesian product of relations $E_1,...,E_n$, where the schema of $E_i$ is $\emph{S}_i$ and $\emph{S}_1 \cap ... \cap \emph{S}_n = \emptyset$.
\end{itemize}
\noindent F-representations help remove redundancies due to functional dependencies inside a hierarchy and independence between hierarchies:

\begin{example}[Hierarchical Data]
  Consider the relation $R = \{(a_1,\\ b_1): 1,(a_1, b_2):1,(a_2, b_3):1,(a_2, b_4):1\}$ over schema $\emph{S} = [A, B]$, with functional dependency $B\to A$.  Its f-representation is: 
  $$(\{(a_1):1\} \times (\{(b_1):1\} \cup \{(b_2):1\}  ))
  \cup ( \{(a_2):1\} \times (\{(b_3):1\} \cup \{(b_4):1\}))$$
\end{example}
\begin{example}[Independent Schemas]
  Consider relation $R_1 = \{(a_1):1,(a_2):1,(a_3):1\}$ over schema $\emph{S}_1 = [A]$ and relation $R_2 = \{(b_1):1,(b_2):1,(b_3):1\}$ over schema $\emph{S}_2 = [B]$.   Their schemas do not overlap, so the join result is quadratic in size (i.e., $9$), whereas its f-representation is linear:
  $$ (\{(a_1):1\} \cup \{(a_2):1\} \cup \{(a_3):1\}) \times (\{(b_1):1\} \cup \{(b_2):1\} \cup \{(b_3):1\})$$
\end{example}


\smallskip\noindent 
\sys develops matrix operations over f-representations, which are decomposed into batches of aggregation queries (\Cref{section:matrxopoverfr}). Here, we introduce the aggregation operator using a \texttt{COUNT}-query example 
and describe when they can be logically pushed through joins. For further background, please refer to~\cite{nikolic2018incremental}.

Consider the aggregation $\gamma_{X_1, ..., X_f,COUNT}(R_1 \Join ... \Join R_n)$, where the schema of the join result is $X_1, ..., X_f, X_{f+1}, ..., X_{m}$.  For tuple t in relation R, the  notation $R[t]$ returns the \texttt{COUNT} for tuple $t$. Then, the aggregation result is:
\begin{align}
Q[(X_1,...,X_f)] = \bigoplus_{X_{f+1}} \hdots  \bigoplus_{X_m} \bigotimes_{i \in [n]}R_i[S_i] \nonumber
\end{align}
where $\bigotimes$ is the join subplan, $\bigoplus_{X}$ is an aggregation that marginalizes over attribute $X$, and $S_i$ is the schema of relation $R_i$. 
$\bigotimes$ and $\bigoplus_X$ are defined as:
\begin{align}
  (R\bigotimes T)[t] =& \HS R[\pi_{S_1} (t)] ∗ T[\pi_{S_2} (t)] \nonumber &\hspace{1em}\forall  t \in D_1  \\
  (\bigoplus_X R)[t] = & \sum \{R[t_1] | \HS t_ 1 \in Dom(S_1) , t = π_{S_1 \backslash  \{X \}} (t_1 )\} \nonumber &\hspace{1em} \forall t \in D_2 
    \end{align}
where $S_1$ and $S_2$ are the schemas for $R$ and $T$, $X \in S_1$, $D_1 = Dom(S_1 \cup S_2)$, and $D_2 = Dom(S_1 \backslash \{X\})$.  Suppose $t=\langle Distinct=Ofla\rangle$.  The first statement says that Ofla's \texttt{COUNT} after the join is equivalent to multiplying the \texttt{COUNT} Ofla records in $R$ and $T$.   The second statement states that marginalizing over $X$ (say, the attribute Year) is computed as the sum of Ofla counts over every year. 

\begin{example}[Join and Aggregation Operators]

Let relations $\emph{R} = \{(a_1,b_1):1,(a_2,b_1):2\}$ over schema $ [A,B]$, and relation $\emph{T}= \{(b_1,c_1):3,(b_1,c_2):4\}$ over schema $[B,C]$.  Consider the query:
\begin{align}
Q[(A,B)] = \bigoplus_{C}  (R[(A,B)] \bigotimes T[(B,C)])  \nonumber
\end{align}
The intermediate result $R_{\Join} = R[(A,B)] \bigotimes T[(B,C)]$ contains:  
$$\{(a_1,b_1,c_1):3, (a_1,b_1,c_2):4, (a_2,b_1,c_1):6, (a_2,b_1,c_2):8 \}$$
$\bigoplus_C$ partitions $R_{\Join}$ by A,B and sums all the counts in each partition to derive $\{(a_1,b_1):7, (a_2,b_1):14 \}$.
\end{example}

\noindent Early marginalization pushes $\bigoplus_C$ down when $C$ is not used 
in the outer query (such as joins):

\begin{example}[Early Marginalization]
Let relations $\emph{R}$, $\emph{T}$ have schemas $[A,B]$ and  $[B,C]$.  Consider the query $ \gamma_{A,COUNT}( \emph{R} \Join  \emph{T})$, where each attribute's domain is $O(n)$.   Both relations are thus $O(n^2)$ and the join result is $O(n^3)$.  Notice that attribute $C$ is not used for the join and can be marginalized early:
\begin{align}
Q[(A)] = \bigoplus_{B} \bigoplus_{C} (R[(A,B)] \bigotimes T[(B,C)])  \nonumber
\end{align}
Thus $\bigoplus_C$ can be pushed through $\bigotimes$ to reduce the join result to $O(n^2)$:
\begin{align}
Q[(A)] = \bigoplus_{B} R[(A,B)] \bigotimes  (\bigoplus_{C} T[(B,C)] ) \nonumber
\end{align}
\end{example}

\section{Approach Overview}\label{s:overview}

\subsection{Problem Definition}

Given relation $\mathbb{R}$ with attributes $\mathbb{A}$, we assume that all the attributes in $\mathbb{R}$ are partitioned into hierarchical dimensions. A dimension's hierarchy $H = [A_1,\ldots,A_k]$ is an ordered list of attributes where there is a functional dependency $A_n\to A_m \forall m < n$. We say that $A_n$ is {\it more specific} than $A_m$ if $m<n$ in the same hierarchy. The last attribute $A_k$ is the most specific attribute in $H$. Note that a dimension's hierarchy may contain a single attribute.

\sys starts with initial view $V = \gamma_{A_{gb}, f(A_{agg})}(\mathbb{R})$, where $A_{gb}\subset \mathbb{A}$,
$A_{agg}\subset \mathbb{A}$, $A_{gb} \cap A_{agg} = \emptyset$,  $f(\cdot)$ is a distributive \cite{gray1997data} aggregation function\footnote{For simplicity, the text assumes a single COUNT aggregation function in $Q$, however \sys supports a general distributive set functions, and queries with multiple aggregation functions, as discussed in \Cref{section:disset}.} such that, given the partition of $\mathbb{R}$ into $J$ subsets $R_1, \cdots, R_J$, there exists function $\emph{G}$: $f(\mathbb{R}) = \emph{G}\HS(f(R_1),\ldots,f(R_J))$. Let $t_i\in V$ be an output tuple and $t_i[agg]$ be $t_i$'s aggregation result.

\begin{example}[Distributive aggregation function]
Count is a distributive aggregation function. Given the partition of $R$ into $\emph{J}$ subsets $\mathbb{R}_1, \cdots, \mathbb{R}_\emph{J}$, we can find $\emph{G}_{count}$ such that $count(\mathbb{R}) = \emph{G}_{count}\HS(\{\\count(\mathbb{R}_1), \cdots, count(\mathbb{R}_\emph{J})\}) =  \sum^{\emph{J}}_{i=1}(count(\mathbb{R}_i))$
\end{example}

User can make complaint about tuple $t_{c} \in V$ in \sys. Define user complaint as a function $\emph{f}_{comp}: t \rightarrow \mathbb{R}$ \footnote{In general, $\emph{f}_{comp}$ may be an expression composed of distributive aggregates in the query.  For instance, \texttt{SUM} can be decomposed into an expression over \texttt{MEAN} and \texttt{COUNT}} which takes tuple as input and output a value that user aims to minimize. This formulation captures common complaints~\citeN{wu2013scorpion, roy2014formal, bailis2017macrobase, miao2019going}, such as $t[agg]$ is too high or too low, or that $t[agg]$ should be a specific value.   For instance, $\emph{f}_{comp}(t) = |t[count]-v|$ states that the output attribute \texttt{count} should have been $v$. 

\sys helps user drill-down from the complaint tuple along different dimensions (e.g., district to village, or from year to month). Given a tuple $t$ in the current view $V = \gamma_{A_{gb}, f(A_{agg})}(\mathbb{R})$, $drilldown(V,t,H)$ adds the next strict attribute in hierarchy $H$ to $A_{gb}$ in $V$ and replaces $\mathbb{R}$ by the provenance of $t$.

\begin{example}[Drill-down]
  \Cref{fig:examplecase} is grouped along geographic and temporal dimensions, with hierarchies $H_{geo}$=[\texttt{District}, \texttt{Village}] and $H_{time} =$ [\texttt{Year}, \texttt{Month}].  \Cref{fig:exampleini} shows the view $V$ that filters and aggregates by \texttt{(District=Ofla, Year)}.  Let $t$ be the tuple for year \texttt{1986}.  $drilldown(V, t, H_{geo})$ would further aggregate the provenance of $t$ by village (\Cref{fig:exampledrill}).
\end{example}

Next, \sys tries to repair tuples in drill-down result. Let $\emph{f}_{repair}: t \rightarrow t$ be a repair function that, given a tuple in the drill-down result, returns a tuple with its expected aggregate statistics. After tuple $t'$ in $V' = drilldown(V,t_{c},H)$ is repaired, the complained tuple's aggregation result is also repaired: $t_{c}' =  \emph{G}(V'/\{t'\} \cup \{\emph{f}_{repair}(t')\}) $

Finally, \sys proposes one hiearchy $H \in \mathbb{H}$ to drill-down, and one tuple $t \in drilldown(V, t_{c},H)$ such that ``fixing'' the $t$'s {\it group statistics} would minimize user complaint $\emph{f}_{comp}(t_{c}')$.  

\begin{problem}[Complaint-based Drill-down]
  Given $t_{c}$, $\emph{f}_{comp}$, $\emph{f}_{repair}$, return the next drill-down hierarchy and tuple $(H^*,t^*)$ where:
\begin{align}
\label{optimizationpro}
H^*,t^* &= \argmin_{H,t} \quad \emph{f}_{comp}(t_{c}') \\
\label{querydefinition}
\textrm{s.t.} \quad  V' &= drilldown(V, t_{c},H),\\
 t_{c}' &=  \emph{G}(V'/\{t\} \cup \{\emph{f}_{repair}(t)\}) \\
 \label{eq:allhiearchies}
  H & \in \mathbb{H} \\
  \label{tupleinquery}
t &\in  V'
  \end{align}
\end{problem}


\begin{example}[Complaint-based Drill-down]
  Given the complaint $t_{c} = (Year: 1986, District: Ofla, count: 62)$ in \Cref{fig:examplecase},  the complaint function is $\emph{f}_{comp}(count) = |count -70|$ (that is, the count of tuples in Ofla in year 1986 should have been 70) and consider the \texttt{Darube} and \texttt{Zata} records after drilling down along $H_{geo}$ (\Cref{fig:exampledrill}).  If the repair function fixes \texttt{Darube}'s count to $15$,  $t_{c}$'s count will update to $67$, and the complaint function returns $f_{comp}(67) = 3$.  In contrast, if \texttt{Zata} is repaired to 72, then its complaint function would return $2$, which is preferable.  
\end{example}

\noindent 
Although the user can easily provide $t_{c}$ and $\emph{f}_{comp}$,
the repair function $\emph{f}_{repair}$ is hard to directly express, yet critical to the problem.   While users are free to provide a custom repair function, \sys  provides a good default: it fits a multi-level model~\cite{gelman2006data} to estimate the expected aggregate statistics for a given drill-down level. Multi-level models are widely used in fields including sociology~\cite{fernandez1981multilevel}, demography~\cite{sacco2005dynamic}, public health~\cite{diez2000multilevel}, and market sectors~\cite{van2015extrinsic} to analyze hierarchical data, and improve on linear models by accounting for both the deviation of observations within a group, and the deviation of a group from the other groups.
\sys also provides APIs to easily tune the model (described next).


\subsection{Model-based Repair}


\sys identifies erroneous groups by comparing their statistics with its expected statistics based on a model.  Models are commonly used to identify and repair numeric errors, and prior works have used log-linear~\cite{sarawagi1998discovery}, and linear regression models~\cite{miao2019going}.  Models provide the flexibility to combine features derived from the drill-down groups, as well as from auxiliary datasets (e.g., satellite sensing data in \Cref{exp:firstexample}).   Below, we illustrate the challenges of a model-based approach using the running example, and then discuss the two techniques \sys uses to address the challenge.

A key challenge is that there may not be enough groups as the result of a drill-down operation (e.g., villages in Ofla in 1998) to fit an accurate model.  We then describe how we use parallel groups to provide more training examples, and use multi-level models to account for variation across the parallel groups.  

\stitle{A Naive Approach}
is to use the results of a candidate drill-down as the training examples for a linear regression model $\mathbi{y} =\mathbi{X} \cdot \boldsymbol{\beta} + \pmb{\epsilon}$.  For instance, after drilling down from Ofla to its villages, $\mathbi{y}$ is the result of the complaint's aggregation function $f(\cdot)$\footnote{In general, the complaint's aggregate can be composed of multiple distributive aggregates.  In this case, we fit separate models for the  distributive aggregates.}  for each village, and $\mathbi{X}$ is the feature matrix derived from village-level information (e.g., population, crops, rainfalls).  The main problem is that Ofla alone may not contain enough villages to train an accurate model.

\stitle{Using Parallel Groups:}
\sys uses the drill-down results for all of the parallel groups (e.g., villages from other districts and years) in the dataset.  In the FIST example, there are 34 years and 295 villages, and using parallel groups increases the number of training examples to 10030.

\stitle{Multi-level Models:}  
The dataset's hierarchical structure naturally clusters the drill-down groups: villages in the arid Tigray region will be dissimilar from villages in the tropical Harari region.  This effect is common in fields such as sociology~\cite{fernandez1981multilevel}, demography~\cite{sacco2005dynamic}, public health~\cite{diez2000multilevel}, and market sectors~\cite{van2015extrinsic}.  Unfortunately, linear models do not take this hierarchical structure into account.

\sys uses multi-level linear models by default.   Multi-level models fit a set of global parameters, as well as separate parameters for each parent group (e.g., year, district)---termed ``clusters'' for convenience---to account for their variations.  
%
%
Suppose we are drilling down from clusters defined by $A_{gb}$ (e.g., year, district) to $A'_{gb}$ (e.g., year ,district, village), and there are $\mathcal{G}$ clusters.  The model for the $i^{th}$ cluster is defined as:
\begin{gather} \label{multimodelformula} 
  \mathbi{y}_\emph{i} =\mathbi{X}_\emph{i} \cdot  \boldsymbol{\beta} + \mathbi{Z}_\emph{i} \cdot  \boldsymbol{b}_\emph{i} + \pmb{\epsilon}_\emph{i}, \emph{i} = 1,..., \mathcal{G} \\
\boldsymbol{b}_\emph{i} \sim \mathcal{N}(\mathbi{0}, \pmb{\Sigma}), \pmb{\epsilon}_\emph{i}  \sim \mathcal{N}(\mathbi{0}, \sigma^2 \mathbi{I} ) \nonumber
\end{gather}
where $\mathbi{y}_\emph{i}$ is the vector of distributive aggregation $f(\cdot)$ result, and $\mathbi{X}_\emph{i}$ is the feature matrix. 
The key difference from the linear model is the additional term $\mathbi{Z}_\emph{i} \cdot \boldsymbol{b}_\emph{i}$, which encodes random effects that vary across clusters. 
It can be interpreted as modeling the residual after fitting the global parameters in $\mathbi{X}_\emph{i}\cdot\boldsymbol{\beta}$.    
$\mathbi{Z}_\emph{i}$ is the random effects, set to $\mathbi{X}_\emph{i}$ by default; $ \boldsymbol{b}_\emph{i}$ is the cluster-specific parameter drawn from a gaussian.  Since $\mathbi{X}_\emph{i}$ contains e.g., village, district, and year level attributes, $\mathbi{Z}_\emph{i}$ may be tuned to only keep attributes relevant within clusters.  
$\pmb{\epsilon}_\emph{i}$ is a cluster-specific error; \mathbi{I} is the identity; $\pmb{\beta}$, $\pmb{\Sigma}$ and $\sigma$ are parameters.  Finally, the full model is constructed by (logically) concatenating each cluster's matrices.

\subsection{Tuning the Repair Function}

An administrator or end user can programmatically tune the repair function by defining features in $\mathbi{X}$.
We first describe default, auxiliary, and custom features for $\mathbi{X}$, and then describe tuning the random effect matrix $\mathbi{Z}$. 
For simplicity, we assume that each attribute directly translates into a feature. The full details are discussed in \Cref{section:featureMatrix} and \Cref{section:mulf}.

\subsubsection{Default Features}
\sys treats all non-aggregation attributes in the drill-down results as categorical.  However, naive featurization by hot-one encoding the attributes would exacerbate the dataset sparsity and leads to low prediction accuracy in practice.  
Instead, we borrow from anomaly detection in multivariate data sets~\cite{laurikkala2000informal} and OLAP data cubes~\cite{sarawagi1998discovery}, and featurize attributes based on their {\it main effects}~\cite{marascuilo1987loglinear}.  We replace each categorical attribute value with the median $\mathbi{Y}$ of records with the value, and we center and normalize numeric attributes.  For instance, if we drill down to (district, village, year) and compute the \texttt{MEAN} statistic for each group, then the feature for the year 1985 would be the median of the mean severities across all villages in 1985.  By default, we treat all hierarchy attributes as categorical.

\subsubsection{Auxiliary Datasets}
Users can reference auxiliary datasets that can be joined with the drill-down results.  When the join is possible, \sys automatically joins and includes the auxiliary measures in the feature matrix.  For instance, the village rainfall data in \Cref{exp:firstexample} is included once \sys drills down to village.  To define an auxiliary dataset, users specify the dataset, join conditions, and its measure attributes.

\subsubsection{Custom Features}
Users can specify custom per-attribute featurizations.  To featurize attribute $A$, the user defines a function $q(A,\mathbi{Y})\to \{(a_i,v) | a_i\in A\land v\in\mathbb{R}\}$ that takes the attribute values and group statistics as input, and outputs a mapping from attribute value $a_i$ to feature value $v$.
This can express geographical clusters, temporal windows, and other distances.  For instance, the previous year's severity may be predictive of this year's.


\subsubsection{Random Effect Matrix}
The random effects $\mathbi{Z}_\emph{i}$ for the $i^{th}$ cluster are modeled using cluster-specific coefficients ${b}_\emph{i}$.  $\mathbi{Z}_\emph{i}$ defines the predictive features to use, and by default uses all features by setting $\mathbi{Z}_\emph{i} = \mathbi{X}_\emph{i}$.  Advanced users can tune the random effect matrix by excluding non-predictive features $\mathbi{Z}_\emph{i}$, and \sys will simply skip those attributes during matrix operations.  
For example, if users believe that rainfall does not vary by district nor year, they can exclude rainfall and all of its derived features will be ignored.

\subsection{Factorised Feature Matrix}
\label{featurematrixdis}

\begin{figure}
     \centering
     \begin{subfigure}[b]{0.11\textwidth}
         \centering
         
         \includegraphics[width=0.8\textwidth]{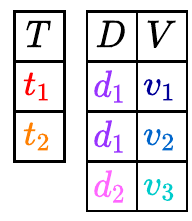}
         
         \caption{}
         \label{fig:hierarchies}
     \end{subfigure}
     \hfill
    \begin{subfigure}[b]{0.11\textwidth}
         \centering
         \includegraphics[width=0.7\textwidth]{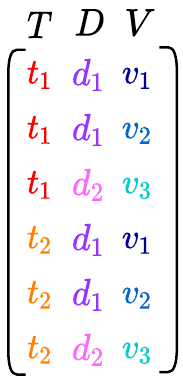}
         \caption{}
         \label{fig:attribute-matrix}
     \end{subfigure}
     \hfill
     \begin{subfigure}[b]{0.11\textwidth}
         \centering
         \includegraphics[width=0.7\textwidth]{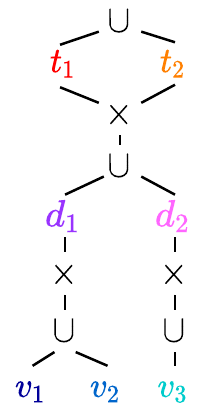}
         \caption{}
         \label{fig:factorisedrep}
     \end{subfigure}
     \hfill
    \caption{Example dataset with (a) Attribute values organized by hierarchy in attribute order, (b) materialized feature matrix, (c) factorised feature matrix.}
    \label{fig:exampledata}
\end{figure}

We now discuss how to construct the feature matrix using the example in \Cref{fig:exampledata}.   Rather than materialize the full matrix, we construct a factorised matrix representation\footnote{For legibility, we use attribute and feature interchangeably.  See \Cref{section:featureMatrix} for details.} in the form of a tree, where each node is node is either an attribute value, union ($\cup$), or cartesian product ($\times$) (see \Cref{backfrep}).  To do so, we must first assign an attribute ordering---matrices expect a fixed column order---that dictates the attributes encoded at each level in the f-representation.     

\stitle{Attribute Ordering:}
We order the attributes by selecting an ordering of the hierarchies, and within each hierarchy, order the attributes from least to most specific.    The specific hierarchy order has no impact on performance, since the f-representation of the matrix can be efficiently translated into a different ordering during matrix multiplications.    The main restriction is that the hierarchy that we are drilling down should be ordered last.  Note that drilling down along different hierarchies will necessitate different attribute orderings; we describe work-sharing optimizations in \Cref{ss:opt-drilldown}.

\Cref{fig:hierarchies} shows data from two hierarchies: Time with attribute $T$ and Geo with attributes District (D) and Village (V).
Suppose the hierarchy ordering is [Time, Geo].  The fully materialized matrix $\mathbi{X}$ (\Cref{fig:attribute-matrix}) is computed as the cross product between the two hierarchy tables.   Note the redundancy across the hierarchies ($t_i$ is replicated), and within the Geo hierarchy ($d_1$ is replicated).

\stitle{Factorised Feature Matrix:}
We now outline the construction of the factorised feature matrix, using \Cref{fig:factorisedrep} as the example.  We refer readers to Olteanu et al.~\cite{olteanu2015size} for a complete procedure\footnote{In their parlance, our ``f-tree'' does not contain branches.}.  At a high level, each attribute corresponds to one level of the tree, and $A_i$'s level is directly above $A_j$'s if $A_i$ directly  precedes $A_j$ in the attribute order.   Each node (e.g., $t_1$) in a level corresponds a distinct attribute value, and levels are connected via operators $\times$ and $\cup$.   The edge structure between levels is dictated by whether the attributes are within the same hierarchy or not.  In \Cref{fig:factorisedrep}, {\it Time} directly  precedes {\it District} but is in a separate hierarchy, thus the {\it District} nodes are unioned ($\cup$) and connected to the {\it Time} level with ($\times$).  In contrast,  attributes within the same hierarchy form a tree structure because villages are strictly partitioned by their district.  Notice that the example has removed the redundant instances of $t_1$, $t_2$, and $d_1$.  \Cref{section:factoriser} describes the detailed implementation.

Matrix operations loop through the matrix row- or column-wise.  These directly correspond to efficient traversals through the f-representation's tree structure.

\section{Details and Optimizations}\label{s:details}

In each iteration, \sys recommends the next drill-down hierarchy and returns the top ranked groups (output tuples of the drill-down query).  For each hierarchy, \sys builds the factorised feature matrix, fits the multi-level model, estimates the expected statistics for each group using the model, and finally ranks the groups by their repair's effects on the user complaint.  In this process, model training is the primary bottleneck.  

Unfortunately, matrix operations do not take factorized matrices as input.  
This section first decomposes matrix operations into collections of aggregation queries efficiently executable over f-representations.  We then develop a suite of work-sharing and caching optimizations to accelerate individual and multi-model training. Our experiments will show that directly operating over f-representations can provide multiple orders of magnitude speedups.

\subsection{EM-based Model Training}
\label{emtrain}
We fit the multi-level model's parameters via maximum likelihood estimation using expectation maximization (EM). EM is widely used to train multi-level models and implemented in statistical packages such as lme~\cite{rmanual} in R and statsmodels~\cite{seabold2010statsmodels} in Python.  In addition, our techniques apply to other algorithms (e.g., Fisher scoring~\cite{aitkin1986statistical}, iterative generalized least squares~\cite{goldstein1986multilevel}).

The EM algorithm (listed in \Cref{section:emalgo}) is composed of 3 types of matrix multiplications---gram matrix ($\mathbi{X}^T \cdot \mathbi{X}$), right multiplication ($\mathbi{X} \cdot \mathbi{A}$), left multiplication ($\mathbi{B} \cdot \mathbi{X}$)---along with their per-cluster counterparts: $\mathbi{X}_{\emph{i}}^T \cdot \mathbi{X}_{\emph{i}}$, $ \mathbi{X}_{\emph{i}} \cdot \mathbi{C}_{\emph{i}}$,  $\mathbi{D}_{\emph{i}} \cdot \mathbi{X}_{\emph{i}}$ for the $i^{th}$ cluster.  Where $\mathbi{A}, \mathbi{B}, \mathbi{C}_{\emph{i}}, \mathbi{D}_{\emph{i}}$ are intermediate matrices, and  $\mathbi{X}$ is the factorized matrix.

To compute these operations, one naive approach is to materialize the full $\mathbi{X}$ matrix and use existing matrix operator implementations, but the matrix can be very large.   Instead, we wish to directly perform matrix operations on the f-representation. Note that the {\it outputs} of Gram matrix, right and left multiplication are materialized as matrices because there is no redundancy to exploit.  

Prior work~\cite{schleich2020lmfao,schleich2016learning} focused on factorized matrices derived from join-only queries, and thus only required gram matrix computations for training.  In join-only queries, $\mathbi{Y}$ can be treated as yet another attribute whose cardinality is independent of the groupby attributes. In contrast, matrices in \sys are derived from join-aggregation queries, so $\mathbi{Y}$ is potentially unique for each of an exponential number of groups.  This requires our extensions to support left and right multiplications.


\subsection{Factorised Matrix Operations}
\label{section:matrxopoverfr}

Prior work~\cite{schleich2016learning,Kobis2017LearningDT,curtin2020rk} decomposes factorized matrix operations into a batch of aggregation queries that can be used to directly compute cells in the output matrix.  We first review the set of decomposed aggregations, and then describe our implementation of left and right multiplication that leverage the data's hierarchical structure.  We describe work-sharing based on early marginalization, and then describe our novel drill-down specific optimizations.


\subsubsection{Decomposed Aggregates}
\begin{figure}
  \centering
      \includegraphics [scale=0.6] {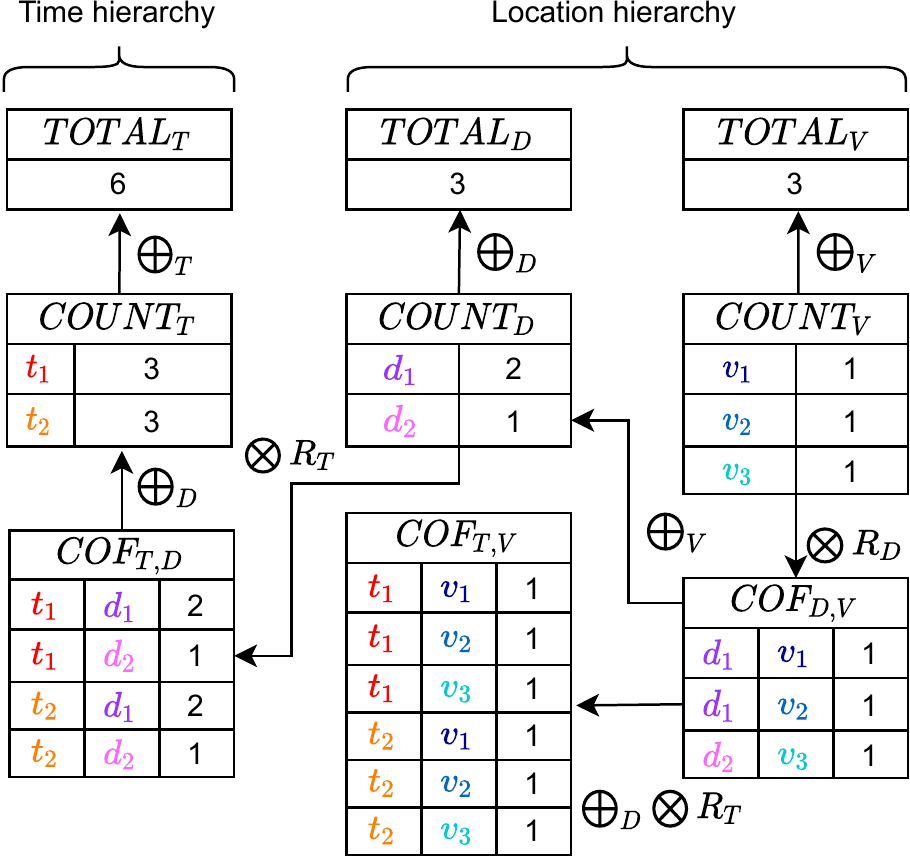}
  \caption{Aggregation results and Multi-query execution}
  \label{fig:multiexe}
\end{figure}

Let us define three classes of count aggregations, $\total_{A_i}$, $\cnt_{A_i}$, $\cof_{A_i,A_j}$, that will be used to define matrix operation outputs.  Recall that the feature matrix orders the attributes $A_n,\ldots,A_1$ by hierarchy, and from least to most specific attribute within each hierarchy.  Using the attribute order \emph{Time (T)}, \emph{District (D)}, \emph{Village (V)} in the running example, \Cref{fig:multiexe} illustrates the outputs of these aggregation queries and their algebraic relationships to each other.  

$\total_{A_i}$ marginalizes over all attributes to the right of $A_i$ (in attribute order), inclusive; it returns a single count value.
$\cnt_{A_i}$ marginalizes all attributes strictly to the right of $A_i$; it returns the count for every unique $A_i$ value.  
$\cof_{A_i,A_j}$ groups by $A_i$ and $A_j$ and computes the count for each group.  Formally:
\begin{align}
  \total_{A_i} = &\bigoplus_{A_{1}} \hdots \bigoplus_{A_{i}} \pi_{A_i}(R_i)\bigotimes_{i \in [i-1]}R_i\nonumber\\
  \cnt_{A_i} =& \bigoplus_{A_{1}} \hdots  \bigoplus_{A_{i-1}} \pi_{A_i}(R_i)\bigotimes_{i \in [i-1]}R_i\nonumber\\
  \cof_{A_i, A_j} = &\bigoplus_{A_{1}} \hdots \bigoplus_{A_{j-1}} \bigoplus_{A_{j+1}} \hdots  \bigoplus_{A_{i-1}} \pi_{A_i}(R_i)\bigotimes_{i \in [i-1]}R_i\nonumber\\
 &i\in[1,n], j\in[1,i-1]\nonumber
\end{align}
These queries can be naively executed by joining the relations together and then computing the aggregation.  The next subsection describes a multi-query optimization that pushes marginalization down, and reuses computation to minimize the intermediate join sizes.

\begin{figure}
     \centering
     \begin{subfigure}[b]{0.5\textwidth}
         \centering
         
         \includegraphics[width=\textwidth]{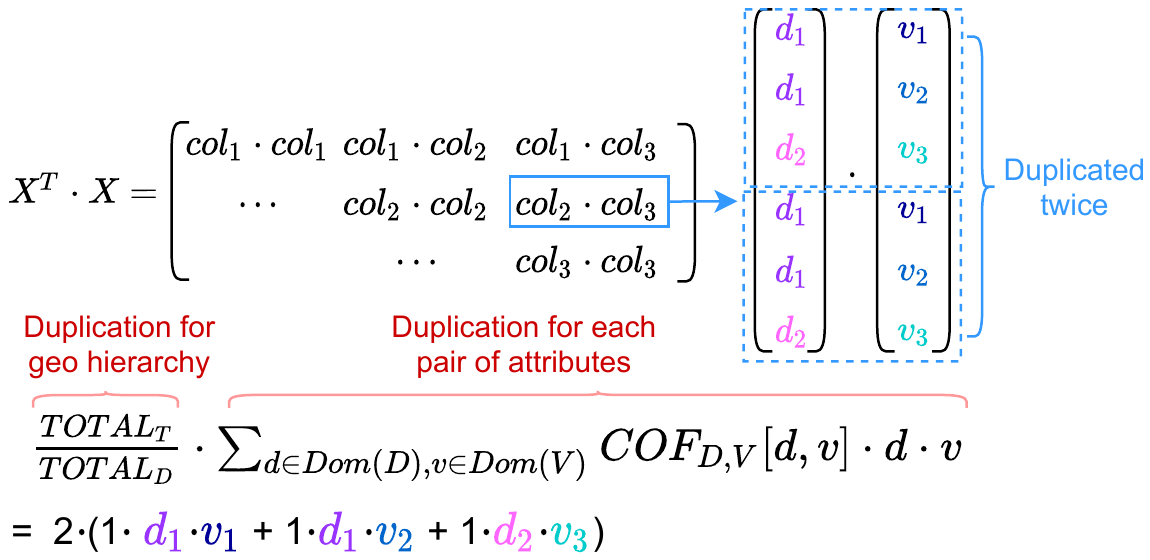}
         
         \caption{Gram Matrix}
         \label{fig:cofactorop}
     \end{subfigure}
     \hfill
     \begin{subfigure}[b]{0.5\textwidth}
         \centering
         \includegraphics[width=\textwidth]{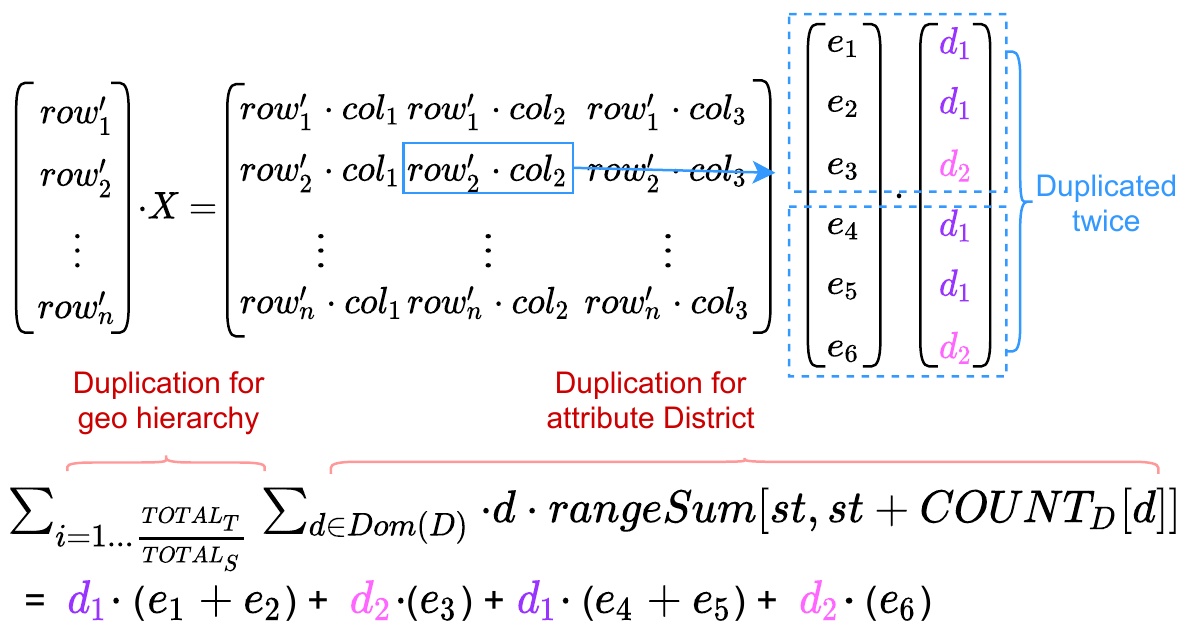}
         \caption{Left Multiplication}
         \label{fig:leftop}
     \end{subfigure}
     \hfill
    \begin{subfigure}[b]{0.45\textwidth}
         \centering
         \includegraphics[width=\textwidth]{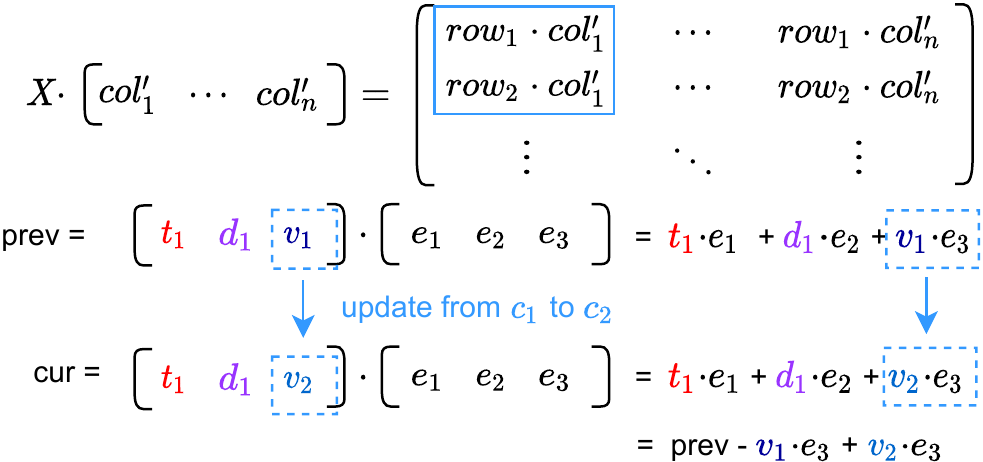}
         \caption{Right Multiplication}
         \label{fig:rightop}
     \end{subfigure}
     \hfill

    \caption{Example matrix operations.}
    \label{fig:exampleops}
\end{figure}

\subsubsection{Matrix Operations Using Decomposed Aggregates}
We now present the intuition for optimizing the most expensive three matrix operations---gram matrix, left and right multiplication---using data from the running example (\Cref{fig:exampledata}).  The key idea is to use the decomposed aggregates above to quantify the redundancy (i.e., duplication) in the vector dot product computations in each output matrix cell.  We also describe the main optimization for their per-cluster variants, and defer details to \Cref{section:matrixops}, and focus on the principles.  We note that the gram matrix implementation is the same as in~\cite{schleich2016learning}, but we introduce an optimization based on the data hierarchies.

\stitle{Gram Matrix:}  \Cref{fig:cofactorop} shows the main idea when computing the dot product between columns $col_2$ and $col_3$ in $\mathbi{X}$.  Since there are two times $t_1$ and $t_2$, the district and village data is duplicated twice.  Instead of recomputing them, we compute $\frac{\total_{T}}{\total_{D}}$ to infer the number of times $col_2\cdot col_3$ is duplicated, and $\cof_{D,V}$ to account for the number of times each pair of district, village values are duplicated.   
Our major optimization is to observe that 
$\cof_{A,B}$ 
is simply a cartesian product that does not need to be materialized when $A$ and $B$ are independent.  This is the case when $A$ and $B$ are from different hierarchies.

\stitle{Left Multiplication:} \Cref{fig:leftop} shows this between a materialized matrix and $\mathbi{X}$; let $row'_2$ contain elements $[e_1,\ldots,e_6]$.  To compute $row'_2 \cdot col_2$, the outer summation iterates over the districts values twice, once for each Times value $t_1,t_2$.  Within each iteration (e.g., $t_1$), each district value is multiplied by the sum of the corresponding elements in $row'_2$.   For instance, $d_1$ is multiplied by $e_1$ and $e_2$, while $d_2$ is multiplied by $e_3$.  Since $row'_2$ will be referenced for every column in $\mathbi{X}$, we preprocess $row'_2$ by computing the prefix sum, to allow for fast range summations (e.g., $rangeSum[0,2]=e_1+e_2$). \emph{st} is used to keep track of the start position of $row'_2$, and is updated for each range summation.


\stitle{Right Multiplication:}   This operation uses rows in $\mathbi{X}$, so cannot benefit from the techniques above.    \Cref{fig:rightop} shows how we leverage the observation that vertically adjacent rows in $\mathbi{X}$ have considerable overlap.  For instance, the only difference between $row_1$ and $row_2$ is the last value ($v_1\to v_2$).  Thus,  output of $row_2\cdot col'_1$ can be incrementally computed from the result of the preceeding row's dot product.


\stitle{Per-cluster Optimizations: }
The per-cluster variants use the same algorithms, albeit for the sub-matrices corresponding to the clusters.  Since clusters correspond to siblings in the f-representation (e.g., districts $d_1$ and $d_2$ in \Cref{fig:factorisedrep}), they are amenable to the same work sharing optimization as for right multiplication.  For instance, the first cluster (rows 1, 2) and second cluster (row 3) in \Cref{fig:factorisedrep} share $t_1$, and can cache $t_1$'s contributes to the matrix operation's output.  The details and experiments are in \Cref{section:matrix cluster}.

\subsection{Multi-Query Optimization}
Early marginalization~\cite{schleich2016learning} is applied to push aggregation operator through joins, and work sharing is used to compute decomposed aggregates \total, \cnt, and \cof. 
For instance, $\total_D$ is simply the sum of counts in $\cnt_D$.  Similarly, $\cof_{D,V}$ can be computed as $\cnt_D\bigotimes R_V$, or as $\cnt_V\bigotimes R_D$.  These relationships are depicted as edges in \Cref{fig:multiexe}.  Given the dependency graph, the aggregations are simply computed in topological order.
We use the same $\cof_{A,B}$ optimization as described for gram matrix above, and avoid materializing the cartesian product for attributes from different hierarchies.

\subsection{Drill-down Optimization}
\label{ss:opt-drilldown}


\begin{figure}
     \centering
     \begin{subfigure}[b]{0.22\textwidth}
         \centering
         
         \includegraphics[width=0.55\textwidth]{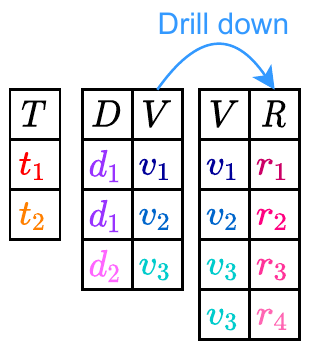}
         
         \caption{Updated Hierarchies}
         \label{fig:drillhie}
     \end{subfigure}
     \hfill
     \begin{subfigure}[b]{0.25\textwidth}
         \centering
         \includegraphics[width=0.8\textwidth]{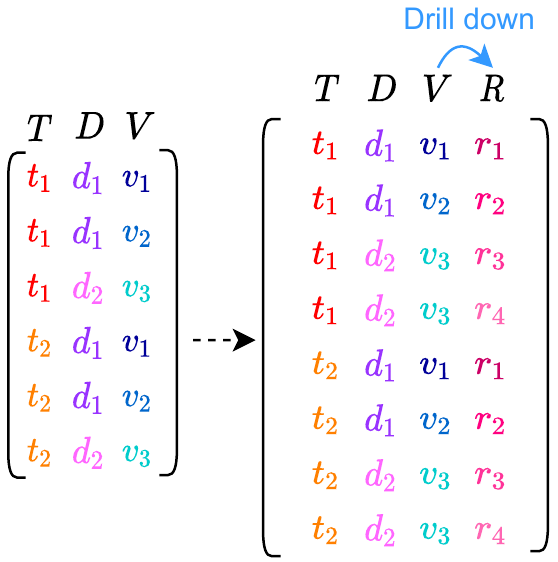}
         \caption{Updated Feature Matrix}
         \label{fig:drillfea}
     \end{subfigure}
     \hfill
    \begin{subfigure}[b]{0.45\textwidth}
         \centering
         \includegraphics[width=\textwidth]{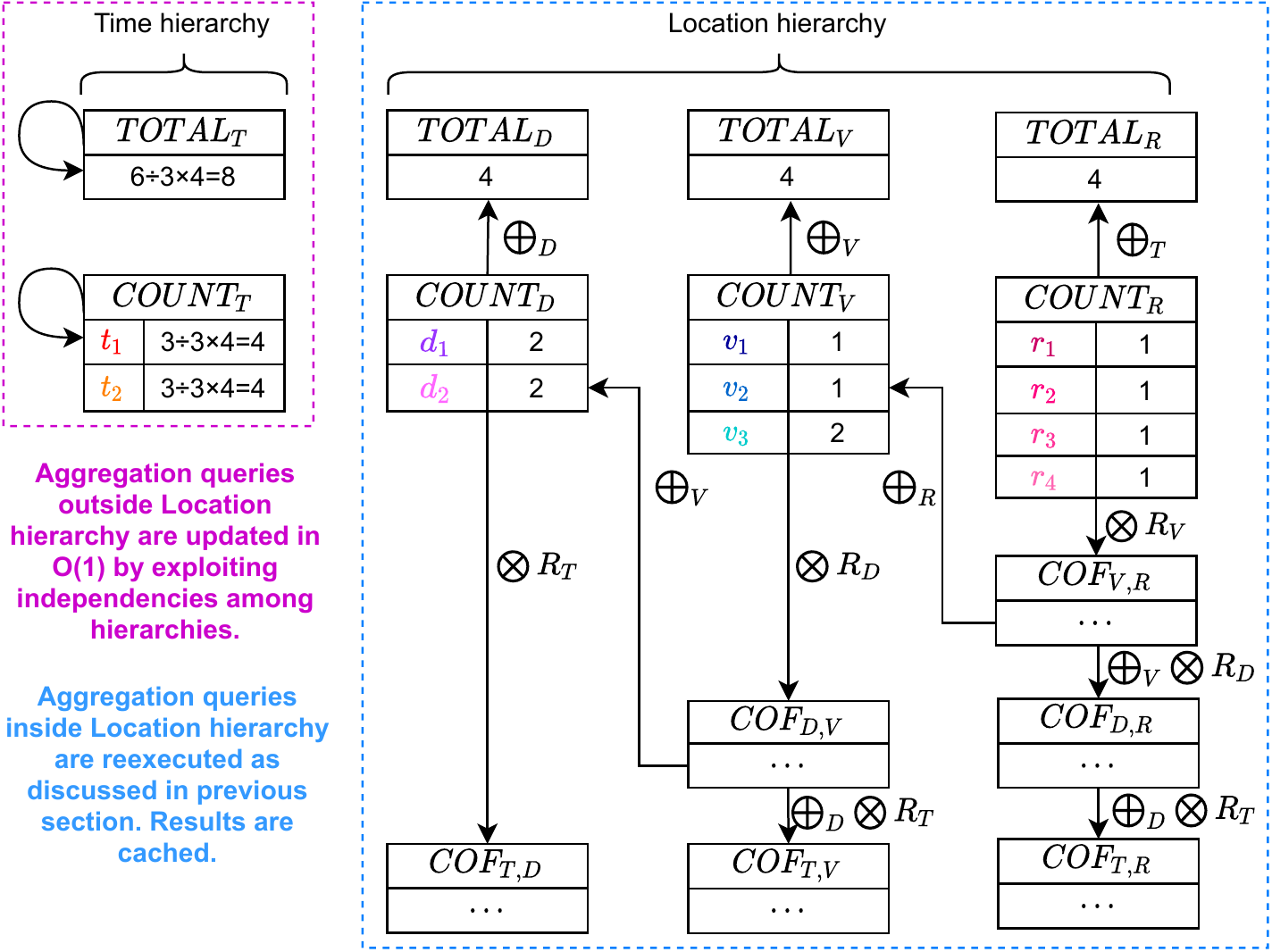}
         \caption{Updated Aggregation Results}
         \label{fig:drillmul}
     \end{subfigure}
     \hfill

    \caption{Example Updates after Drill-Down.}
    \label{fig:drilldownupdate}
\end{figure}

Equations \ref{querydefinition} and \ref{eq:allhiearchies} in the problem statement requires drilling down each hierarchy, and each drill-down augments the factorised feature matrix with the additional columns corresponding to the next attribute in the hierarchy.  For instance in \Cref{fig:drillhie}, the user further drills down along the geography hierarchy from Village (V) to Road (R), which expands the feature matrix (\Cref{fig:drillfea}).
Notice that after drilling down to Road, the multiplicities of the preceding attributes change---$t_1$ is duplicated 4 rather than 3 times, and $v_3$ is duplicated twice for each $r_i$ value.  Although the decomposed aggregates for the attributes in the drill-down hierarchy (D, V, R) need to be recomputed (using the multi-query optimizations in the previous subsection), we can update each of the decomposed aggregates for attributes in the other hierarchies in O(1).  For instance, the multiplicity for $t_1$ can be updated by dividing by the current $\total_D$ (e.g., 3) and multiplying by the updated $\total_D$ after the drill-down (e.g., 4).  This is based on the observation that attributes between different hierarchies are independent.    \Cref{fig:drillmul} depicts the updated aggregates in the example.

The user will ultimately pick one drill-down hierarchy that \sys recommends (e.g., Time).  However, the next call to \sys would need to re-evaluate all hierarchies again, and we cache decomposed aggregates to accelerate this case. See \Cref{section:drilldown} for full details.

Drilldown optimization differs from incremental view maintenance (IVM) for f-representations~\cite{nikolic2018incremental}. IVM updates query outputs assuming that the input update size $O(\Delta)$ is smaller than the relation size $O(n)$.  However, during drill-down, the decomposed aggregates of the other attributes all change due to the new attribute. Thus, $\Delta\approx n$ and does not benefit from IVM.

\subsection{Putting It All Together}

\sys performs the following operations in each iteration to recommend the most promising drill-down results that will repair the user's complaint.  For each candidate hierarchy $H$, it 1) constructs the factorised feature matrix after drill-down, 2) recomputes the decomposed aggregates for the attributes in $H$ with multi-query optimizations, 3) updates each of the remaining decomposed aggregates in constant time, 4) translates EM into matrix operations that are executed until parameter convergence, 5) repairs each drill-down group based on the model prediction and incrementally updates the complaint to check the extent it is resolved.

\section{Experiments}
\label{sec:exp}

We now evaluate the effectiveness of our optimizations, and assess \sys's ability to identify group-wise data errors such as missing data or systematic corruptions.
One challenge with proposing a new interactive cleaning method is the lack of existing benchmarks.  Thus we evaluate runtimes using both synthetic data and complaints, and real-world case studies. The first case study is based on known and resolved errors in COVID-19 data, and the second is based on an expert user study with FIST data and team members.

\sys is implemented in C++ \footnote{https://github.com/zachary62/Dynamic-F-tree}.
 All experiments are run single-threaded on a Macbook Pro with 1.4 GHz Quad-Core Intel Core i5, 8 GB 2133 Mhz LPDDDR3 memory, and 256GB SSD. 
All the experiments fit and run in memory. 

\subsection{Performance Evaluation}
Given a complaint, \sys enumerates and drills down on each hierarchy, computes decomposed aggregates, builds the (factorized) feature matrix, trains the multi-level model, and ranks the groups.    We first evaluate individual steps---the effectiveness of factorized matrix operations (\Cref{exp:matrixop}), the cost of computing decomposed aggregates as compared to prior work (\Cref{exp:mulquerydownopt}), and the drill-down optimizations (\Cref{exp:drilldownopt})---and then evaluate end-to-end run times on two real-world datasets (\Cref{sec:endtoend}).

\stitle{Default Setup:}
The attributes in the input relations are organized into hierarchies.  The synthetic datasets vary the number of hierarchies (default: $\emph{d} = 3$) and number of attributes in each hierarchy (default: $\emph{t} = 3$).  By default, each attribute contains $\emph{w} = 10^6$ unique values.
The data is in BCNF and sorted.  Since we only report runtimes, we run \sys to completion and return a random group.


\subsubsection{Factorized Matrix Operations}
\label{exp:matrixop}

The number of hierarchies $d$ dictates the size of the feature matrix \mathbi{X}: exponential in the number of rows, and linear in the number of columns.  Thus, the matrix materialization cost and gram matrix costs are exponential in $d$.  In contrast, the factorized representation is linear in $d$.

We measure runtimes for matrix materialization, gram matrix, and left and right multiplication.  The former compares the full and factorized matrix construction, while the latter three compares the Lapack\cite{laug} implementations over the full feature matrix with \sys's factorized implementation. Lapack is a heavily optimized and widely used linear algebra library.  To minimize the effects of our multi-query optimizations,  each hierarchy is configured with only one attribute that has cardinality $\emph{w} = 10$.
Thus, the shape of $\mathbi{X}$ is $\emph{w}^\emph{d} \times \emph{t} \cdot \emph{d} = 10^\emph{d} \times 3 \cdot \emph{d}$.

\begin{figure}
  \centering
    \includegraphics[width=\columnwidth]{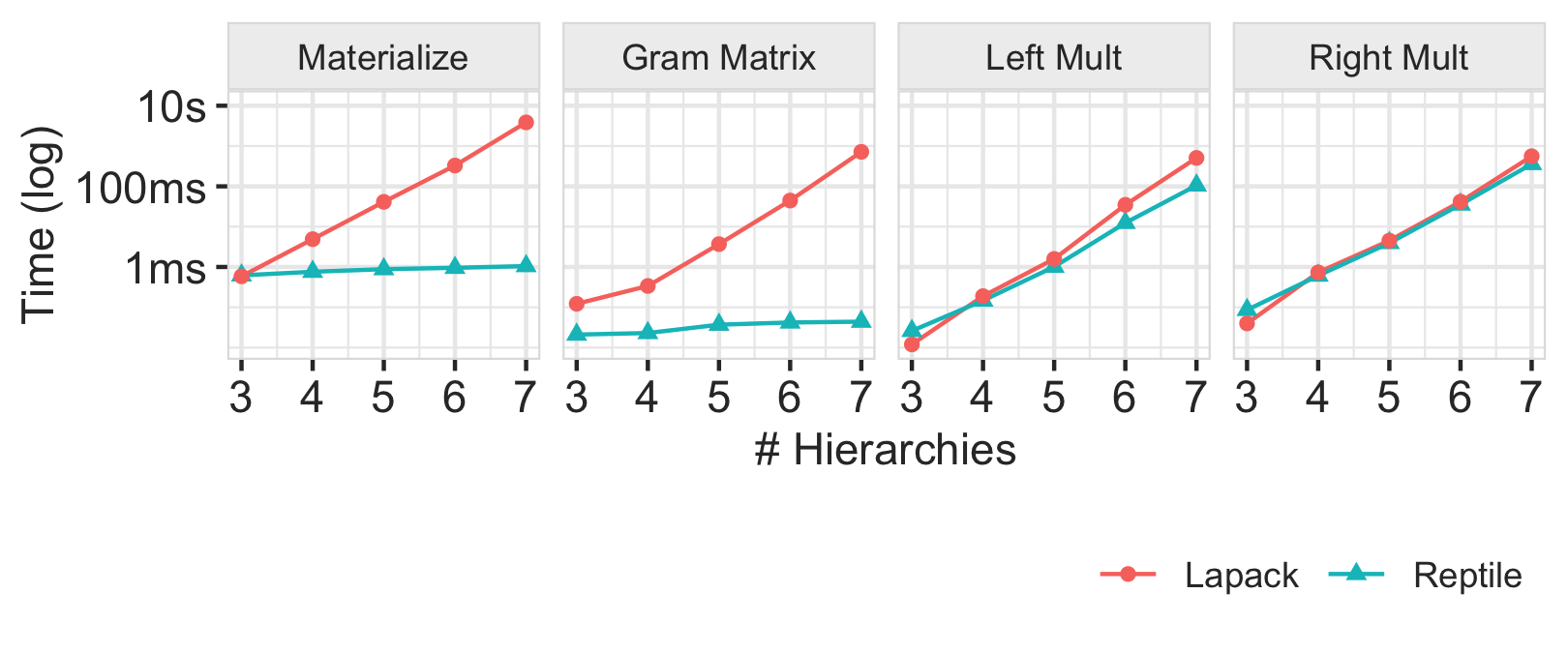}
  \caption{Matrix operation runtimes compared to Lapack-based implementation.}
  \label{fig:matrixops}
\end{figure}

\Cref{fig:matrixops} reports runtimes in log scale.   Materialization and gram matrix are exponential as a consequence of the matrix size, and the factorized implementations reduce the costs to linear.  For left multiplication, we use a random $1\times10^\emph{d}$ matrix as input; the size of the random matrix dominates the cost, thus both methods increase exponentially.   At 7 hierarchies, \sys is $5\times$ faster by exploiting redundancies in the matrix and the range sum optimization.  For right multiplication, we use a random $3\cdot \emph{d} \times 1$ matrix. The runtime again grows exponentially due to the size of the output matrix (which is fully materialized due to the lack of inherent redundancy).  At 7 hierarchies, \sys is $1.6\times$ faster by exploiting overlaps between vertically adjacent rows.

\subsubsection{Multi-query execution}
\label{exp:mulquerydownopt}
We now evaluate the benefits of our work-sharing multi-query optimizations for computing the decomposed aggregates $\cnt$, $\cof$, and $\total$.    We compare against \texttt{LMFAO}\cite{schleich2020lmfao}, which is the state-of-art factorised batch aggregation engine implemented in C++\footnote{https://github.com/fdbresearch/LMFAO}. Its current implementation only supports computing $\cnt$ and $\cof$ (as a by-product of computing the gram matrix), thus we use $\cnt$ and gram matrix in the benchmark. In addition, LMFAO computes $\cnt$ and the gram matrix serially, while \sys shares their work, however this is simply an implementation detail that has minor benefits.  $\total$ is quickly computed by scanning $\cnt$ so we disregard it.

\begin{figure}
  \centering
      \includegraphics [width=\columnwidth] {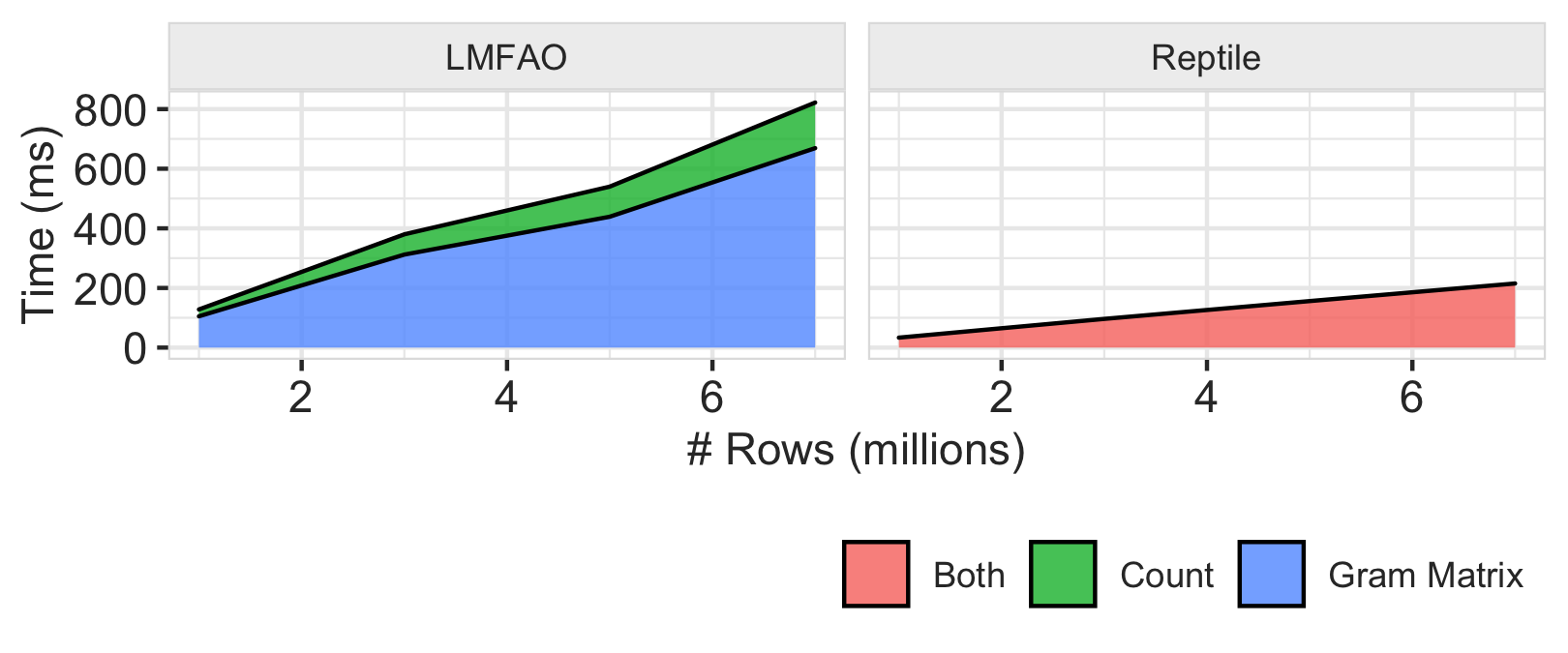}
	  \caption{Multi-query execution}
  \label{fig:queryexe}
\end{figure}

Since join cost is the bottleneck, we vary the cardinality for the attributes along the x-axis.    \Cref{fig:queryexe} shows that \sys is over $4\times$ faster than \texttt{LMFAO}.  The primary reduction is due to our optimizations based on independence between hierarchies.

\subsubsection{Drill-Down Optimization}
\label{exp:drilldownopt}
We test the work-sharing of multi-query optimizations between multiple invocations of \sys.
\sys uses hierarchy independence to update the non-drill-down hierarchies in constant time. Thus, two hierarchies, $A=[A_1,\ldots,A_6]$ and $B=[B_1,\ldots,B_6]$ are sufficient to characterize the drill-down costs and optimizations.   
In addition, the number of decomposed aggregates to compute is quadratic in the number of attributes that have already been drilled down upon.  For instance, if we drill down from $A_2$ to $A_3$, after already drilling down to $B_2$.   

For these reasons, we measure the cost of computing decomposed aggregates for each hierarchy by invoking \sys three times, where we pick $A$ to drill down each time.    We assume that for hierarchy A, we have already drilled down to $A_3$, and for hierarchy B, we have already drilled down $\emph{n}=3,4,5$ attributes.  We compare \texttt{Static}, which recomputes decomposed aggregates for each query, \texttt{Dynamic} which exploits the independence between hierarchies, and \texttt{Cache + Dynamic}, which further reuses cached results from hierarchies not drilled down.
 
\begin{figure}
  \centering
      \includegraphics [width=\columnwidth] {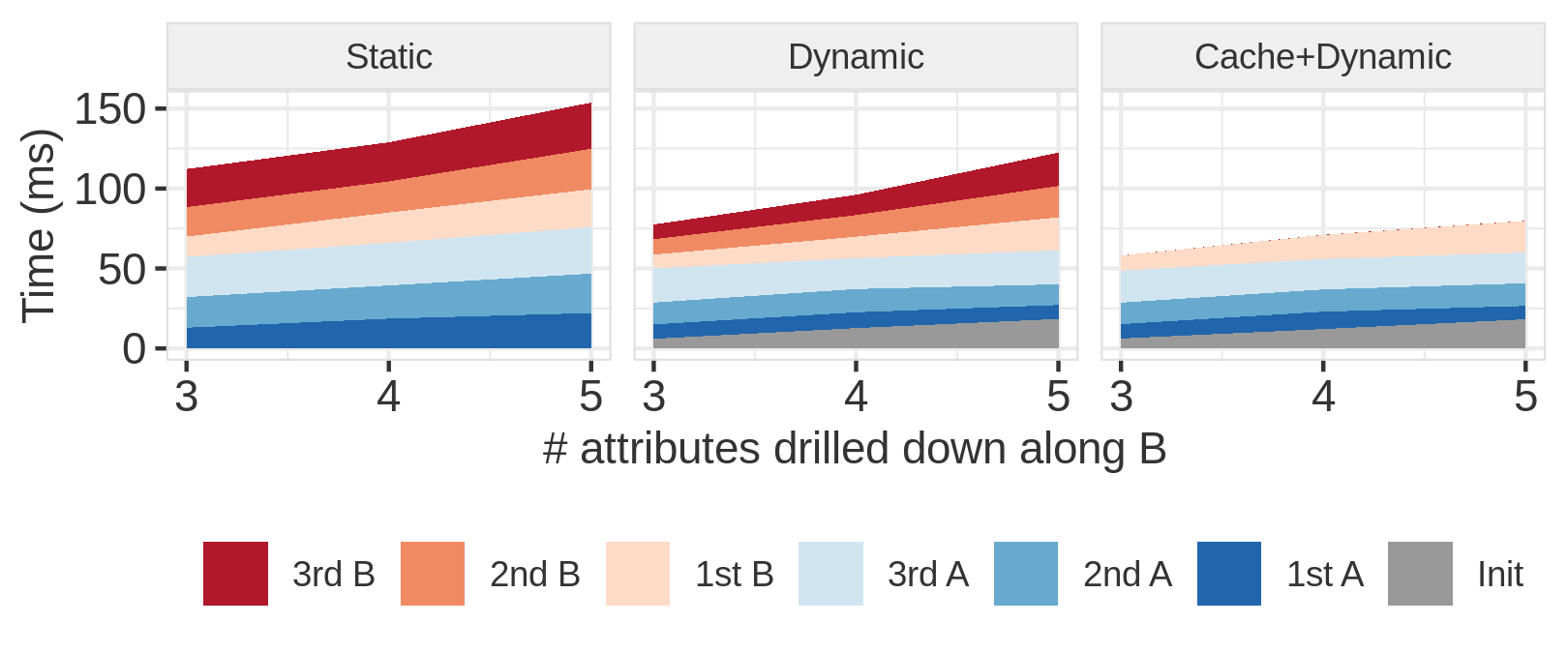}
  \caption{Drill-Down Optimization}
  \label{ftreeupdate}
\end{figure}

\Cref{ftreeupdate} varies the number of attributes already drilled down along hierarchy B in the x-axis.  For the areas, $3rd B$ means the cost of update hierarchy B's decomposed aggregates during the third invocation of \sys.  The gray area corresponds to the initial cost of computing the aggregates.  The lines are stacked to show total runtimes to run \sys.  \texttt{Dynamic} is $>1.2\times$ faster than \texttt{Static} by updating independent hierarchies more efficiently, while adding caching eliminates the cost of $2nd B$ and $3rd B$, since their aggregates were computed and cached in the first \sys invocation.


\begin{figure}[h]
  \centering
      \includegraphics [width=0.8\columnwidth] {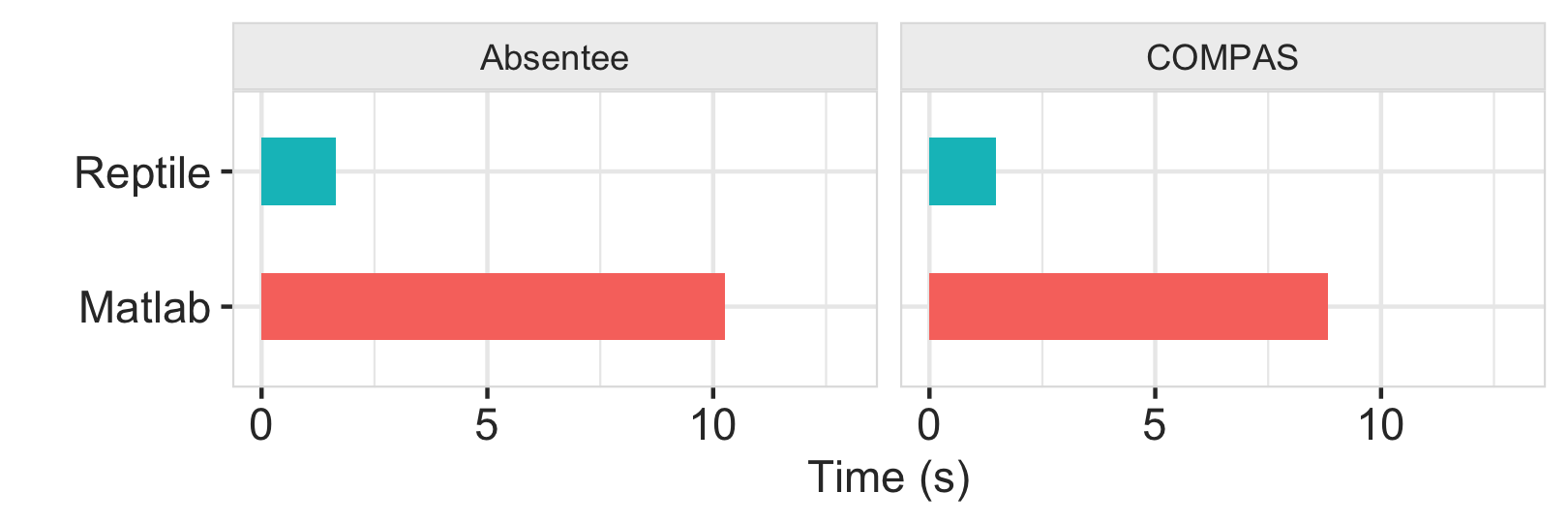}
  \caption{ End to end runtime evaluation on real datasets.}
  \label{fig:endtoend}
\end{figure}

\subsubsection{End-to-end Runtime Evaluation}
\label{sec:endtoend}
Finally, we report end-to-end runtime using two popular real-world analysis datasets.

\textbf{Absentee}\footnote{ https://www.ncsbe.gov/results-data/absentee-data}: 
there are 179K records of North Carolina absentee voting data for 2020.  We explore 4 hierarchies with one attribute each: county (100 unique values), party (6), week (53), gender (3). 
We invoke \sys 4 times.  Since we focus on runtime and not accuracy, we arbitrarily pick a sequence of drill-down attributes: county, party, week, gender.

\textbf{COMPAS}\footnote{ https://www.propublica.org/datastore/dataset/compas-recidivism-risk-score-data-and-analysis}: 
there are 60,843 records of defendant recidivism risk scores.
We explore 4 hierarchies. Time hierarchy has 3 attributes (year, month and day; 704 unique days in total), and the remaining have one attribute each: age (3 ranges), race (6), and charge degree (3).  We invoke \sys 6 times, in the arbitrary drill-down attribute order: year, month, day, age range, race, charge degree.

For both datasets, the initial complaint is that the overall \texttt{COUNT} is too high, and the models are trained using 20 EM iterations. 
Figure \ref{fig:endtoend} shows \sys is over $6\times$ faster than \matlab~\cite{MATLAB:2010}, which internally uses \texttt{Lapack} to train over the full materialized feature matrix.  This is largely due 
to avoiding full feature matrix materialization and exploiting its high degree of redundancy through work sharing optimization. 

These results also illustrate the limitations of using f-representations in ML.  Although matrix materialization and gram matrix are linear in the number of attributes, left and right multiplication remain exponential because the predicted variable $\mathbi{y}$ is an aggregate statistic that varies by group.  Our evaluation studies the worst-case scenario where the parallel groups include all exponential number of drill-down groups (even empty groups).  A simple optimization may be to only sample or truncate the number of parallel groups to train over, or to prune empty groups.

%

\subsection{
Explanation Accuracy: Synthetic Data}
\label{ss:accsynth}

\sys is unique in that it leverages complaints, hierarchical data, and multi-level models to identify group-wise data errors.
We now evaluate and show the value of each of these design decisions via an ablation study, and also compare against two alternative approaches based on prior work.
We use synthetic data to tune the problem difficulty and ensure a ground truth error.

\subsubsection{Setup}

In each \sys invocation, the user picks the group statistic and \sys selects top groups from the set of potential drill-downs.   Thus, we designed the minimal experiment to evaluate how accurately \sys can pick from the set of candidate drill-down groups.  We generate a dataset with one dimension attribute (i.e., one hierarchy) that has $100$ unique values (and thus 100 groups), and one measure for computing aggregates.  The number of rows in each group is drawn from a normal distribution $\mathcal{N}(100, 20)$, and each measure value is drawn from $\mathcal{N}(100, 20)$.  In each experiment,  we generate 1000 datasets and report the average accuracy of the top group.

\stitle{Auxiliary Data:}
\sys is able to combine auxiliary data provided by domain experts which has a (potentially weak) correlation to the correct aggregate statistics.  
We simulate this by generating one auxiliary table for each aggregate statistic (\texttt{COUNT}, \texttt{MEAN}, \texttt{STD}), where \texttt{STD} is only used when evaluating the \texttt{Raw} condition described below.  The auxiliary table contains the same dimension attribute in original dataset, and one measure which is correlated ($\rho\in [0.6-1.0]$) with the aggregate statistic. To generate correlated random variables, we use the procedure proposed by Iman and Conover \cite{iman1982distribution}.


\stitle{Error Generation:}
We randomly chose a group to be erroneous, and introduced different classes of errors: missing/duplicate records to change the \texttt{COUNT} statistics, and data drift~\cite{barddal2017survey} to change the \texttt{MEAN} statistic.
For the former, half of rows are deleted (\texttt{Missing}) or duplicated (\texttt{Dup}). For data drift, we either increase ($\uparrow$) or decrease ($\downarrow$)  all measure values in the group by 5 to simulate a subtle systematic value error.  We consider each error type individually, and in combination (\texttt{\bf Missing + $\downarrow$} and \texttt{\bf Dup + $\uparrow$)}).
We submit \texttt{COUNT} and \texttt{MEAN} complaints for the individual \texttt{COUNT} and \texttt{MEAN} errors; we use $\texttt{SUM} = \texttt{MEAN} \times \texttt{COUNT}$  complaints for the combination errors.

\stitle{Approaches:}
Five approaches are used to identify the erroneous group: \sys, \texttt{Outlier}, \texttt{Raw}, 
\texttt{Sensitivity} \cite{wu2013scorpion} and \texttt{Support}. \sys utilizes the auxiliary data to repair the dataset and recommends one group which, after repaired, best resolves the complaint.
\texttt{Outlier} ignores the complaint and simply returns the group whose statistics most deviates from the model's prediction.
\texttt{Raw} is a record-level bottom up approach based on Winsorization~\cite{lien2005regression}.  For each drill-down group, it computes the mean and standard deviation of the measure attribute within the group, and clips each input row's measure to [\texttt{MEAN} - \texttt{STD}, \texttt{MEAN} + \texttt{STD}].  \texttt{Raw} then returns the group whose clipping-based repairs best resolves the complaint. 
Finally, we compare with two prior explanation approaches: \texttt{Sensitivity} is based on interventional deletions~\cite{wu2013scorpion}, and recommends the group which, after deleting all rows, best resolves the complaint.
\texttt{Support} is a density-based approach that returns the fraction of rows in a drill-down row.  It is commonly used as a pruning criterion in explanation systems~\cite{abuzaid2020diff}.  In this experiment, this amounts to recommending the group with the largest \texttt{COUNT} (i.e., support) as there is only one dimension attribute.

\begin{figure}
  \centering
      \includegraphics [width=\columnwidth] {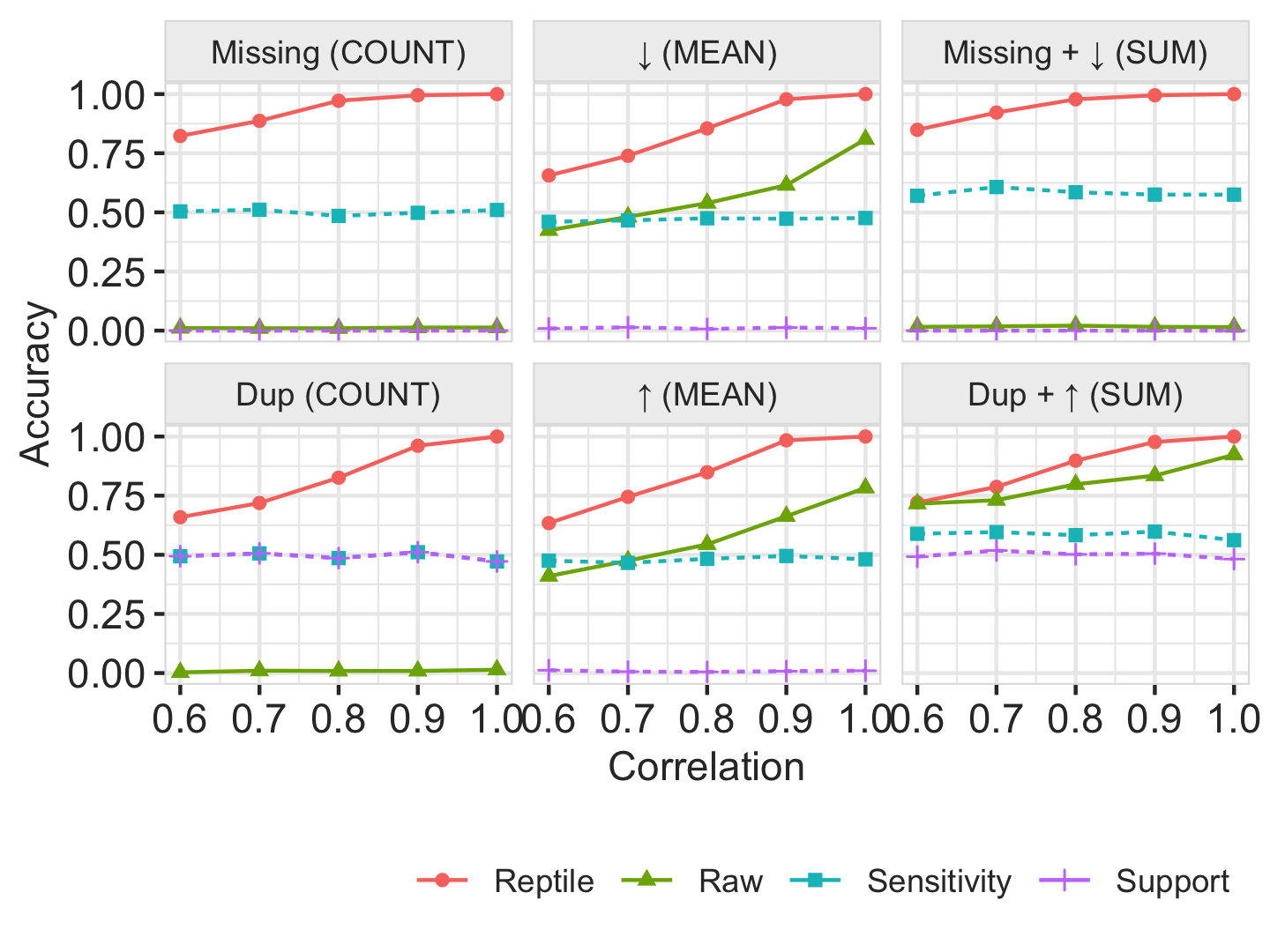}
      \caption{Accuracy comparison with naive approaches and prior work.  $\uparrow$ is Increase, $\downarrow$ is Decrease, and Dup is Duplication. The complained aggregation is in the parentheses.}
  \label{fig:crit}
\end{figure}

\subsubsection{Baselines}
\label{exp:expapproach}
We first evaluate \sys against \texttt{Raw}, and the two prior approaches (\texttt{Outlier} is defered next).
\Cref{fig:crit} varies the correlation of the auxiliary data (x-axis) for the different error types (columns).
\texttt{Raw} fails to detect missing/duplicated records because repairing at raw data level can’t capture these errors. \texttt{Raw} performs well for Duplication + Increase but poorly for Missing + Decrease because \texttt{SUM} is sensitive to group with a larger number of rows: \texttt{Raw} searches for group which after drifting its values back best resolves the complaint, and for group with duplicated rows, the drift has more impact to \texttt{SUM} so that it is more likely to be recommended. 
\texttt{Sensitivity} and \texttt{Support} are flat because they do not leverage auxiliary data. 
\texttt{Sensitivity} fails to leverage auxiliary data, so its recommendation is less reliable. 
\texttt{Support} only performs well under duplication because it is density-based and is designed for the complaint ``\texttt{COUNT} is high''.   
\sys is considerably and consistently more accurate, and successfully leverages the auxiliary data even when the correlation is weak.

\begin{figure}
  \centering
      \includegraphics [width=\columnwidth] {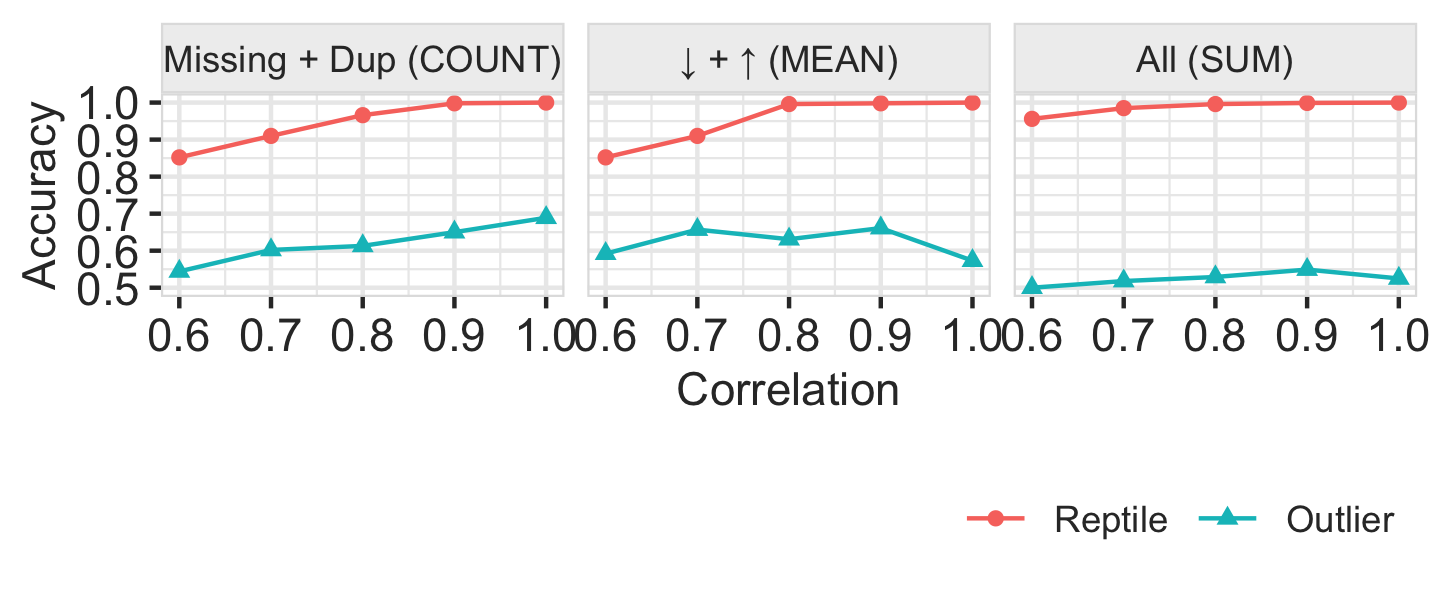}
      \caption{Accuracy comparison for multiple errors.}
  \label{fig:crit_multiple}
\end{figure}

\subsubsection{Complaint Ablation}
\label{exp:comp}

We now compare \texttt{Outlier} with \sys to study the value of leveraging complaints to distinguish between different possible errors.
To do so, we choose two groups whose error affects the complaint (true errors), and one group whose error does not (false positive).   We generate three conditions: 
\texttt{\bf Missing + Duplication} corrupts two groups with missing records (the correct errors), and one group with duplication. The complaint is ``\texttt{COUNT} is low''. 
 \texttt{\bf $\downarrow$ + $\uparrow$} introduces data drift to two groups by decreasing their measure values  (the correct errors) and one group by increasing the measure values. The complaint is ``\texttt{MEAN} is low''. 
 \texttt{\bf All} corrupts two groups by decreasing the measure values and causing missing records. For the false-positive group, we increase the measure values and introduce duplicates. The complaint is ``SUM is low''. 

\Cref{fig:crit_multiple} shows that the complaint direction is critical to distinguishing between multiple error candidates. As expected, increasing the correlation of the auxiliary dataset helps better predict the true group statistics, however \texttt{Outlier} hovers between 50-70\% accuracy because, while \texttt{Outlier} is able to identify three imputed groups, it cannot distinguish between them. Given that only two of them are correct, the accuracy of \texttt{Outlier} is bounded by 66\%.

\subsection{Case Study: COVID-19}
The COVID-19 data~\cite{dong2020interactive} maintained by the Johns Hopkins University Center for Systems Science and Engineering (JHU CSSE) contains two datasets.
The US data contains 1,175,680 rows, location (state, county) and time (day) hierarchies, and count measures for confirmed infections and deaths.
The global data 96,096 rows, location (country, state) and time (day) hierarchies, and measures for confirmed infections, deaths, and recoveries. Most statistics are reported at the country level (state is null), however large countries excluding the U.S. (e.g., Australia, Canada, China) report province/state-level statistics.




\stitle{Setup:}
The dataset's Github issues report a variety of data errors that have been confirmed and resolved, and we use errors resolved between \texttt{12/2/2020} and \texttt{1/27/2021} as ground truths for evaluation.  We generate a corrupt dataset for each issue, submit a complaint, and compare methods to identify the erroneous location.

Most issues are due to missing data on a specific day, which causes underreporting. Others may be due to backlogged reports (e.g., on days $n-m$) that are totaled and reported on day $m+1$, or due to changes in a location's reporting methodology\footnote{\tiny\url{https://www.azdhs.gov/preparedness/epidemiology-disease-control/infectious-disease-epidemiology/covid-19/dashboards/}}.
We use 16 (14) issues from the US (global) datasets and construct a complaint for each issue.  To do so, we first filter by the complaint's day, aggregate the total statistics at the immediately higher geographical level (e.g, if the error is in New York, then we compute the US aggregate), and then specify whether the result is too high or too low based on the ground truth.   For instance,  Texas under-reported infections on 1/21/2021, thus the complaint is that the total US cases on that day is too low. 
We compare \sys with \texttt{Sensitivity}~\cite{wu2013scorpion} and \texttt{Support} (described in \Cref{ss:accsynth}).

\begin{figure}
     \centering
     \begin{subfigure}[b]{0.23\textwidth}
         \centering
         
         \includegraphics[width=1\textwidth]{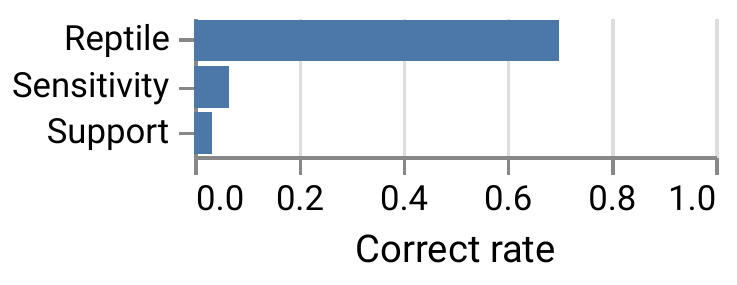}
         
         \caption{\centering Avg. accuracy of top result.}
         \label{fig:covidcorrect}
     \end{subfigure}
     \hfill
     \begin{subfigure}[b]{0.23\textwidth}
         \centering
         \includegraphics[width=1\textwidth]{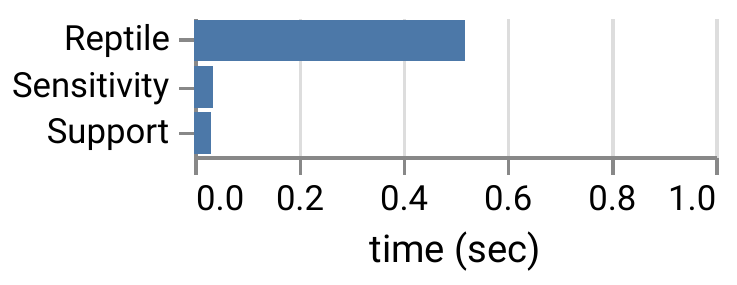}
         \caption{\centering Average Runtime}
         \label{fig:covidtime}
     \end{subfigure}
     \hfill
     \hfill
    \caption{COVID-19 Case Study}
    \label{fig:covidcase}
\end{figure}

\stitle{Results and Error Analysis:}
\Cref{fig:covidcase} shows that although \sys takes longer in order to fit models (${\approx}0.5s$), it is considerably more accurate ($70\%$) as compared to the baselines ($6.6\%$ for \texttt{Sensitivity}, $3.3\%$ for \texttt{Support}).   We conducted an error analysis of the 9 errors that \sys did not identify.  
The first type (5 issues) is due to minor data drift across several weeks (e.g. a missing data source) that is later fixed. For instance, Quebec's death statistics between 3/17/2020 to 1/27/2021 were all increased by a small amount, however the date range included all Quebec data in the experiment.
The second type (4 issues) are subtle issues smaller than the natural variation in the data, and unlikely to result in complaints. For example, on 12/18/2020, Washinton state submitted 21,308 instead of 21,038.    \Cref{section:coviddetail} discusses the experiment in more detail.


\subsection{Case Study: FIST}
The Columbia University Financial Instruments Sector Team (FIST) group collects Ethiopian farmer-reported drought data to design drought insurance~\cite{osgood2018farmer}. 
The data contains geography (Region, District, Village) and time (Year) hierarchies, and a severity measure from 1 (low severity) to 10 (high). The FIST group historically performed manual data cleaning based on domain expertise and by cross-referencing (noisy) external data sources (e.g., satellite estimates).  
We recruited 3 FIST team members\footnote{We attempted to recruit novice users, however they could not interpret the domain-specific data.} to use the system to submit complaints based on their experience, help verify the correctness of the results, and provide qualitative feedback (see \Cref{section:userstudydetail} for screenshots, protocol, and further details).  \textbf{Overall, \sys correctly identified errors for 20 out of 22 complaints.}


\stitle{Protocol and Complaints:}
Users are shown visualizations of annual Region-level statistics (count, mean, standard deviation).  They click on suspicious statistics to create a complaint.  \sys recommends drill-downs and highlights the candidate group in the drill-down results.  They can continue this process until they examine individual records to conclude whether the recommendations were correct.  We ask users to follow a think-aloud protocol and share their interpretation throughout the cleaning process. 
Example complaints (and their rationale) include:  ``the \texttt{MEAN} in Tigray 2009 should be much higher because I remember farmers argued about this year (P1)'', and ``the \texttt{STD} in Medebay Zana 2018 is too high compared to other years (P2)''.  

\stitle{Results and Failure Analysis:} The users accepted 20 out of the 22 submitted complaints.  These errors revealed issues such as: farmers that confuse planting and harvesting
years (e.g. plant in one year, but harvest in the next year), misremember the events, report non-drought years as highly severe, and more.   One failed complaint was due to inherent ambiguity and team members disagreed about the causes.  The second was because a unique combination of two districts needed to be fixed together, but \sys only return one of the two. See more details in \Cref{section:userstudydetail}.

\stitle{Qualitative Results and Discussion:} FIST users said that \sys ``is valuable to clean and make sense of this massive data (P3).'',  ``is helping to save the day for the project in Ethiopia during this year of Covid and civil strife (P1).'' 
A major benefit is to automate group-level inspection and cross-reference with external data sources.  
P3 stated that ``previously, we only had 5 villages in the Amhara region ... and data is cleaned manually using excel spreadsheet ... Now the project has scaled and we have 173 villages in Amhara. It is not possible to visit all these villages (P3).''
Finally, users suggested that ``it would be great [to] understand why the model makes certain prediction (P1),'' and  ``I hope there are more flexible visualizations that display different satellite data in one geographic map (P2).''

\section{Related Work}\label{s:related}


\stitle{Error Detection:}
Error detection traditionally uses integrity constraints~\cite{chu2013discovering} to find violations, while quantitative error detection often relies on statistical methods (e.g., outlier detection~\cite{bailis2017macrobase,Rousseeuw2011RobustSF,Liu2008IsolationF,hodge2004survey} or explicit error-prediction models~\cite{mahdavi2019raha,heidari2019holodetect,liu2020picket}).   \sys combines a complaint-based approach~\cite{wu2013scorpion,roy2014formal,chalamalla2014descriptive,abuzaid2020diff,miao2019going} based on how detected errors affect output complaints, with a model-based error prediction approach to identify candidate repairs.

\stitle{Data Repair: }
Data repair is an optimization problem that satisfies a set of violated constraints over the database instance~\cite{chu2013discovering,yakout2013don}, and can leverage signals (e.g., past repairs~\cite{volkovs2014continuous}, knowledge bases~\cite{chu2015katara}).   The solution space is large, so human involvement is crucial.  Spreadsheet interfaces~\citeN{raman2001potter,kandel2011wrangler} and visualizations~\cite{kandel2011wrangler,wu2013scorpion,Luo2020VisCleanIC} help users identify potential errors and help guide repairs.  Interfaces such as Profiler~\cite{kandel2012profiler} run a library of common error detectors and embed the errors directly in the visualization interface.

Model-based repairs estimate the correct value of an error. ERACER~\cite{mayfield2010eracer} uses graphical models that combine convolution and regression models to  repair raw data tuples. Active learning approaches~\cite{yakout2011guided,thirumuruganathan2017uguide,mahdavi2020baran} ask users to verify whether repair candidates are correct.  Daisy~\cite{giannakopoulou2020cleaning} uses categorical histograms to identify and repair errors in join attributes.  In \sys, the user submits a single complaint over an aggregate query result, and the system trains multi-level models.   Finally, techniques such as unknown unknows~\cite{chung2018estimating} can be viewed as repairing group-wise missing record errors under species estimation assumptions.

\stitle{Complaint-based Explanation:}
This class of problems follows the framework where, given a complaint over query results, they search for a good explanation from a candidate set (e.g., predicates, tuples, etc). They primarily differ in the ranking metric (e.g., sensitivity-based~\cite{wu2013scorpion,roy2014formal,chalamalla2014descriptive,abuzaid2020diff}, density-based~\cite{joglekar2015smart, ruhl2018cascading,sarawagi1999explaining}, counterbalance~\cite{miao2019going}), and typically focus on deletion-based interventions.  
\sys ranks drill-down groups based on how much repairing their aggregate statistics, as learned by a multi-level model, would resolve the complaint.  
Repairing at aggregation level enables \sys to combine predictive signals and uncover a broader range of errors like missing records which previous metrics fail to detect.

The hierarchical density attribution problem~\cite{Fagin2005MultistructuralD,sarawagi1999explaining} returns a set of non-overlapping subgroups that account for the largest mass of the total density.  Ruhl et al.~\cite{ruhl2018cascading} extends this from a single hierarchy to a product of trees (i.e., overlapping hierarchies) and shows that this problem is NP-hard.
Joglekar et al.~\cite{joglekar2015smart} is restricted to count-based densities and leverages its submodular structure to design a greedy solution to recommending sets of drill-down groups. \sys is designed for a single hierarchy and supports more complex aggregation functions, and returns a ranked list of drill-down groups rather than an optimal set.


\stitle{Factorised Representation: }
Factorised Representation~\cite{olteanu2015size} reduces redundancies due to functional dependencies, and has been used to optimize model training (linear regression~\cite{schleich2016learning}, decision tree~\cite{Kobis2017LearningDT} and Rk-mean~\cite{curtin2020rk}) over factorised matrices derived from join queries.  
\sys extends prior work~\cite{schleich2020lmfao,schleich2016learning} to matrices based on join-aggregation queries that exhibit fewer redundancies, supports extra operations including right and left multiplication, and further exploits the hierarchical structure for optimization.

\section{Conclusions}
We presented \sys, which helps users iteratively identify and repair errors in the output of aggregation queries.  \sys supports the ``Overview, zoom, details-on-demand'' analysis pattern common in visual analysis by recommending drill-down operations and highlighting groups in the drill-down results that most contributed to the user's reported data error.  \sys trains a model to estimate each group's expected aggregate statistics, and  measures the extent that the complaint is resolved by repairing the group statistic to its expectation.  Our implementation leverages a factorised matrix representation, and we developed factorised matrix operations as well as optimizations that leverage the data's hierarchical structure.  Our optimizations reduce end-to-end runtimes by over $6\times$ as compared to a Matlab-based implementation.  \sys identified 21 out of 30 data errors in John Hopkin's COVID-19 data, and identified 20 out of 22 complaints in a user study with Columbia University's Financial Instruments Sector Team based on their data collected from Ethipoian farmers.


\bibliographystyle{abbrv}
\bibliography{paper}

\begin{thebibliography}{10}

\bibitem{abuzaid2020diff}
F.~Abuzaid, P.~Kraft, S.~Suri, E.~Gan, E.~Xu, A.~Shenoy, A.~Ananthanarayan,
  J.~Sheu, E.~Meijer, X.~Wu, et~al.
\newblock Diff: a relational interface for large-scale data explanation.
\newblock {\em The VLDB Journal}, pages 1--26, 2020.

\bibitem{aitkin1986statistical}
M.~Aitkin and N.~Longford.
\newblock Statistical modelling issues in school effectiveness studies.
\newblock {\em Journal of the Royal Statistical Society: Series A (General)},
  149(1):1--26, 1986.

\bibitem{akaike1998information}
H.~Akaike.
\newblock Information theory and an extension of the maximum likelihood
  principle.
\newblock In {\em Selected papers of hirotugu akaike}, pages 199--213.
  Springer, 1998.

\bibitem{laug}
E.~Anderson, Z.~Bai, C.~Bischof, S.~Blackford, J.~Demmel, J.~Dongarra,
  J.~Du~Croz, A.~Greenbaum, S.~Hammarling, A.~McKenney, and D.~Sorensen.
\newblock {\em {LAPACK} Users' Guide}.
\newblock Society for Industrial and Applied Mathematics, Philadelphia, PA,
  third edition, 1999.

\bibitem{bailis2017macrobase}
P.~Bailis, E.~Gan, S.~Madden, D.~Narayanan, K.~Rong, and S.~Suri.
\newblock Macrobase: Prioritizing attention in fast data.
\newblock In {\em Proceedings of the 2017 ACM International Conference on
  Management of Data}, pages 541--556, 2017.

\bibitem{barddal2017survey}
J.~P. Barddal, H.~M. Gomes, F.~Enembreck, and B.~Pfahringer.
\newblock A survey on feature drift adaptation: Definition, benchmark,
  challenges and future directions.
\newblock {\em Journal of Systems and Software}, 127:278--294, 2017.

\bibitem{burnham2004multimodel}
K.~P. Burnham and D.~R. Anderson.
\newblock Multimodel inference: understanding aic and bic in model selection.
\newblock {\em Sociological methods \& research}, 33(2):261--304, 2004.

\bibitem{chalamalla2014descriptive}
A.~Chalamalla, I.~F. Ilyas, M.~Ouzzani, and P.~Papotti.
\newblock Descriptive and prescriptive data cleaning.
\newblock In {\em Proceedings of the 2014 ACM SIGMOD international conference
  on Management of data}, pages 445--456, 2014.

\bibitem{chu2013discovering}
X.~Chu, I.~F. Ilyas, and P.~Papotti.
\newblock Discovering denial constraints.
\newblock {\em Proceedings of the VLDB Endowment}, 6(13):1498--1509, 2013.

\bibitem{chu2015katara}
X.~Chu, J.~Morcos, I.~F. Ilyas, M.~Ouzzani, P.~Papotti, N.~Tang, and Y.~Ye.
\newblock Katara: reliable data cleaning with knowledge bases and
  crowdsourcing.
\newblock {\em Proceedings of the VLDB Endowment}, 8(12):1952--1955, 2015.

\bibitem{chung2018estimating}
Y.~Chung, M.~L. Mortensen, C.~Binnig, and T.~Kraska.
\newblock Estimating the impact of unknown unknowns on aggregate query results.
\newblock {\em ACM Transactions on Database Systems (TODS)}, 43(1):1--37, 2018.

\bibitem{curtin2020rk}
R.~Curtin, B.~Moseley, H.~Ngo, X.~Nguyen, D.~Olteanu, and M.~Schleich.
\newblock Rk-means: Fast clustering for relational data.
\newblock In {\em International Conference on Artificial Intelligence and
  Statistics}, pages 2742--2752, 2020.

\bibitem{diez2000multilevel}
A.~V. Diez-Roux.
\newblock Multilevel analysis in public health research.
\newblock {\em Annual review of public health}, 21(1):171--192, 2000.

\bibitem{dong2020interactive}
E.~Dong, H.~Du, and L.~Gardner.
\newblock An interactive web-based dashboard to track covid-19 in real time.
\newblock {\em The Lancet infectious diseases}, 20(5):533--534, 2020.

\bibitem{Fagin2005MultistructuralD}
R.~Fagin, R.~Guha, R.~Kumar, J.~Novak, D.~Sivakumar, and A.~Tomkins.
\newblock Multi-structural databases.
\newblock In {\em PODS '05}, 2005.

\bibitem{fernandez1981multilevel}
R.~M. Fernandez and J.~C. Kulik.
\newblock A multilevel model of life satisfaction: Effects of individual
  characteristics and neighborhood composition.
\newblock {\em American Sociological Review}, pages 840--850, 1981.

\bibitem{gelman2006data}
A.~Gelman and J.~Hill.
\newblock {\em Data analysis using regression and multilevel/hierarchical
  models}.
\newblock Cambridge university press, 2006.

\bibitem{giannakopoulou2020cleaning}
S.~Giannakopoulou, M.~Karpathiotakis, and A.~Ailamaki.
\newblock Cleaning denial constraint violations through relaxation.
\newblock In {\em Proceedings of the 2020 ACM SIGMOD International Conference
  on Management of Data}, pages 805--815, 2020.

\bibitem{goldstein1986multilevel}
H.~Goldstein.
\newblock Multilevel mixed linear model analysis using iterative generalized
  least squares.
\newblock {\em Biometrika}, 73(1):43--56, 1986.

\bibitem{gray1997data}
J.~Gray, S.~Chaudhuri, A.~Bosworth, A.~Layman, D.~Reichart, M.~Venkatrao,
  F.~Pellow, and H.~Pirahesh.
\newblock Data cube: A relational aggregation operator generalizing group-by,
  cross-tab, and sub-totals.
\newblock {\em Data mining and knowledge discovery}, 1(1):29--53, 1997.

\bibitem{heidari2019holodetect}
A.~Heidari, J.~McGrath, I.~F. Ilyas, and T.~Rekatsinas.
\newblock Holodetect: Few-shot learning for error detection.
\newblock In {\em Proceedings of the 2019 International Conference on
  Management of Data}, pages 829--846, 2019.

\bibitem{hodge2004survey}
V.~Hodge and J.~Austin.
\newblock A survey of outlier detection methodologies.
\newblock {\em Artificial intelligence review}, 22(2):85--126, 2004.

\bibitem{iman1982distribution}
R.~L. Iman and W.-J. Conover.
\newblock A distribution-free approach to inducing rank correlation among input
  variables.
\newblock {\em Communications in Statistics-Simulation and Computation},
  11(3):311--334, 1982.

\bibitem{joglekar2015smart}
M.~Joglekar, H.~Garcia-Molina, and A.~Parameswaran.
\newblock Smart drill-down: A new data exploration operator.
\newblock In {\em Proceedings of the VLDB Endowment International Conference on
  Very Large Data Bases}, volume~8, page 1928. NIH Public Access, 2015.

\bibitem{kandel2011wrangler}
S.~Kandel, A.~Paepcke, J.~Hellerstein, and J.~Heer.
\newblock Wrangler: Interactive visual specification of data transformation
  scripts.
\newblock In {\em Proceedings of the SIGCHI Conference on Human Factors in
  Computing Systems}, pages 3363--3372, 2011.

\bibitem{kandel2012profiler}
S.~Kandel, R.~Parikh, A.~Paepcke, J.~M. Hellerstein, and J.~Heer.
\newblock Profiler: Integrated statistical analysis and visualization for data
  quality assessment.
\newblock In {\em Proceedings of the International Working Conference on
  Advanced Visual Interfaces}, pages 547--554, 2012.

\bibitem{Kobis2017LearningDT}
L.~Kobis.
\newblock Learning decision trees over factorized joins.
\newblock 2017.

\bibitem{laurikkala2000informal}
J.~Laurikkala, M.~Juhola, E.~Kentala, N.~Lavrac, S.~Miksch, and B.~Kavsek.
\newblock Informal identification of outliers in medical data.
\newblock In {\em Fifth international workshop on intelligent data analysis in
  medicine and pharmacology}, volume~1, pages 20--24. Citeseer, 2000.

\bibitem{lien2005regression}
D.~Lien and N.~Balakrishnan.
\newblock On regression analysis with data cleaning via trimming,
  winsorization, and dichotomization.
\newblock {\em Communications in Statistics—Simulation and
  Computation{\textregistered}}, 34(4):839--849, 2005.

\bibitem{Liu2008IsolationF}
F.~Liu, K.~Ting, and Z.~Zhou.
\newblock Isolation forest.
\newblock {\em 2008 Eighth IEEE International Conference on Data Mining}, pages
  413--422, 2008.

\bibitem{liu2020picket}
Z.~Liu, Z.~Zhou, and T.~Rekatsinas.
\newblock Picket: Self-supervised data diagnostics for ml pipelines.
\newblock {\em arXiv preprint arXiv:2006.04730}, 2020.

\bibitem{Luo2020VisCleanIC}
Y.~Luo, C.~Chai, X.~Qin, N.~Tang, and G.~Li.
\newblock Visclean: Interactive cleaning for progressive visualization.
\newblock {\em Proc. VLDB Endow.}, 13:2821--2824, 2020.

\bibitem{mahdavi2020baran}
M.~Mahdavi and Z.~Abedjan.
\newblock Baran: effective error correction via a unified context
  representation and transfer learning.
\newblock {\em Proceedings of the VLDB Endowment}, 13(12):1948--1961, 2020.

\bibitem{mahdavi2019raha}
M.~Mahdavi, Z.~Abedjan, R.~Castro~Fernandez, S.~Madden, M.~Ouzzani,
  M.~Stonebraker, and N.~Tang.
\newblock Raha: A configuration-free error detection system.
\newblock In {\em Proceedings of the 2019 International Conference on
  Management of Data}, pages 865--882, 2019.

\bibitem{marascuilo1987loglinear}
L.~A. Marascuilo and P.~L. Busk.
\newblock Loglinear models: A way to study main effects and interactions for
  multidimensional contingency tables with categorical data.
\newblock {\em Journal of Counseling Psychology}, 34(4):443, 1987.

\bibitem{MATLAB:2010}
MATLAB.
\newblock {\em version 7.10.0 (R2010a)}.
\newblock The MathWorks Inc., Natick, Massachusetts, 2010.

\bibitem{mayfield2010eracer}
C.~Mayfield, J.~Neville, and S.~Prabhakar.
\newblock Eracer: a database approach for statistical inference and data
  cleaning.
\newblock In {\em Proceedings of the 2010 ACM SIGMOD International Conference
  on Management of data}, pages 75--86, 2010.

\bibitem{miao2019going}
Z.~Miao, Q.~Zeng, B.~Glavic, and S.~Roy.
\newblock Going beyond provenance: Explaining query answers with pattern-based
  counterbalances.
\newblock In {\em Proceedings of the 2019 International Conference on
  Management of Data}, pages 485--502, 2019.

\bibitem{nikolic2018incremental}
M.~Nikolic and D.~Olteanu.
\newblock Incremental view maintenance with triple lock factorization benefits.
\newblock In {\em Proceedings of the 2018 International Conference on
  Management of Data}, pages 365--380, 2018.

\bibitem{North2000SnaptogetherVA}
C.~North and B.~Shneiderman.
\newblock Snap-together visualization: a user interface for coordinating
  visualizations via relational schemata.
\newblock In {\em AVI '00}, 2000.

\bibitem{olteanu2015size}
D.~Olteanu and J.~Z{\'a}vodn{\`y}.
\newblock Size bounds for factorised representations of query results.
\newblock {\em ACM Transactions on Database Systems (TODS)}, 40(1):1--44, 2015.

\bibitem{osgood2018farmer}
D.~Osgood, B.~Powell, R.~Diro, C.~Farah, M.~Enenkel, M.~E. Brown, G.~Husak,
  S.~L. Blakeley, L.~Hoffman, and J.~L. McCarty.
\newblock Farmer perception, recollection, and remote sensing in weather index
  insurance: An ethiopia case study.
\newblock {\em Remote Sensing}, 10(12):1887, 2018.

\bibitem{rmanual}
{R Core Team}.
\newblock {\em R: A Language and Environment for Statistical Computing}.
\newblock R Foundation for Statistical Computing, Vienna, Austria, 2013.

\bibitem{raman2001potter}
V.~Raman and J.~M. Hellerstein.
\newblock Potter's wheel: An interactive data cleaning system.
\newblock In {\em VLDB}, volume~1, pages 381--390, 2001.

\bibitem{Rousseeuw2011RobustSF}
P.~Rousseeuw and M.~Hubert.
\newblock Robust statistics for outlier detection.
\newblock {\em Wiley Interdisciplinary Reviews: Data Mining and Knowledge
  Discovery}, 1, 2011.

\bibitem{roy2014formal}
S.~Roy and D.~Suciu.
\newblock A formal approach to finding explanations for database queries.
\newblock In {\em Proceedings of the 2014 ACM SIGMOD international conference
  on Management of data}, pages 1579--1590, 2014.

\bibitem{ruhl2018cascading}
M.~Ruhl, M.~Sundararajan, and Q.~Yan.
\newblock The cascading analysts algorithm.
\newblock In {\em Proceedings of the 2018 International Conference on
  Management of Data}, pages 1083--1096, 2018.

\bibitem{sacco2005dynamic}
J.~M. Sacco and N.~Schmitt.
\newblock A dynamic multilevel model of demographic diversity and misfit
  effects.
\newblock {\em Journal of Applied Psychology}, 90(2):203, 2005.

\bibitem{sarawagi1999explaining}
S.~Sarawagi.
\newblock Explaining differences in multidimensional aggregates.
\newblock In {\em VLDB}, volume~99, pages 7--10. Citeseer, 1999.

\bibitem{sarawagi1998discovery}
S.~Sarawagi, R.~Agrawal, and N.~Megiddo.
\newblock Discovery-driven exploration of olap data cubes.
\newblock In {\em International Conference on Extending Database Technology},
  pages 168--182. Springer, 1998.

\bibitem{schleich2020lmfao}
M.~Schleich and D.~Olteanu.
\newblock Lmfao: An engine for batches of group-by aggregates.
\newblock {\em arXiv preprint arXiv:2008.08657}, 2020.

\bibitem{schleich2016learning}
M.~Schleich, D.~Olteanu, and R.~Ciucanu.
\newblock Learning linear regression models over factorized joins.
\newblock In {\em Proceedings of the 2016 International Conference on
  Management of Data}, pages 3--18, 2016.

\bibitem{seabold2010statsmodels}
S.~Seabold and J.~Perktold.
\newblock statsmodels: Econometric and statistical modeling with python.
\newblock In {\em 9th Python in Science Conference}, 2010.

\bibitem{thirumuruganathan2017uguide}
S.~Thirumuruganathan, L.~Berti-Equille, M.~Ouzzani, J.-A. Quiane-Ruiz, and
  N.~Tang.
\newblock Uguide: User-guided discovery of fd-detectable errors.
\newblock In {\em Proceedings of the 2017 ACM International Conference on
  Management of Data}, pages 1385--1397, 2017.

\bibitem{van2015extrinsic}
S.~Van~de Walle, B.~Steijn, and S.~Jilke.
\newblock Extrinsic motivation, psm and labour market characteristics: A
  multilevel model of public sector employment preference in 26 countries.
\newblock {\em International Review of Administrative Sciences},
  81(4):833--855, 2015.

\bibitem{volkovs2014continuous}
M.~Volkovs, F.~Chiang, J.~Szlichta, and R.~J. Miller.
\newblock Continuous data cleaning.
\newblock In {\em 2014 IEEE 30th international conference on data engineering},
  pages 244--255. IEEE, 2014.

\bibitem{wu2013scorpion}
E.~Wu and S.~Madden.
\newblock Scorpion: Explaining away outliers in aggregate queries.
\newblock {\em Proc. VLDB Endow.}, 6(8):553–564, June 2013.

\bibitem{yakout2013don}
M.~Yakout, L.~Berti-{\'E}quille, and A.~K. Elmagarmid.
\newblock Don't be scared: use scalable automatic repairing with maximal
  likelihood and bounded changes.
\newblock In {\em Proceedings of the 2013 ACM SIGMOD International Conference
  on Management of Data}, pages 553--564, 2013.

\bibitem{yakout2011guided}
M.~Yakout, A.~K. Elmagarmid, J.~Neville, M.~Ouzzani, and I.~F. Ilyas.
\newblock Guided data repair.
\newblock {\em arXiv preprint arXiv:1103.3103}, 2011.

\end{thebibliography}

\clearpage

\appendix
\section{Problem Definition}

\stitle{Distributive Set of Functions}
\label{section:disset}
\sys supports complaint over the results of a distributive set of aggregation functons. We extend the definition of distributive function~\cite{gray1997data} to a set of functions.
A set of $\emph{I}$ aggregation functions $\mathbb{F}_{agg} = \{F_{agg_1},  \cdots,F_{agg_\emph{I}}\}$ is distributive if, given the partition of $R$ into $\emph{J}$ subsets, and the aggregation results $\mathbb{F}_{agg}(R_1), \cdots, \mathbb{F}_{agg}(R_\emph{J})$ after applying $\mathbb{F}_{agg}$ to $\emph{J}$ subsets, there exists function $\emph{G}$ such that: $\mathbb{F}_{agg}(R) = \emph{G}\HS(\{R_1, \cdots, R_\emph{J}\}))$

For example, consider the following distributive set of aggregation functions: Mean, Count and Standard deviation. Given a set of $\emph{J}$ aggregation results $\mathbb{F}_{agg}(R_1), \cdots, \mathbb{F}_{agg}(R_\emph{J})$, there exists function $\emph{G} =  \{\emph{G}_{mean}, \emph{G}_{count}, \emph{G}_{std}\}$ such that:
\begin{align}\emph{G}_{mean}(\mathbb{F}_{agg}(R_1), \cdots, \mathbb{F}_{agg}(R_\emph{J})) =  & \frac{\sum_{j = 1}^{\emph{J}} F_{count} (R_\emph{j}) \cdot F_{mean} (R_\emph{j})}{\sum_{j = 1}^{\emph{J}} F_{count} (R_\emph{j})} \nonumber \\
\emph{G}_{count}(\mathbb{F}_{agg}(R_1), \cdots, \mathbb{F}_{agg}(R_\emph{J})) =  & \sum_{j = 1}^{\emph{J}} F_{count} (R_\emph{j})\nonumber\\
\emph{G}_{std}(\mathbb{F}_{agg}(R_1), \cdots, \mathbb{F}_{agg}(R_\emph{J})) =   & \nonumber\\
\span\sqrt{\frac{\sum_{j = 1}^{\emph{J}} (F_{count} (R_\emph{j}) - 1) \cdot F_{std}^2 (R_\emph{j}) + \sum_{j = 1}^{\emph{J}} F_{count} (R_\emph{j})\cdot (\emph{G}_{mean} -F_{mean} (R_\emph{j}) )^2}
{\emph{G}_{count} - 1}}\nonumber\end{align}

\section{Feature Matrix}
\label{section:featureMatrix}

In this section, we discuss, given all registered features, how to build feature matrix.

\begin{figure}
     \centering
     \begin{subfigure}[b]{0.08\textwidth}
         \centering
         
         \includegraphics[width=0.8\textwidth]{images/approach_hierarchies.pdf}
         
         \caption{}
         \label{fig:hierarchies2}
     \end{subfigure}
     \hfill
     \begin{subfigure}[b]{0.12\textwidth}
         \centering
         \includegraphics[width=0.9\textwidth]{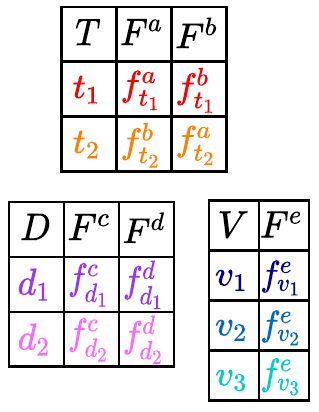}
         \caption{}
         \label{fig:features2}
     \end{subfigure}
     \hfill
    \begin{subfigure}[b]{0.1\textwidth}
         \centering
         \includegraphics[width=0.7\textwidth]{images/approach_attributematrix.pdf}
         \caption{}
         \label{fig:attribute matrix2}
     \end{subfigure}
     \hfill
     \begin{subfigure}[b]{0.12\textwidth}
         \centering
         \includegraphics[width=\textwidth]{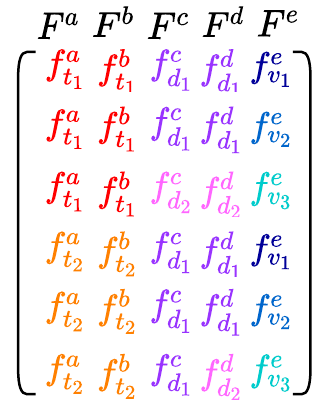}
         \caption{}
         \label{fig:feature matrix2}
     \end{subfigure}
     \hfill
    \caption{Example dataset with (a) hierarchies, (b) features, (c) attribute matrix, and (d) feature matrix.}
    \label{fig:exampledata2}
\end{figure}

\stitle{Attribute matrix:} We first define attribute matrix, which helps us build feature matrix. Attribute matrix is built from the query result $\emph{Q} = \gamma_{A_{gb}', f(A_{agg})}(\mathbb{R})$ projected out aggregation function $f$ and ordered by the attribute order (the same as feature matrix in \Cref{featurematrixdis}). In \Cref{fig:exampledata2}, given hierarchies in \Cref{fig:hierarchies2} with order: Time and Location, attribute matrix is shown in \Cref{fig:attribute matrix2}.

\stitle{Feature matrix:} We then discuss how to derive feature matrix from attribute matrix. For feature registered with attribute $\emph{A}$, given current view $V' = \gamma_{A_{gb}', f(A_{agg})}(prov(t_{c}))$, this feature is applicable if $\emph{A} \in A_{gb}'$. Given all applicable features and attribute matrix, feature matrix $\mathbi{X}$ is derived by replacing each attribute value in attribute matrix with feature values. Continue with examples in \Cref{fig:exampledata2}, all applicable features are shown in \Cref{fig:features2} and feature matrix is shown in \Cref{fig:feature matrix2}.

\stitle{Optimization:} As an optimization during model training, feature matrix is not directly used. Instead,  we isolate attribute matrix from feature matrix in aggregation queries.
Because the mapping between attribute and  feature is one-to-one, we can computate aggregation queries over attribute matrix, and infer the aggregation queries over feature matrix by mapping the value from attribute to feature. For example, in \Cref{fig:exampledata2}, suppose we want to compute the sum of feature $F^a$ in feature matrix (whose result is $3f^a_{t_1} + 3f^a_{t_2}$).
We can first compute the count of each value $\cnt_{T}$ for attribute Time (T) ( whose result is $\{t_1 : 3, t_2 : 3\}$).  Suppose $f^a(\cdot)$ maps attribute Time (T) to feature $F^a$, then the sum of feature $F^a$ can be computed by $\sum_{a\in Dom(T)} \cnt_{T}(a) \cdot f^a(a) =  3f^a_{t_1} + 3f^a_{t_2}$. The isolation of attribute from feature can simplify the problem and improve performance because attribute matrix is smaller than feature matrix.

\section{Factorizer}
\label{section:factoriser}

In this section, we discuss the implementation and interface of {\it Factorizer} in \sys. Given input relations, {\it Factorizer} in \sys stores the factorised feature matrix, and presents an interface.

\subsection{Storage}
We discuss how {\it Factorizer} stores factorised feature matrix. {\it Factorizer} first exploits the one to one mapping between attribute value and feature value to store the factorised attribute matrix and the feature mapping separately. Then, to store factorised attribute matrix, 
\sys doesn't materialize all unary relations and  algebraic expressions in f-representation. Instead, \sys exploits the independence between different hierarchies and functional dependencies inside each hierarchy, and stores factorised attribute matrix with relations implemented as sorted map. Sorted map makes it easy for \sys to iterate through data.
Given input relations, {\it Factorizer} normalizes relations to BCNF, sort them according to attribute order, and stores relations using sorted map.
For example data in \Cref{fig:exampledata2} with attribute order "Time (T), District (D), City (V)", {\it Factorizer} stores normalized relations $R[T]$ for Time hierarchy which enumerate attribute values, and $R[D,V]$ for geography hierarchy which maps attribute District to Village. ($R[D,V]$ is implemented as sorted map with key D and value V). {\it Factorizer} records the dependency among relations. Now consider the marginalization operation: $\bigoplus_{V} R[D,V]$. This marginalization is implemented by iterate through each District (key) and sum the count of its Villages (value).

\subsection{Interface}

{\it Factorizer} presents interface of relation and row iterator. 

\stitle{Relation:}
Given attribute, {\it Factorizer} returns relations, which are used to compute decomposed aggregates discussed in \Cref{section:matrxopoverfr} to exploit redundancy in columns of attribute matrix. Given attribute $A_i$, if it is the least strict attribute in its hierarchy, factorizer returns $R_i[A_i]$ which enumerates all attribute values in $A_i$. Otherwise, factorizer returns $R_i[A_{i+1}, A_i]$ which connects $A_i$ with the next less strict attribute $A_{i+1}$. For example, for example data in \Cref{fig:exampledata2},  {\it Factorizer} returns $R_T[T]$, $R_R[R]$, and $R_V[D,V]$.

\stitle{Row Iterator:} For row iterator, {\it Factorizer} exploits the fact that the rows in attribute matrix are sorted by the attribute order and the difference between rows is relatively small. {\it Factorizer}  iterates each row and only returns the difference between rows. To build row iterator, we first build the set $\textit{end}$ for each attribute to determine when iterator should propagate the change.  For attribute $A_i$, let $itr_{A_i}$ be the iterator of attribute value in ascending order. The intuition behind the set $\textit{end}$ is that, when $itr_{A_i}$ iterates over any value in set $\textit{end}$, $itr_{A_{i+1}}$ should also increment. For example, in \Cref{fig:hierarchies}, city 2 and city 3 are in $\textit{end}$ because when city iterator $itr_{C}$ iterates over them, state iterator $itr_{S}$ should also increment. 

Algorithm \ref{alg:rowit} implements the iterator of attribute rows. Notice that, instead of returning the row values of attribute matrix, it returns the difference between current row and previous row.

\begin{algorithm}[H]
\KwResult{Update to the previous row for attributes from $A_i$}
$itr_{A_i}$ := current iterator for attribute $A_i$\;
nextValue := $itr_{A_i}$.next()\;
update[{$A_i$}] := nextValue\;
\If{current attribute value $\in$ end $\textbf{and}$ $A_i \neq root$ }{
    next(parent($A_i$), update)\;
}
\If{$!itr_{A_i}.hasNext()$}{
    $itr_{A_i}$ := new itr()\;
}
return  update\;
 \caption{Row iterator next($A_i$, \&update) algorithm}
 \label{alg:rowit}
\end{algorithm}

\section{Expectation Maximization Algorithm}
\label{section:emalgo}

We write the multilevel-model in matrix form where $\mathbi{y}$,$\mathbi{X}$,$\boldsymbol{\beta}$,$\mathbi{b}$, and $\boldsymbol{\epsilon}$ are vertical concatenations of their row-wise vectors/matrices, and $\mathbi{Z}$ is a diagonal matrix with 
$\mathbi{X}_i$ along the diagonal:
\begin{equation}\mathbi{y} =\mathbi{X} \cdot \boldsymbol{\beta} + \mathbi{Z} \cdot \boldsymbol{b} + \pmb{\epsilon} \end{equation}
EM iterates between two steps. 
The expectation step uses the estimates $\hat{\boldsymbol{\beta}}$, 
$\hat{\pmb{\Sigma}}$, $\hat{\sigma^2}$ to find the expected value of 
$\hat{\boldsymbol{b}}_{\emph{i}}$, and $\hat{\boldsymbol{b}}_{\emph{i}} \cdot\hat{\boldsymbol{b}}_{\emph{i}}^T$:
\begin{align}\mathbi{V}_{\emph{i}} = & \HS(\frac{\mathbi{X}_{\emph{i}}^T \cdot \mathbi{X}_{\emph{i}}}{\hat{\sigma^2}} + \hat{\pmb{\Sigma}}^{-1})^{-1}\\
\pmb{\mu}_{\emph{i}} = & \HS\frac{\mathbi{V}_{\emph{i}} \cdot \mathbi{X}_{\emph{i}}^T \cdot (\mathbi{y}_{\emph{i}} - \mathbi{X}_{\emph{i}} \cdot \hat{\boldsymbol{\beta}})}{\hat{\sigma^2}}\\
\hat{\boldsymbol{b}} _{\emph{i}} = & \HS \pmb{\mu}_{\emph{i}}\\
\hat{\boldsymbol{b}}_{\emph{i}} \cdot\hat{\boldsymbol{b}}_{\emph{i}}^T =&\HS \mathbi{V}_{\emph{i}} + \pmb{\mu}_{\emph{i}} \cdot \pmb{\mu}_{\emph{i}}^T
\end{align}
The maximization step uses the current estimate of $\hat{\boldsymbol{b}}$ to estimate the $\hat{\boldsymbol{\beta}}$, $\hat{\pmb{\Sigma}}$, $\hat{\sigma^2}$ with maximum likelihood:
\begin{align} \label{alg:beta} \hat{\boldsymbol{\beta}}= &\HS(\mathbi{X}^T \cdot \mathbi{X})^{-1} \cdot \mathbi{X}^T \cdot (\mathbi{y} - \mathbi{Z} \cdot \hat{\boldsymbol{b}}) \\
\hat{\pmb{\Sigma}}= &\HS\frac{1}{ \mathcal{G}} \cdot \sum^{ \mathcal{G}} _{\emph{i}=1}\hat{\boldsymbol{b}} _{\emph{i}} \cdot \hat{\boldsymbol{b}}_{\emph{i}}^T\\
\hat{\sigma^2}= &\HS\frac{1}{\emph{n}}((\mathbi{y} - \mathbi{X} \cdot \hat{\boldsymbol{\beta}})^T \cdot (\mathbi{y} - \mathbi{X} \cdot \hat{\boldsymbol{\beta}}) + \sum^{ \mathcal{G}}_{\emph{i}=1}\mathbi{Tr}(\mathbi{X}_{\emph{i}}^T \cdot \mathbi{X}_{\emph{i}} \cdot \boldsymbol{b} _{\emph{i}} \cdot \boldsymbol{b} _{\emph{i}}^T) \nonumber \\
& -2\cdot (\mathbi{y} - \mathbi{X} \cdot \hat{\boldsymbol{\beta}})^T \cdot (\mathbi{Z} \cdot \hat{\boldsymbol{b}})) \end{align}
where $\mathbi{Tr}(\cdot)$ is the trace (sum of main diagonal elements) of a matrix.

\stitle{Vertical Concatenation:} Notice that $\mathbi{Z}$ has shape $\emph{n} × \emph{m}\cdot \mathcal{G}$ where $ \mathcal{G}$ is the number of clusters (typically exponential in the depth of the attribute in its hierarchy).  $\mathbi{Z}$ is non-zero along the diagonal, thus its sparsity can be exploited by computing $\mathbi{Z} \cdot \hat{\boldsymbol{b}}$ with vertical concatenation without fully materializing $\mathbi{Z}$: 
\begin{align} \mathbi{Z} \cdot \hat{\boldsymbol{b}} = vertcat(\mathbi{X}_\emph{1} \cdot \hat{\boldsymbol{b}}_1,\mathbi{X}_\emph{2} \cdot \hat{\boldsymbol{b}}_2,\ldots,\mathbi{X}_ \mathcal{G} \cdot \hat{\boldsymbol{b}}_ \mathcal{G}) \nonumber\end{align}

\stitle{Multiplication Order:}  Associative law of matrix multiplication can be exploited to avoid large intermediate result. For example, in equation \ref{alg:beta}, if matrix chain multiplications are from left to right, there will be an intermediate result with shape $m \times n$:
\begin{align}\hat{\boldsymbol{\beta}}= \HS \underset{\textcolor{red}{m\times n}}{\underbrace{((
\underset{m\times n}{\mathbi{X}^T} \cdot \underset{n\times m}{\mathbi{X}})^{-1} \cdot \underset{m\times n}{\mathbi{X}^T}})} \cdot \underset{n\times 1}{\underbrace{(\underset{n\times 1}{\mathbi{y}} - \underset{n\times m \mathcal{G}}{\mathbi{Z}} \cdot \underset{m \mathcal{G}\times 1}{\hat{\boldsymbol{b}}})}} \nonumber \end{align}

This could be avoided by reordering matrix multiplications:
\begin{align}\hat{\boldsymbol{\beta}}= \HS \underset{m\times m}{\underbrace{(
\underset{m\times n}{\mathbi{X}^T} \cdot \underset{n\times m}{\mathbi{X}})^{-1}} }\cdot \underset{m\times 1}{\underbrace{\underset{m\times n}{(\mathbi{X}^T} \cdot (\underset{n\times 1}{\mathbi{y}} - \underset{n\times m \mathcal{G}}{\mathbi{Z}} \cdot \underset{m \mathcal{G}\times 1}{\hat{\boldsymbol{b}}}))}} \nonumber \end{align}

\stitle{Bottleneck:} The EM updates above are primarily bottlenecked by six types of matrix multiplication operations: $\mathbi{X}^T \cdot \mathbi{X}$, $ \mathbi{X} \cdot \mathbi{A}$, $\mathbi{B} \cdot \mathbi{X}$, $\mathbi{X}_{\emph{i}}^T \cdot \mathbi{X}_{\emph{i}}$, $ \mathbi{X}_{\emph{i}} \cdot \mathbi{C}_{\emph{i}}$,  $\mathbi{D}_{\emph{i}} \cdot \mathbi{X}_{\emph{i}}$ for $\emph{i} = 1,..., \mathcal{G}$, where $\mathbi{A}, \mathbi{B}, \mathbi{C}_{\emph{i}}, \mathbi{D}_{\emph{i}}$ are intermediate matrices and $\mathcal{G}$ is the number of clusters.  We can precompute $\mathbi{X}^T \cdot \mathbi{X}$ and  $\mathbi{X}_{\emph{i}}^T \cdot \mathbi{X}_{\emph{i}}$. We need to perform each other operation once during each iteration.

All of these operations involve $\mathbi{X}$, which is the factorised feature matrix.  A naive approach is to materialize the full $\mathbi{X}$ matrix and use existing matrix operator implementations, but the matrix can be very large.   Instead, we wish to directly perform matrix operations on the f-representation. 


\section{Matrix operations}
\label{section:matrixops}
In this section, we provide formal algorithms to compute matrix operations through aggregation queries. We assume that the total number of rows in the relations of each hierarchy is $O(\emph{w})$, the number of attributes is $\emph{d}$, the number of columns in feature matrix is $\emph{m}$ and the number rows in feature matrix is $\emph{n}$. 
For simplicity, we assume that feature matrix is the same as attribute matrix. The extension to customized feature matrix is trivial by mapping attribute value to feature value during operations.



\stitle{Gram Matrix:} First consider gram matrix $\mathbi{X}^T\cdot\mathbi{X}$.
The naive multiplication $\mathbi{X}^T\cdot\mathbi{X}$ has time complexity $O(\emph{n} \cdot \emph{m}^2)$. In \Cref{fig:attribute matrix2}, the columns in attribute matrix have a lot of redundancy, and, given two columns, we can iterate all attribute values and leverage $\cof$ to derive how many times two attribute values are duplicated. 
Note that gram matrix is symmetrical, so we only need to calculate half of the matrix. Let $c_i$ be the $\emph{i}th$ column and  $r_i$ be the $\emph{i}th$ row of attribute matrix.

\begin{algorithm}[H]
\label{alg:cofactor}
\KwResult{$c_i \cdot c_j$}
$A_p$ := attribute of $c_i$\;
$A_q$ := attribute of $c_j$\;
 \eIf{$A_p == A_q$}{
    return $\frac{\total_{A_{\emph{d}}}}{\total_{A_p}}  \cdot$ \\ $\sum_{a_p \in Dom(A_p)} \cnt_{A_p}[a_p] \cdot a_p \cdot a_p $ \;
   }{
   return $\frac{\total_{A_{\emph{d}}}}{\total_{A_p}} \cdot \sum_{a_p \in Dom(A_p), a_q \in Dom(A_q)} \cof_{A_p, A_q}[a_p, a_q] \cdot a_p \cdot  a_q $\;
  }

 \caption{Gram matrix algorithm}
\end{algorithm}

Algorithm \ref{alg:cofactor} is used to compute each element of gram matrix $c_i \cdot c_j$ where $i \leq j$.  The time complexity to compute each element is $O(\emph{w}^2)$ and the whole gram matrix is $O(\emph{m}^2\cdot\emph{w}^2)$. Even if attribute matrix has height $\emph{n}$ exponential in the number of attributes, we can use algorithm \ref{alg:cofactor} to compute gram matrix in time polynomial in $\emph{m}$.

\stitle{Left Multiplication:} Next consider left multiplication $\mathbi{A} \cdot \mathbi{X}$,
where the shape of $\mathbi{A}$ is $\emph{q} \times \emph{n}$. 
The naive matrix multiplication $\mathbi{A} \cdot \mathbi{X}$ has time complexity $O(\emph{q} \cdot \emph{n} \cdot \emph{m})$. Similar to gram matrix, we exploit the fact that the columns in attribute matrix has a lot of redundancy. For each column, we leverage $\cnt$ to infer the times each attribute value is duplicated. For ith row $r_i'$ in $\mathbi{A}$, we precompute the prefix sum of $r_i'$ in $O(\emph{n})$ to get range sum of $r_i'$ in $O(1)$. 

\begin{algorithm}[H]
\KwResult{$r_i'\cdot c_j$}
 result := 0\;
 start := 0\;
$A_p$ := attribute of $c_j$\;
 \For{ k:= 0; k < $\frac{\total_{A_{\emph{d}}}}{\total_{A_p}}$; k++:}
 {
     \For{ $a_p \in Dom(A_p)$ in ascending order }{ 
            rangeSum := $sum(r_i'[start:start + \cnt_{A_p}[a_p]]) $\;
            result+= $rangeSum\cdot  a_p$\;
            start+= $\cnt_{A_p}[a_p]$\;
      }
 }
 return  result\;
 \caption{Left multiplication algorithm}
 \label{alg:left}
\end{algorithm}

Algorithm \ref{alg:left} is used to compute each element of left multiplication $c_i \cdot c_j$. Note that the input size is $O(\emph{q} \cdot \emph{n})$ so that the lower bound of the time complexity of algorithm \ref{alg:left} is $O(\emph{q} \cdot \emph{n})$. For each $r_i'$, the first attribute only needs to iterate over attribute values and compute multiplication result in $O(\emph{w})$, while the last attribute can't utilize the prefix sum and have to iterate  $r_i'$ in  $O( \emph{w}^\emph{m})$. The total time complexity of  algorithm \ref{alg:left} is $O(\emph{q} \cdot (\emph{n} + \emph{w} + \emph{w}^2 + ... + \emph{w}^\emph{m}))  = O(\emph{q} \cdot\emph{n})$, which is optimal. 

\stitle{Right Multiplication:} Then consider right multiplication $ \mathbi{X} \cdot \mathbi{A}$,
where the shape of $\mathbi{A}$ is $\emph{n} \times \emph{p}$. The naive matrix multiplication $ \mathbi{X} \cdot \mathbi{A}$ has time complexity $O(\emph{p} \cdot \emph{n} \cdot \emph{m})$.  Algorithm \ref{alg:rightm} uses the row iterator in {\it Factorizer} to implement right multiplication by updating multiplication result from previous row. Similar to left multiplication, the output size is $O(\emph{p} \cdot \emph{n})$ so that the lower bound of the time complexity of algorithm \ref{alg:rightm} is $O(\emph{p} \cdot \emph{n})$. For each row iterator, the first attribute is updated $O(\emph{w})$ times, while the last attribute is updated $O( \emph{w}^\emph{m})$ times. The total time complexity of  algorithm \ref{alg:rightm} is $O(\emph{p} \cdot (\emph{n} + \emph{w} + \emph{w}^2 + ... + \emph{w}^\emph{m})) = O(\emph{p} \cdot\emph{w}^\emph{m}) = O(\emph{p} \cdot\emph{n})$, which is optimal. 
 
\begin{algorithm}[H]
\KwResult{$r_1\cdot c_j', r_2\cdot c_j', \cdots, r_n\cdot c_j'$}
$r_{prev}$ := the first values for all attributes\;
 $r_1\cdot c_j'$ = $r_{prev}\cdot c_j'$\;
 \For{ k:= 2; k <= n ; k++:}
 {
    A := last attribute in attribute order\;
    update := new map()\;
    update = RowItr.next(A, update)\;
    $r_n\cdot c_j'$ = $r_{n-1}\cdot c_j'$ \;
     \For{attribute $A_i$, value $v \in$ update}
    {
            
            $r_n\cdot c_j' \mathrel{-}= r_{prev}[i]\cdot c_j'[i]$\;
            $r_n\cdot c_j' \mathrel{+}= v\cdot c_j'[i]$\;
            $r_{prev}[i] = v$\;
    }
 }
 \caption{Right multiplication algorithm}
 \label{alg:rightm}
\end{algorithm}

         

\section{Matrix operations over clusters}
\label{section:matrix cluster}

We study the matrix operations over each cluster of attribute matrix  ($\mathbi{X}_{\emph{i}}^T \cdot \mathbi{X}_{\emph{i}}$, $ \mathbi{X}_{\emph{i}} \cdot \mathbi{C}_{\emph{i}}$,  $\mathbi{D}_{\emph{i}} \cdot \mathbi{X}_{\emph{i}}$ where $\emph{i} = 1,..., \mathcal{G}$) in this section. Given the initial view $V = \gamma_{A_{gb}, F_{agg}(A_{agg})}(\mathbb{R})$, we call $A_{gb}$ inter cluster attributes. After user drill-down to a hierarchy, the additional attribute $\emph{S}$ in $A_{gb}'$ is called intra cluster attribute. Because we previously require that intra cluster attribute is placed last in the attribute order, the rows in the same cluster are adjacent, so we can reuse the row iterator to iterate through clusters. We exploit the fact that, for each cluster, inter cluster attributes have the same value and reuse the row iterator to only calculate the difference between clusters. We update the previous matrix according to the difference. We also assume that attribute matrix is the same as attribute matrix for simplicity.

\stitle{Gram Matrices:} First consider gram matrices for all clusters $\mathbi{X}_{\emph{i}}^T \cdot \mathbi{X}_{\emph{i}}$ for $i = 1,..., \mathcal{G}$. The naive implementation takes $O(\emph{m}^2 \cdot \emph{w} \cdot\mathcal{G}) = O(\emph{n} \cdot \emph{m}^2)$. Algorithm \ref{alg:clustercof} computes the gram matrix for each cluster by iterating over each cluster and updating the difference. Notice that, the updates are in place and the outputs are read-only except for the last output. Even if we reuse the same matrix, the matrix is yielded $\mathcal{G}$ times and each time at least $O(\emph{m})$ elements need to be changed, so the lower bound of time complexity is $O(\emph{m} \cdot \mathcal{G})$. The first inter cluster attribute is updated $O(\emph{w})$ and the last is updated $O(\emph{w}^{m-1})$. Each update involve $O(m)$ changes in the matrix. Change for intra attributes takes $O(\emph{m})$ for $O(\mathcal{G})$ times. Therefore, the total time complexity is $O(\emph{m} \cdot (\emph{w} + ... + \emph{w}^{m-1} + \mathcal{G})) = O(\emph{m} \cdot \mathcal{G})$ which is optimal.



\begin{algorithm}[H]
\KwResult{Gram matrix for each cluster}
\SetNoFillComment
$r_{inter}$ := values of inter cluster attributes in the first cluster\;
$r_{intra}$ := value sums of intra cluster attributes in the first cluster\;
gram := compute gram matrix for the first cluster naively\;
prevSize := number of tuples in the first cluster\;
yield gram\;
\For{ k:= 2; k <= n ; k++:}{
    A := last attribute among inter cluster attributes\;
    update := new map()\;
    update = RowItr.next(A, update)\;
    curSize := number of tuples in kth cluster\;
    
     \For{attribute $A_i$, value $v \in$ update}
    {

        \For{j:= 1; j <= number of inter cluster attributes; j++:}{
            $gram[i,j] \mathrel{/}= r_{inter}[i]$\;
            $gram[i,j]  \mathrel{*}= v$\;
            $gram[i,j] \mathrel{*}= curSize / prevSize$\; 
            }

            $r_{inter}[i] = v$\;
        
    }
    
    \tcc{Cache given intra cluster attribute value}
    \For{each pair of intra cluster attributes}
    {Update corresponding gram matrix elements naively\;}
    
      \For{attribute $a_i \in$ intra cluster attributes}
    { 
        
        sum := sums of values of attribute $a_i$ in kth cluster\;
        \For{j:= 1; j <= number of inter cluster attributes; j++:}{
            $gram[i,j] \mathrel{/}= r_{intra}[i]$\;
            $gram[i,j]  \mathrel{*}= sum$\;
        }
        $r_{intra}[i] = sum$\;
    }
    prevSize :=  curSize\;
    yield gram\;
}

\caption{Cluster gram matrix iterator algorithm}
\label{alg:clustercof}
\end{algorithm}
 
\stitle{Left Multiplication:} Next consider left multiplication for all clusters $\mathbi{A}_{\emph{i}} \cdot \mathbi{X}_{\emph{i}}$ for $i = 1,..., \mathcal{G}$, where the shape of  $\mathbi{A}_{\emph{i}}$ is $\emph{q} \times \emph{n}_\emph{i}$. The naive implementation takes $O(\emph{q} \cdot\emph{n} \cdot \emph{m})$. Algorithm \ref{alg:clusterl} computes the left multiplication for each cluster by iterating over each cluster and updating the difference. The input size is $O(\emph{q} \cdot \emph{n})$ so that the lower bound of the time complexity is $O(\emph{q} \cdot \emph{n})$. Each row in each $\mathbi{A}_{\emph{i}}$ takes $O(\emph{m} + \emph{w})$. Therefore, the total time complexity is $O(\emph{q} \cdot \mathcal{G} \cdot (\emph{m} + \emph{w})) = O(\emph{q} \cdot \emph{m} \cdot \mathcal{G} + \emph{q} \cdot \emph{n})$.

\begin{algorithm}[H]
\KwResult{Left multiplication with $r_{i,k}'$ for $\emph{k}th$ cluster}
\SetNoFillComment
$r_{inter}$ := values of inter cluster attributes in the first cluster\;
result := compute result for the first cluster naively\;
yield result\;
\For{ k:= 2; k <= n ; k++:}{
    A := last attribute in inter cluster attributes\;
    update := new map()\;
    update = RowItr.next(A, update)\;
    rowSum := sum($r_{i,k}'$)\;
    
     \For{attribute $A_i$, value $v \in$ update}
    {
            
            
        $r_{inter}[i] = v$\;
    }
    
    \For{each inter cluster attribute $A_i$}
    {
        $result[i] \mathrel = r_{inter}[i] * rowSum $\;
    }
    
    \For{attribute $A \in$ intra cluster attributes}
    {Update corresponding result matrix elements naively\;}
    yield result\;
}
 \caption{Cluster left multiplication iterator algorithm}
 \label{alg:clusterl}
\end{algorithm}

\stitle{Right Multiplication:} Finally, consider right multiplication for all clusters $\mathbi{X}_{\emph{i}} \cdot \mathbi{A}_{\emph{i}}$ for $i = 1,..., \mathcal{G}$, where the shape of  $\mathbi{A}_{\emph{i}}$ is $\emph{m} \times \emph{p}$. The naive implementation takes $O(\emph{p} \cdot\emph{n} \cdot \emph{m})$. Algorithm \ref{alg:clusterr} computes the right multiplication for each cluster. The output size is $O(\emph{p} \cdot \emph{n})$ and, for each cluster, we can always find $\mathbi{A}_{\emph{i}}$ such that all elements in the output have to change. Therefore the lower bound of the time complexity is $O(\emph{p} \cdot \emph{n})$. Each column in each $\mathbi{A}_{\emph{i}}$ takes $O(\emph{m} + \emph{w})$. Therefore, the total time complexity is $O(\emph{p} \cdot \mathcal{G} \cdot (\emph{m} + \emph{w})) = O(\emph{p} \cdot \emph{m} \cdot \mathcal{G} + \emph{p} \cdot \emph{n})$.

\begin{algorithm}[H]
\KwResult{Right multiplication with $c_{i,k}'$ for $\emph{k}th$ cluster}
$r_{inter}$ := values of inter cluster attributes in the first cluster\;
result := compute result for the first cluster naively\;
yield result\;
\For{k:= 2; k <= n ; k++:}{
    A := last attribute in inter cluster attributes\;
    update := new map()\;
    update = RowItr.next(A, update)\;
    \For{attribute $A_i$, value $v \in$ update}
    {
            
        $r_{inter}[i] = v$\;
    }
    base = $\sum_{j=0}^{number\HS of \HS inter\HS cluster\HS attributes} r_{inter}[j] \times  c_{i,k}'[j]$;
    \For{j:= 1; j <= $\emph{n}_k$; j++:}{
        value := 0\;
        
         \For{attribute $A_i \in$ intra cluster attributes}{
            value $\mathrel{+}= jth\HS value\HS of\HS A_i \HS in\HS \emph{k}th\HS cluster \times c_{i,k}'[i]$
        }
        result[j] := base + value\;
    }
    yield result\;
}
 \caption{Cluster right multiplication iterator algorithm}
  \label{alg:clusterr}
\end{algorithm}

\begin{figure}
  \centering
      \includegraphics [width=\columnwidth] {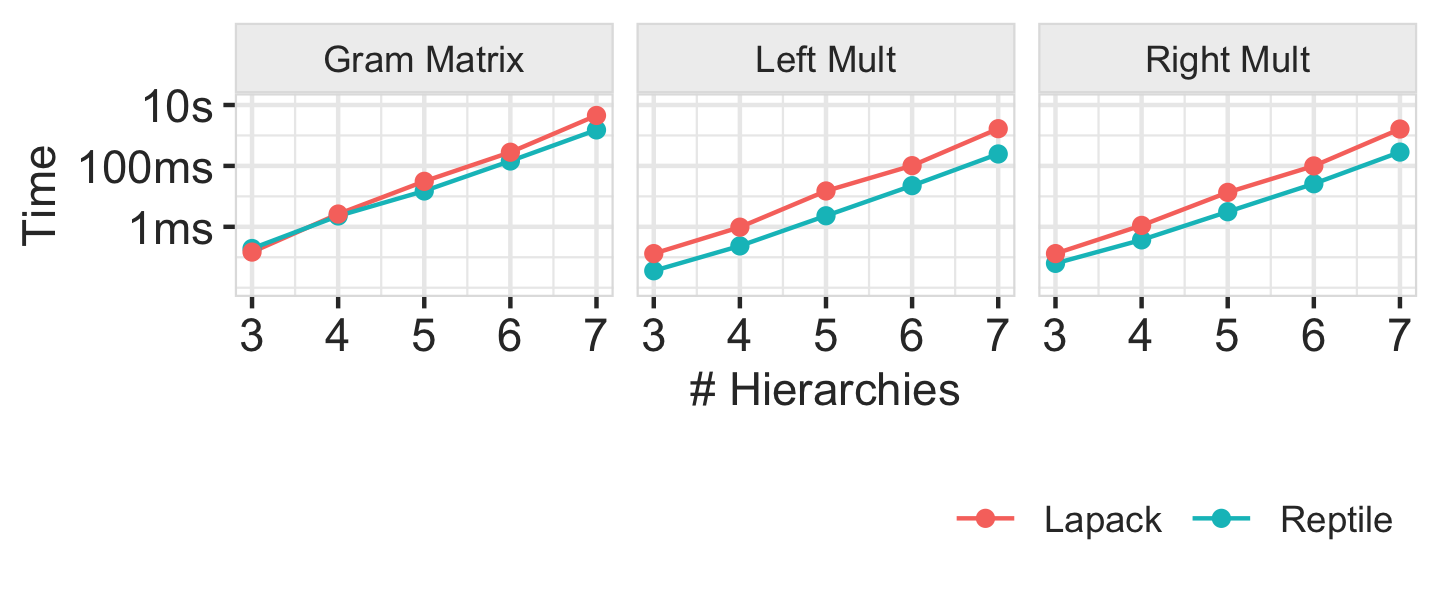}
  \caption{Matrix operation over clusters runtimes compared to Lapack-based implementation.}
  \label{fig:clustermatrixops}
\end{figure}

\stitle{Evaluation:} We evaluate the performance of matrix operation using synthetic datasets with $\emph{d}$ hierarchies. For each hierarchy, there are three attributes. Each attribute contains $\emph{w} = 10$ unqiue values. Given $\emph{d}$ attributes, the total number of rows $\emph{n} = 10^\emph{d}$. $\mathbi{X}$ has the shape $10^\emph{d} \times 3 \cdot \emph{d}$ and each cluster $\mathbi{X}_{\emph{i}}$ has the shape $10 \times 3\cdot \emph{d}$ for $i = 1,...,\mathcal{G}$. There are $10^{\emph{d}-1}$ clusters in total. For right Multiplications over clusters $ \mathbi{X}_{\emph{i}} \cdot \mathbi{C}_{\emph{i}}$, $\mathbi{C}_{\emph{i}}$ has the shape $3\cdot \emph{d} \times 1$. For each Left Multiplications over clusters  $\mathbi{D}_{\emph{i}} \cdot \mathbi{X}_{\emph{i}}$, $\mathbi{D}_{\emph{i}}$ has the shape $1 \times 10$. We randomly generate matrix to be multiplied.

\Cref{fig:clustermatrixops} reports runtimes in log scale. At 7 hierarchies, \sys is $3\times$ faster for gram matrix, $5.8\times$ faster for left multiplication, and $6.9\times$ faster for right multiplication. Overall, \sys outperforms Lapack.


\section{Matrix operation over general factorised representation}
\label{section:matrixgeneral}

In this section, we briefly discuss how to extend Matrix operations over general factorised representation. 

Given general F-tree, we first need to determine attribute order. We require that for any pair of attributes in attribute order, attribute before doesn't transitively depends on attribute after. This requirement is to ensure that row iterator can work properly. Iterator for attribute after should increment first and propagate the change to iterator for attribute before. 

The first extension is, for relation without functional dependency,  the join operator and aggregation operator need to record the order of tuples even if tuples have same value. Consider the following example:

\begin{example}[Order in operator]
Given relation $ R = [(a_1,b_1),\\(a_1, b_2),(a_2, b_1)]$ over schema $\emph{S} = [A, B]$, where there is no functional dependency. After marginalize out attribute A, the result is an ordered map $\bigoplus_A R = \{b_1: 2, b_2: 1\}$. However, the information that $b_2$ is in middle of two $b_1$ is lost, which is necessary during multiplication as we need to infer the positions. One solution would be that, aggregation operator returns an ordered list: $\bigoplus_A R = [b_1: 1, b_2: 1, b_1: 1]$.
\end{example}

The second extension is to redefine the aggregation query. Given a set of attributes $S$, let $\emph{dep}(S)$ be the dependency set of S. Dependency set $\emph{dep}(S)$ is the set with minimum size that, for each attribute in $\emph{dep}(S)$, its dependency is also in $\emph{dep}(S)$. Let $\emph{rel}(S)$ be  the set of relations whose schema contains any attribute in $S$.  Let $\emph{after}(A)$ be the set of attributes after $A$ in attribute order including $A$.

Then aggregation queries are redefined as:
\begin{align}
\total_{A_k} = &\bigoplus_{\emph{dep}(\emph{after}(A))} \bigotimes_{\emph{rel}(\emph{dep}(\emph{after}(A)))}R \nonumber\\
\cnt_{A_k} =& \bigoplus_{\emph{dep}(\emph{after}(A))/\{A_k\}} \bigotimes_{\emph{rel}(\emph{dep}(\emph{after}(A)))}R \nonumber\\
\cof_{A_k, A_j} = &\bigoplus_{\emph{dep}(\emph{after}(A))/\{A_k,A_j\}} \bigotimes_{\emph{rel}(\emph{dep}(\emph{after}(A)))}R\nonumber
\end{align}

In general, the dependency set $\emph{dep}(\emph{after}(A)))$ may include all attributes if all attributes have no-empty dependency. Because operator needs to store the order of attribute value, in the worst case, the join result may be as large as the fully joined relation even if attributes  are marginalized early.

\section{Multi-attribute features}
\label{section:mulf}

In this section, we discuss the extension to multi-attribute features. If the number of attributes is a constant, previous time complexity analyses still apply. In the worst case, if all the features are multi-attribute features related to all the attributes, there would be no redundancy in feature matrix, and our solution would be the same as the naive solution.

For multi-attribute external feature, user may have dataset that maps multiple attributes to feature values. Given a list of attribute $\mathbb{A}$ with $\emph{k}$ attributes, tuple of attribute value $(\emph{a}_1,..., \emph{a}_\emph{k})\in Dom(\mathbb{A})$ and aggregation function $F_{agg}$, assume that there is an external dataset $R_{external}$ which maps $(\emph{a}_1,..., \emph{a}_\emph{k})$ to its feature value, the external feature is then:
\begin{align}
feature_{external}[(\emph{a}_1,..., \emph{a}_\emph{k})] = R_{external}[(\emph{a}_1,..., \emph{a}_\emph{k})] \nonumber
\end{align}

To register multi-attribute external feature, user needs to provide external dataset $R_{external}$, a list of attributes $\mathbb{A}$ and target aggregation function $F_{agg}$ . 

For multi-attribute custom feature, given a list of attribute $\mathbb{A}$ with $\emph{k}$, tuple of attribute value $(\emph{a}_1,..., \emph{a}_\emph{k})\in Dom(\mathbb{A})$ and aggregation function $F_{agg}$, the custom feature is then:
\begin{align}
feature[(\emph{a}_1,..., \emph{a}_\emph{k})] = & \sigma_{\emph{pred}}(\gamma_{\mathbb{A}, \emph{F}'}(\emph{Q})) \nonumber
\end{align}
where $\emph{pred}$ is the predicate of selection, and $\emph{F}'$ is the aggregation function which user can customize. User needs to provide the predicate $\emph{pred}$, aggregation function $\emph{F}'$,  a list of attributes $\mathbb{A}$ and the target aggregation function $F_{agg}$ to register derived variable as feature. 

For multi-attribute feature registered with a list of attributes $\mathbb{A}$ and target aggregation function $F$, given the view $V' = \gamma_{A_{gb}', F_{agg}(A_{agg})}(R)$, this feature is applicable if $\mathbb{A} \subseteq A_{gb}' \wedge F_{agg} = \emph{F}$.

For feature matrix, all the multi-attribute features are appended to end of columns. Hierarchy order, attribute order and attribute matrix remain unchanged.

\begin{algorithm}[H]
\label{alg:cofactormulti}
\KwResult{$c_i \cdot c_j$}
 $f_i$ := feature for $c_i$\;
 $f_j$ := feature for $c_j$\;
 $\mathbb{A}_p$ := list of attribute of $f_i$\;
 $\mathbb{A}_q$ := list of attribute of $f_j$\;
 $\mathbb{A}$ := $\mathbb{A}_q \cup \mathbb{A}_p$\;
 $\emph{k}$ := size of $\mathbb{A}$ \;
 $A_{first}$ := first attribute in $\mathbb{A}$\;
 $A_{last}$ := last attribute in $\mathbb{A}$\;
 $\mathbb{A}_{all}$ := all attributes before $A_{last}$ \;
 $\cof = \bigoplus_{\mathbb{A}_{all}/\mathbb{A}}    \pi_{A_{last}}(R_{last})\bigotimes_{i \in [_{last} - 1]}R_i$ \;
 
 return $\frac{\total_{A_{\emph{d}}}}{\total_{A_{first}}}  \cdot$  $\sum_{(\emph{a}_1,..., \emph{a}_\emph{k}) \in Dom(\mathbb{A})} \cof\HS[(\emph{a}_1,..., \emph{a}_\emph{k})] \cdot f_i(\sigma_{\mathbb{A}_p}(\emph{a}_1,..., \emph{a}_\emph{k})) \cdot  f_j(\sigma_{\mathbb{A}_q}(\emph{a}_1,..., \emph{a}_\emph{k})) $ \;

 \caption{gram matrix algorithm for multi-attribute features}
\end{algorithm}
For matrix operations, first consider gram matrix. Algorithm \ref{alg:cofactormulti} computes gram matrix element and Algorithm \ref{alg:leftmulti} computes left multiplication for multi-attribute features. We assume that attributes in $\mathbb{A}_p$, $\mathbb{A}_q$ and $\mathbb{A}$ are ordered by the attribute order. Right  multiplication is similar to algorithm \ref{alg:rightm}, except that, for each update, the change is $f_{idx}((\emph{a}_1,..., \emph{a}_\emph{k}))\cdot c_j'[idx]$ instead of $f_{idx}(\emph{a})\cdot c_j'[idx]$.

\begin{algorithm}[H]
\KwResult{$r_i'\cdot c_j$}
 result := 0\;
 start := 0\;
 $f_j$ := feature of $c_j$\;
 $\mathbb{A}_p$ := list of attribute of $f_j$\;
 $\emph{k}$ := size of $\mathbb{A}_p$ \;
 $A_{first}$ := first attribute in $\mathbb{A}_p$\;
 $A_{last}$ := last attribute in $\mathbb{A}_p$\;
 $\mathbb{A}_{all}$ := all attributes before $A_{last}$ \;
 $\cof = \bigoplus_{\mathbb{A}_{all}/\mathbb{A}_p}    \pi_{A_{last}}(R_{last})\bigotimes_{i \in [_{last} - 1]}R_i$ \;
 
 \For{ k:= 0; k < $\frac{\total_{A_{\emph{d}}}}{\total_{A_first}}$; k++:}
 {
     \For{ $(\emph{a}_1,..., \emph{a}_\emph{k}) \in Dom(\mathbb{A}_p)$ in ascending order }{ 
            rangeSum := $sum(r_i'[start:start + \cof\HS[(\emph{a}_1,..., \emph{a}_\emph{k})]]) $\;
            result+= $rangeSum\cdot  f_j((\emph{a}_1,..., \emph{a}_\emph{k}))$\;
            start+= $\cof\HS[(\emph{a}_1,..., \emph{a}_\emph{k})]$\;
      }
 }
 return  result\;
 \caption{Left multiplication algorithm for multi-attribute features}
 \label{alg:leftmulti}
\end{algorithm}

\section{Multi-query execution}
\label{section:multiquery}

Suppose there are $\emph{d}$ attributes in attribute order. For each model training, there are $2\emph{d} + \frac{\emph{d}(\emph{d}-1)}{2}$ queries to execute. One naive way to execute these queries is to join all relations together and apply aggregation function.

We can rewrite the queries such that these quries can reuse results from other queries:
\begin{align}
\cnt_{A_1} =&\HS \pi_{A_1}(R_1)\nonumber\\
\cnt_{A_k+1} =& \bigoplus_{A_{k}} \cof_{A_{k+1}, A_k} \HS for\HS k = 2,...,\emph{d} -1\nonumber\\
\total_{A_k} =& \bigoplus_{A_{k}} \cnt_{A_k}\HS for\HS k = 1,...,\emph{d}\nonumber\\
\cof_{A_{k}, A_{k-1}} =& \pi_{A_k}(R_{k}) \bigotimes R_{k-1} \bigotimes \cnt_{A_k}\HS for\HS k = 2,...,\emph{d}\nonumber\\
\cof_{A_k, A_j} =& \bigoplus_{A_{k-1}} \pi_{A_k}(R_{k}) \bigotimes R_{k-1} \bigotimes \cof_{A_{k-1}, A_j} \nonumber\\
& \HS for \HS k = 1,...,\emph{d}, j = 1,...,\emph{d}, k > j + 1 \nonumber
\end{align}

Algorithm \ref{alg:mulquery} leverages the dependency to compute query results. The naive solution materializes the join result and apply aggregation functions with total time complexity $O({\emph{d}}^2 \cdot \emph{w}^{\emph{d}})$. For algorithm \ref{alg:mulquery}, attributes in join results are marginalized as soon as possible when they are no longer used in the future queries. The join results are also stored in factorised representations. For \cof \space between different hierarchies, we are computing the Cartesian Products. \sys exploits the independence by storing factorised representation. For implementation, only pointers to two relations are stored in O(1). Because we assume that the total number of rows in the relations of each hierarchy is $O(\emph{w})$, join operator between attributes in the same hierarchy takes $O(\emph{w})$. Suppose there are $O(|\mathbb{H}|)$ hierarchies, each with $O(\emph{t})$ attributes and $O(|\mathbb{H}|\cdot \emph{t}) = O(\emph{d})$. The total time complexity for algorithm \ref{alg:mulquery} is $O(|\mathbb{H}|^2 \cdot \emph{t}^2  + |\mathbb{H}|\cdot \emph{t}^2 \cdot \emph{w})$. 
If $\emph{w}$ is much larger than $|\mathbb{H}|$, the time complexity is then 
 $O(|\mathbb{H}|\cdot \emph{t}^2 \cdot \emph{w})$.

\begin{algorithm}[H]
\label{alg:multiquery}
\KwResult{Query results}
 $\cnt_{A_1} := \pi_{A_1}(R_1)$\;
 $\total_{A_1} = \bigoplus_{A_{1}} \cnt_{A_1}$\;
 $\cof_{A_{2}, A_{1}} = \pi_{A_2}(R_{2}) \bigotimes R_{1} \bigotimes \cnt_{A_1}$
 
 \For{i := 3; i <= $\emph{d}$; i++}
 {
     $\cof_{A_{i}, A_{1}} =  \bigoplus_{A_{i-1}} \pi_{A_i}(R_{i}) \bigotimes R_{i-1} \bigotimes \cof_{A_{i-1}, A_1}$\;
 }
 
 \For{i := 2; i <= $\emph{d}$; i++}
 {
    $\cnt_{A_i} = \bigoplus_{A_{i-1}} \cof_{A_{i}, A_{i-1}}$\;
    $\total_{A_i} = \bigoplus_{A_{i}} \cnt_{A_i}$\;
    \If{i < $\emph{d}$}
    {$\cof_{A_{i+1}, A_{i}} = \pi_{A_{i+1}}(R_{i+1}) \bigotimes R_{i} \bigotimes \cnt_{A_i}$\;}

    \For{j := i+2; j <= $\emph{d}$; j++}
    {
     $\cof_{A_{j}, A_{i}} =  \bigoplus_{A_{j-1}} \pi_{A_j}(R_{j}) \bigotimes R_{j-1} \bigotimes \cof_{A_{j-1}, A_i}$\;
    }
 }
 \caption{Mutiple query plan}
 \label{alg:mulquery}
\end{algorithm}

\section{Drill-down}
\label{section:drilldown}

Drilling down an attribute involves two steps:

1. append the attribute to the corresponding hierarchy.

2. move the hierarchy to the end of hierarchy order. 

After the drill-down operation, F-tree has an additional attribute, and all attributes in one hierarchy is moved to the bottom of the tree. One naive way to implement drill-down operation is to rebuilt the F-tree and recompute all aggregation results from scratch. 
Assume that $\emph{w}$ is much larger than $|\mathbb{H}|$,  the time complexity to drill-down all hierarchies is then $O(|\mathbb{H}|^2\cdot \emph{t}^2 \cdot \emph{w})$.

We then introduce optimization to reuse the aggregation results from the previous drill-down. The main property we exploit is the independence between hierarchies. Given the aggregation query over the Cartesian's Product, we can marginalize each relation before join:
\begin{align}\bigoplus_{\emph{A} \in \emph{S}} \bigotimes_{i \in [k]} R_\emph{i}[\emph{S}_\emph{i}] =  \bigotimes_{i \in [k]} (\bigoplus_{\emph{A} \in \emph{S}_{i}} R_\emph{i}[\emph{S}_\emph{i}]) \nonumber
\end{align}
where k is the number of relations, $\emph{S}_\emph{i}$ is the schema of $R_\emph{i}$ and $\emph{S} = \bigcup_{i \in [k]}\emph{S}_\emph{i}$ is the schma of the join result. For $i \neq j$, $\emph{S}_\emph{i} \cap \emph{S}_\emph{j} = \emptyset$.

Notice that, between different hierarchies, we need to compute cartesian product. Suppose that there are $\emph{t}$ hierarchies. For each hierarchy $D_\emph{i}$ for $\emph{i} = 1, ..., \emph{t}$, let $[D_\emph{i}]$ be the set of indices of attributes under this hierarchy. Given hierarchy order $D_\emph{t},...,D_1$, define:

\begin{align}
\total_{D_\emph{k}} =  \bigoplus_{A_\emph{i}: \emph{i} \in [D_\emph{k}]} \bigotimes_{\emph{i} \in [D_\emph{k}]} R_\emph{i} \nonumber
\end{align}
for $\emph{k} = 1,...,\emph{t}$. $\total_{D_\emph{k}}$ outputs the number of tuples in the hierarchy $D_\emph{k}$.

Therefore, we can rewrite all the queries to exploit the independence between hierarchies. Assume that attribute $A_k$ is in hierarchy $D_s$, attribute $A_j$ is in $D_v$ and $k>j$:
\begin{flalign}
& \total_{A_k} && \nonumber \\
=&  \bigoplus_{A_{1}} \hdots \bigoplus_{A_{k}} \pi_{A_k}(R_k)\bigotimes_{\emph{i} \in [k-1]}R_\emph{i} && \nonumber \\
=&\HS(\bigoplus_{A_\emph{i}: \emph{i} \in [D_\emph{d}]} \bigotimes_{\emph{i} \in [D_\emph{d}]}R_i) \bigotimes  \hdots (\bigoplus_{A_\emph{i}: \emph{i} \in [D_\emph{\emph{s}-1}]} \bigotimes_{\emph{i} \in [D_\emph{\emph{s}-1}]}R_i)
\bigotimes && \nonumber \\
& \HS(\bigoplus_{A_\emph{i}: \emph{i} \in [D_\emph{\emph{s}}] \wedge \emph{i} \leq k } \pi_{A_k}(R_k) \bigotimes_{\emph{i} \in [D_\emph{\emph{s}}] \wedge \emph{i} < k }R_i) && \nonumber \\
=& \bigotimes_{\emph{i} \in [s-1]}\total_{D_\emph{i}} \bigotimes (\bigoplus_{A_\emph{i}: \emph{i} \in [D_\emph{\emph{s}}] \wedge \emph{i} \leq k} \pi_{A_k}(R_k)\bigotimes_{\emph{i} \in [D_\emph{\emph{s}}] \wedge \emph{i} < k  }R_i)&& \nonumber \\ 
&\HS for\HS \emph{k} = 1,...,\emph{d}&& \nonumber
\end{flalign}

\begin{flalign}
&\cnt_{A_k} &&\nonumber \\
=& \bigotimes_{\emph{i} \in [s-1]}\total_{D_i}\bigotimes (\bigoplus_{A_\emph{i}: \emph{i} \in [D_\emph{\emph{s}}] \wedge \emph{i} < k } \pi_{A_k}(R_k)\bigotimes_{\emph{i} \in [D_\emph{\emph{s}}] \wedge \emph{i} < k }R_i)&& \nonumber \\ 
& for\HS \emph{k} = 1,...,\emph{d} &&\nonumber
\end{flalign}

\begin{flalign}
& \cof_{A_k, A_j} &&\nonumber \\
=& \bigotimes_{i \in [v-1]}\total_{D_i} \bigotimes (\bigoplus_{A_\emph{i}: \emph{i} \in [D_\emph{\emph{v}}] \wedge \emph{i} < \emph{j} } \pi_{A_j}(R_j)\bigotimes_{\emph{i} \in [D_\emph{\emph{v}}] \wedge i < j }R_i) && \nonumber\\ 
&\bigotimes_{i \in [v+1,s-1]}\total_{D_i} \bigotimes (\bigoplus_{A_\emph{i}: \emph{i} \in [D_\emph{\emph{s}}] \wedge \emph{i} < \emph{k} } \pi_{A_k}(R_k)\bigotimes_{\emph{i} \in [D_\emph{\emph{s}}] \wedge \emph{i} < \emph{k} }R_i)&&\nonumber \\
& \HS for \HS \emph{k} = 1,...,\emph{d}, \emph{j} = 1,...,\emph{d}, \emph{k} > \emph{j} && \nonumber
\end{flalign}

After rewriting the queries, we can exploit the fact that, when drill-down attribute $A_k$ is in hierarchy $D_s$, for $\total_{A_i}$, $\cnt_{A_i}$ and $\cof_{A_i, A_j}$ where $A_i$ and $A_j$ are not in hierarchy $D_s$, only the parts $ \bigotimes \total_{D}$ are affected, which are scalars. For join operator $\bigotimes$, when multiplied by scalar, we don't need to apply the multiplication to each tuple. We can maintain a scalar for each relation as the zoom value so that multiplication by scalar is in O(1).


Algorithm \ref{alg:drilldown} shows how to update the aggregation results after drill-down. For aggregation results involved with only the attributes in the hierarchy to drill-down, we have to recompute them in $O( \emph{t}^2 \cdot \emph{w})$. However, for other attributes, the updates can be done in $O(1)$. The total time complexity would be $O(\emph{t}^2 \cdot \emph{w})$. 

In algorithm \ref{alg:drilldown}, we also cache \total', \cnt', \cof', and $\total_{D_v}'$ involved with only attributes in the hierarchy to drill-down because, given the query in equation \ref{querydefinition} and drill-down hierarchy $H$, these aggregation results will always be the same independent of the current view. Consider a scenario when users make complaint twice. For the first complaint, \sys drills down each hierarchy in $ \mathbb{H}$ and selects one optimal hierarchy $H^* \in \mathbb{H}$ as in \Cref{optimizationpro}. The time complexity for the first complaint is $O(|\mathbb{H}|\cdot \emph{t}^2 \cdot \emph{w})$. 
For the second complaint, without cache, \sys needs to recompute each hierarchy in $\mathbb{H}$ and the total time is also $O(|\mathbb{H}|\cdot \emph{t}^2 \cdot \emph{w})$.  If all the attributes not selected in the first complaint $\mathbb{H} / \{H^*\}$ are cached, each cached hierarchy can be updated in $O({\emph{d}}^2)$ for the second complaint. So the total time for the second complaint would be $O(\emph{t}^2 \cdot \emph{w})$.

\begin{algorithm}[H]
\KwResult{Updated query results after drill-down}

$A_{new}$:= Attribute in $D_v$ to drill-down\;
$A_u,\cdots,A_{u+t}$ := Attributes in $D_v$ from lowest to highest level\;
\SetNoFillComment 
 \tcc{Cache \total', \cnt', \cof', and $\total_{D_v}'$}
Compute updated \total', \cnt' and \cof' involved with only attributes $A_{new}, A_u, \cdots, A_{u+t}$ using Algorithm~\ref{alg:multiquery}\;
$\total_{D_v}' =  \bigoplus_{A_\emph{i}: \emph{i} \in [D_\emph{v}]} \bigotimes_{\emph{i} \in [D_\emph{v}]}R_i$\;

\For{Attribute $A_k \notin \{A_{new}, A_u, \cdots, A_{u+t}\}$}
{\eIf{$k < u$}
{$\cnt_{A_k}' = \cnt_{A_k} \bigotimes \total_{D_v}'$\;
$\total_{A_k}' =\total_{A_k} \bigotimes \total_{D_v}'$\;}
{$\cnt_{A_k}' = \cnt_{A_k} \bigotimes ( \total_{D_v}'/  \total_{D_v}$)\;
$\total_{A_k}' =\total_{A_k} \bigotimes (\total_{D_v}'/ \total_{D_v}$)\;}

\For{Attribute $A_j \in \{A_{new}, A_u, \cdots, A_{u+t}\}$}
{$\cof_{A_k, A_j} = \cnt_{A_k}' \bigotimes  \cnt_{A_j}'/ \total_{D_v}'$}

\For{Attribute $A_j \notin \{A_{new}, A_u, \cdots, A_{u+t}\} \wedge k>j $}{
    \eIf{$u > k > j$}
    {$\cof_{A_k, A_j}' = \cof_{A_k, A_j} \bigotimes \total_{D_v}'$}
    {$\cof_{A_k, A_j}' = \cof_{A_k, A_j} \bigotimes (\total_{D_v}' / \total_{D_v})$}
}
}

 \caption{drill-down hierarchy $D_v$}
 \label{alg:drilldown}
\end{algorithm}

\section{Quality of multi-level model}
\label{modelev}
\begin{figure}
  \centering
      \includegraphics [scale=0.65] {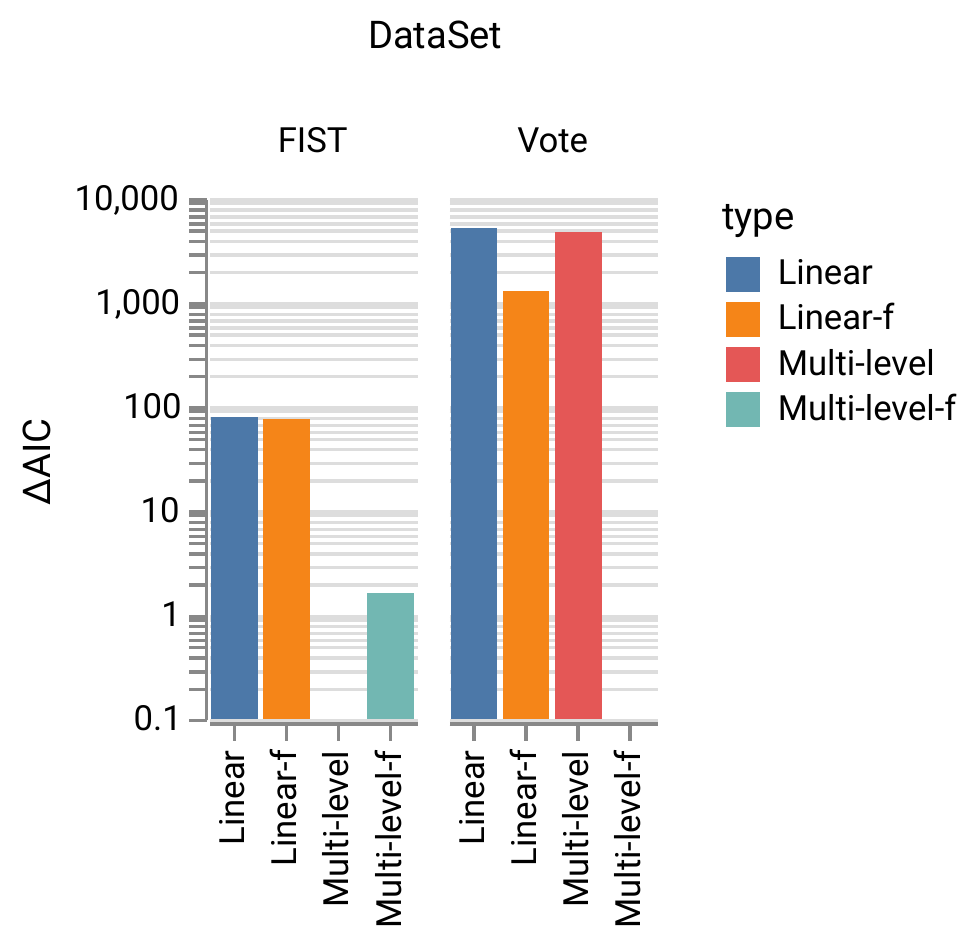}
  \caption{Model evaluation}
  \label{fig:modelaic}
\end{figure}

We conduct model evaluation between linear regression model and multi-level with default features only, and with external features. The following two datasets are considered:

\textbf{FIST}: 
This dataset contains the farmer reported drough severity at different villages in different years in Ethiopia. There are 2 hierarchies: year (one attribute with 36 values), location (three attributes: region, district and village, with 161 village values). Sensing data of rainfall are available each year for each village, which are used as external feature. The mean drought severity has been estimated using different models.

\textbf{Vote}: 
This dataset contains the 2020 presidential election vote results at different counties in the United States. There are 1 hierarchies: location (two attributes: state, and county, with 3147 county values). 2016 presidential election vote results at different counties are available, which are used as external feature. The percentage of votes for Donald Trump has been estimated using different models.

To evaluate model performance, we use Akaike information criterion (AIC) \cite{akaike1998information}, which estimates the qualities of a collections of models. AIC makes the trade-off between both goodness of fit of the model and the simplicity of the model. Given the same set of data, model with low AIC scores are considered to be relatively better. For each dataset, the difference of AIC: $\Delta AIC_i = AIC_i - AIC_{min}$ for $\emph{i}$th model is shown, where $AIC_{min}$ is the lowest AIC among the collection of models. As a rule of thumb, for the same dataset, one model is considered to be substantially better than the other if the difference of AIC is larger than 10 \cite{burnham2004multimodel}.

The result of model evaluation is shown in \Cref{fig:modelaic}. Linear is the linear regression model with only default features. Linear-f is the linear regression model with both default and auxiliary features. Multi-level is the multi-level model with default features only. Multi-level-f is the multi-level model with both default and auxiliary features. For FIST dataset, multi-level models are substantially better than linear regression models. For Vote dataset, multi-level model with auxiliary feature is substantially better than linear regression model with auxiliary feature. Because the vote results in 2016 are strong predictors of vote results in 2020, models with auxiliary feature are substantially better than models without auxiliary feature.

\section{Case Study Details: COVID-19}
\label{section:coviddetail}

In this section, we discuss the details of COVID-19 case study.

We first discuss the basic setting of \sys. COVID-19 dataset includes global data and United States data. For global data, because of the large number of countries, we further cluster countries by regions. \sys use 1 day and 7 day lag features for trend and seasonality. Given data cleaning issues, we create complaint at the higher level of geographical hierarchy. For instance, given issue that the total confirmed cases in Texas is under reported on Jan 21 2021, we complain that the total confirmed cases is too low in the United States on that day. (It is also possible to make complaint at the higher level of time hierarchy. In the experiment, we only make complaint about one day because, for COVID-19 dataset, people tend to focus on daily number across different locations. For all issues we studied, people make complaint about different locations on one specific day instead of the whole month/year.)
We then check if different approaches can successfully recommend the cluster with data cleaning problems. 

\begin{table}[h!]
\begin{center}
\setlength{\tabcolsep}{0.2em} 
\begin{tabular}{ | m{2.5em} | m{18em} | m{10pt}| m{10pt} | m{10pt} |} 
\hline
ID & Issue & RP & ST & SP \\ 
\hline
3572 & Texas confirmed missing reports & \checkmark  &  & \\ 
\hline
3521 & Arizona death methodology altered & \checkmark  &  & \\ 
\hline
3482 & Washington missing reports & \checkmark  &  & \\ 
\hline
3476 & $\bigstar$ Utah missing source  &   &  & \\ 
\hline
3468 & New York death missing reports & \checkmark  &  & \\ 
\hline
3466 & Montana missing reports & \checkmark  &  & \\ 
\hline
3456 & North Dakota confirmed backlog & \checkmark  &  & \\ 
\hline
3451 & Iowa death missing reports & \checkmark  &  & \\ 
\hline
3449 & Arizona test over reported & \checkmark  &  & \\ 
\hline
3448 & Washington death wrongly reported & \checkmark  &  & \\ 
\hline
3441 & $\bigstar$ Albany confirmed day shift &  &  & \\ 
\hline
3438 & Ohio confirmed backlog & \checkmark  &  & \\ 
\hline
3424 & Massachusetts confirmed backlog &  &  & \\ 
\hline
3416 & Nevada death over reported & \checkmark  &  & \\ 
\hline
3414 & Eureka death over reported & \checkmark  &  & \\ 
\hline
3402 & Washington confirmed typo &  &  & \\ 
\hline
\end{tabular}

\end{center}
\caption{List of COVID-19 issues in the US. RP is \sys, ST \texttt{Sensitivity}, and SP is \texttt{Support}. Prevalent errors are highlighted with $\bigstar$.}
\label{table:covid}
\end{table}

\begin{table}[h!]
\begin{center}
\setlength{\tabcolsep}{0.2em} 
\begin{tabular}{ | m{2.5em} | m{18em} | m{10pt}| m{10pt} | m{10pt} |} 
\hline
ID & Issue & RP & ST & SP \\ 
\hline
3623 & Germany recovered over reported & \checkmark &  &\\ 
\hline
3618 & $\bigstar$ Quebec death missing source  &  &  & \\ 
\hline
3578 & US recovery nullified & \checkmark  & \checkmark & \\ 
\hline
3567 & India confirmed missing reports & \checkmark  &  & \\ 
\hline
3546 & $\bigstar$ Thailand confirmed missing source &   &  & \\ 
\hline
3538a & Mexico confirmed definition altered & \checkmark  &  & \\ 
\hline
3538b & Mexico confirmed missing reports & \checkmark  &  & \\ 
\hline
3518 & $\bigstar$ Sweden death missing source &   &  & \\ 
\hline
3498 & $\bigstar$ Alberta missing source &   &  & \\ 
\hline
3494 &  UK death missing reports & \checkmark  &  & \\ 
\hline
3471 & Turkey confirmed definition altered & \checkmark & \checkmark & \checkmark\\ 
\hline
3423 & Afghanistan confirmed wrongly reported &  &  & \\ 
\hline
3413 & France missing reports & \checkmark  &  & \\ 
\hline
3408 & Kazakhstan confirmed over reported & \checkmark  &  & \\ 
\hline

\end{tabular}

\end{center}
\caption{List of global COVID-19 issues.}
\label{table:globalcovid}
\end{table}

The details of issues for US and global are in \Cref{table:covid} and \Cref{table:globalcovid} respectively. We highlight one type of errors: prevalent errors. Prevalent errors are defined as errors widespread across all time or locations. For example, some sources of confirmed and death are missing over the course of the pandemic in Utah, which affects data all the time and makes result inconsistent with official report on Dec 18 2020. The other non-prevalent common issues  are missing report (e.g. the reports of confirmed cases in Texas are missing on Jan 15 2021), data backlog (e.g. confirmed cases are not fully updated for North Dakota, and spike on Dec 9 2020), change of definition (e.g., Arizona updated guidance for identifying deaths, which causes abnormally high deaths on Jan 5 2021), etc.


Overall, \sys outperforms \texttt{Sensitivity} and \texttt{Support} because \texttt{Sensitivity} and \texttt{Support} only recommend outliers. For example, given complaint that the \texttt{COUNT} of confirmed cases is too high, \texttt{Sensitivity}  and \texttt{Support} always choose the location with the highest \texttt{COUNT} of confirmed cases, disregarding the fact that these locations have the highest population and the high \texttt{COUNT} is normal.

Next, we discuss issues which \sys fail to identify. \sys fail to detect all prevalent errors. Since prevalent errors repeat across large number of clusters, \sys is unable to tell if these clusters are all normal or all problematic. Besides prevalent errors, \sys is unable to identify errors whose effects are not strong enough and are masked by noises from other clusters. For issue 3424, there is a backlog of 680 confirmed cases in Massachusetts on Dec 18 2020, which is relatively small given that there are 290578 total confirmed cases and 4853 new cases in Massachusetts on that day. For issue 3423 there is a decrease of confirmed case from 46980 to 46718 on Dec 3 2020 which is relatively small. For issue 3402, there is a typo for the number of confirmed cases in Washington on Dec 18 2020, whose difference is relatively small.

\section{Case Study Details: FIST}
\label{section:userstudydetail}
In this section, we show the user interface and discuss two complaints that our system fail to identify all causes.

\begin{figure*}
  \centering
      \includegraphics [scale=0.45] {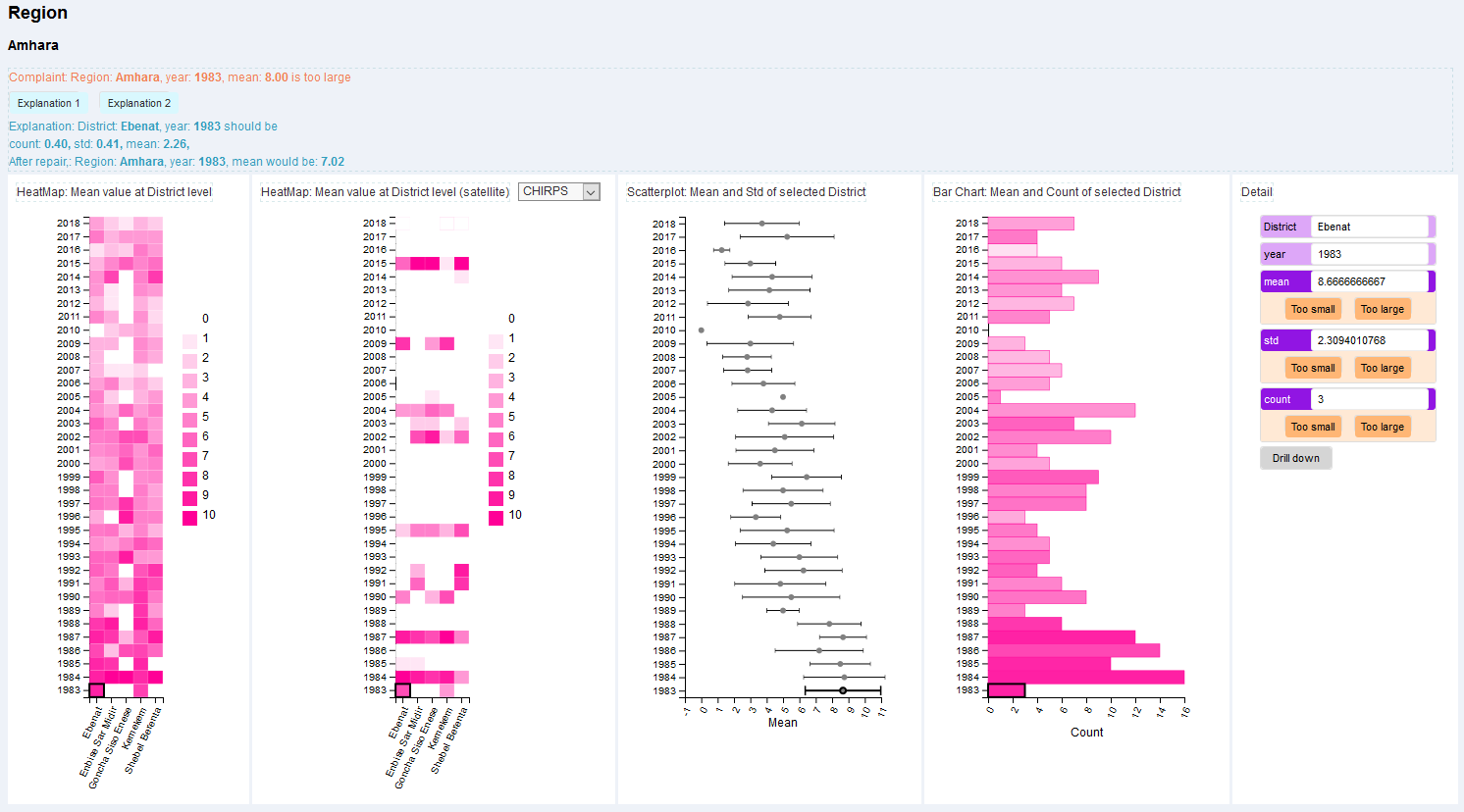}
  \caption{User interface of \sys}
  \label{fig:userstudy}
\end{figure*}

Figure \ref{fig:userstudy} shows the user interface for the study. Here, participant has made a complaint about \texttt{Region} Amhara. At the top, two explanations are generated that highlights two \texttt{Districts} which, if their aggregation results are repaired, can resolve complaint. The first heatmap shows drought severity for \texttt{Districts} in Amhara. The second heatmap shows the remote sensing. The scatterplot and barchart visualize aggregations results (\texttt{AVG}, \texttt{STD} and \texttt{COUNT}). Participant can further make complaint at \texttt{District} level.

For the first complaint, one team member recalls that one year is a severe year and complains that the mean severity of one region is too low. However, it turns out that all the districts in this region have low mean severity. For sensing data, some of them indicate that this year is severe, but some of them don't. Different team members also hold different opinions about this year. More investigations are needed to understand this complaint.

For the second complaint, one team member complains that the standard deviation of one regions is too high. The error is caused by two districts, but our system only identify one district. The failure is because of the property of the standard deviation. When the complaints are caused by multiple clusters, repair only one cluster may not cause the standard deviation closer to the true value. Consider the following minimum example: 

Suppose there are three same values $\emph{n}$ initially. The initial mean is $\emph{n}$ and variation is 0. Suppose we corrupt first two values by adding $\Delta$: $\emph{n} + \Delta$, $\emph{n} + \Delta$, and $\emph{n}$. The mean becomes $\emph{n} + \frac{2}{3}\Delta$ and variation becomes $\frac{2}{3}\Delta^2$. Suppose we fix the first corruption: $\emph{n}$, $\emph{n} + \Delta$, and $\emph{n}$. The mean becomes $\emph{n} + \frac{1}{3}\Delta$ and variation becomes $\frac{2}{3}\Delta^2$. Notice that variation is the same as that before fix. Suppose user complains about the high standard deviation, fixing any of two corrupted values wouldn't resolve user's complaint. Suppose we fix the first corruption partly to $\Delta'$: $\emph{n} + \Delta'$, $\emph{n} + \Delta$, and $\emph{n}$. The mean becomes $\emph{n} + \frac{1}{3}\Delta + \frac{1}{3}\Delta'$ and variation becomes $\frac{2}{3}(\Delta^2 - \Delta\Delta' + \Delta'^2)$. Let variation be a function of $\Delta'$: $\emph{f}(\Delta') = \frac{2}{3}(\Delta^2 - \Delta\Delta' + \Delta'^2)$. This is a parabola with turning point $(\frac{1}{2}\Delta,\frac{1}{2}\Delta^2)$. That is, the minimum standard deviation is achieved by fixing half of the corruption.

One solution is that, for equation \ref{tupleinquery} in the optimization problem, we search for a set tuples $\alpha \subseteq \emph{Q}$ instead of one tuple. This makes the optimization problem NP-hard because, given $\emph{n}$ tuples in $\emph{Q}$, there are $2^\emph{n}$ possible subsets of tuples. Joglekar et al. \cite{joglekar2015smart} exploits the property of submodularity to greedily search the optimal solution, while in our case, submodularity can't be guaranteed. Another solution is to relax the boolean constraint for the optimization problem and allows the repaired aggregation values to be within the range of $[\emph{n} + \Delta', \emph{n} + \Delta]$.  In future work, we plan to further study this problem.




\section{Case Study: Vote}
\label{casevote}
We conduct case study of Vote dataset, which is introduced in \Cref{modelev}. We consider a distributive set of two aggregation functions: Percentage of Votes for Donald Trump and total votes. We study Georgia state, which is one of the swing states and Joe Biden wins by a margin about 0.25\%. Given the complaint that the Percentage of Votes in the whole state is too low, \sys is used to find which counties contribute to the loss.

\begin{figure*}
     \centering
     \begin{subfigure}[b]{0.24\textwidth}
         \centering
         \includegraphics[width=\textwidth]{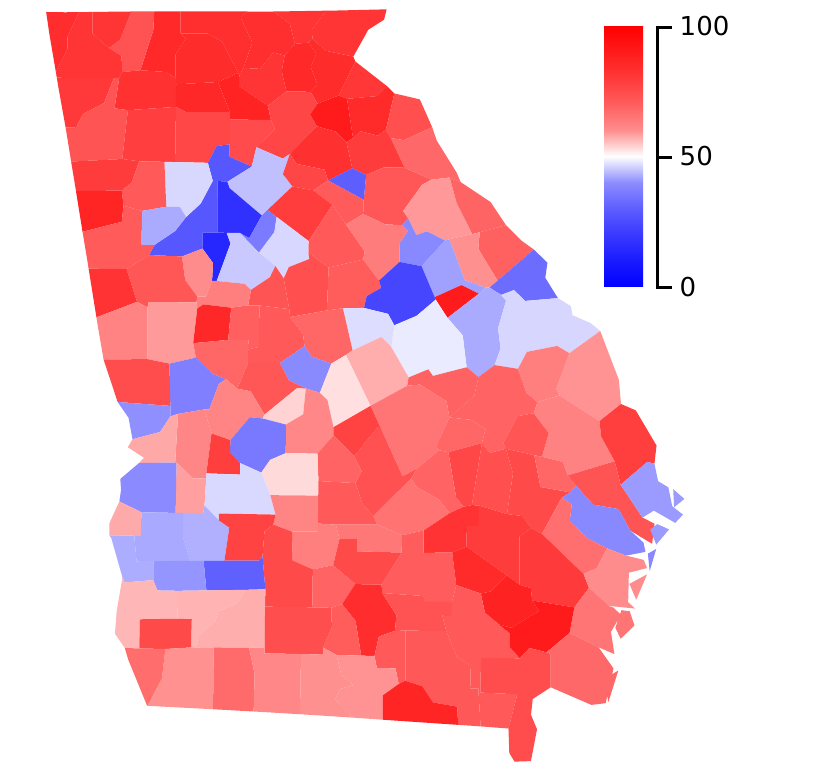}
         
         \caption{Percentage of Votes in 2016}
         \label{fig:vote2016}
     \end{subfigure}
     \hfill
     \begin{subfigure}[b]{0.24\textwidth}
         \centering
         \includegraphics[width=\textwidth]{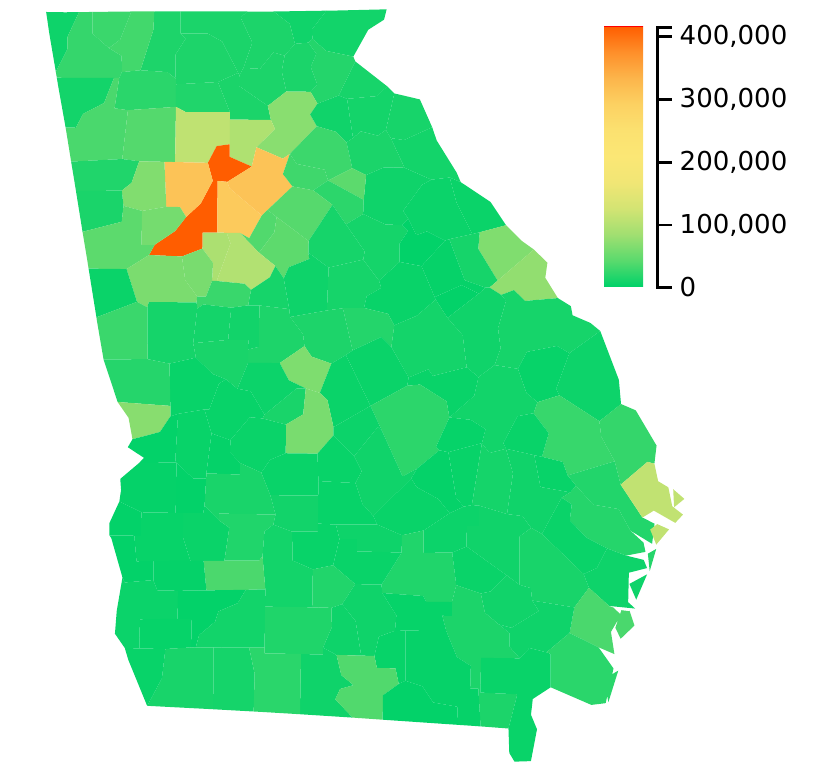}
         \caption{Total Votes in 2016}
         \label{fig:total2016}
     \end{subfigure}
     \hfill
    \begin{subfigure}[b]{0.24\textwidth}
         \centering
         \includegraphics[width=\textwidth]{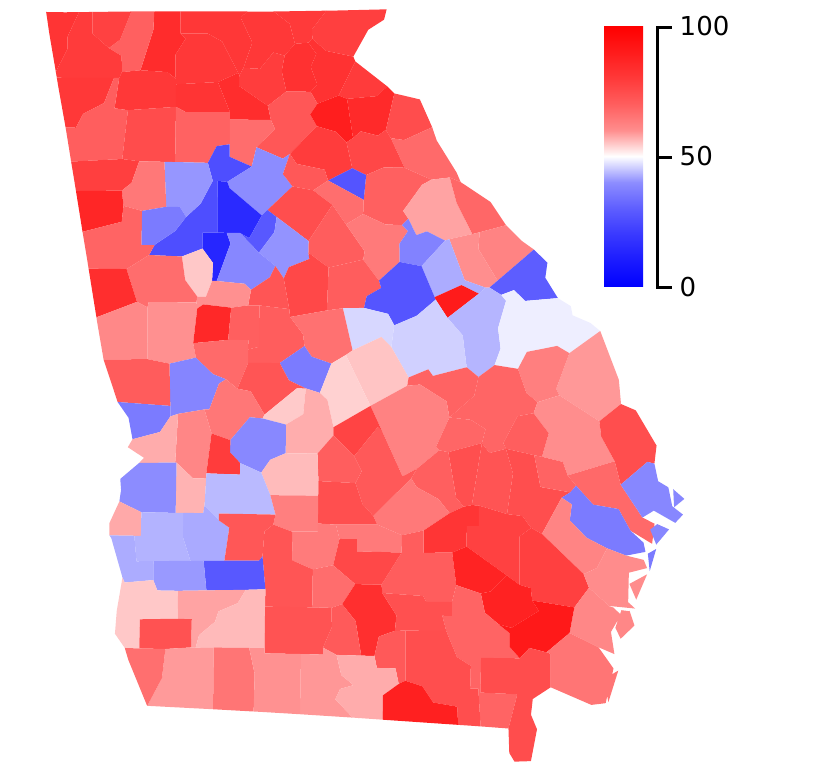}
         \caption{Percentage of Votes in 2020}
         \label{fig:vote2020}
     \end{subfigure}
     \hfill
     \begin{subfigure}[b]{0.24\textwidth}
         \centering
         \includegraphics[width=\textwidth]{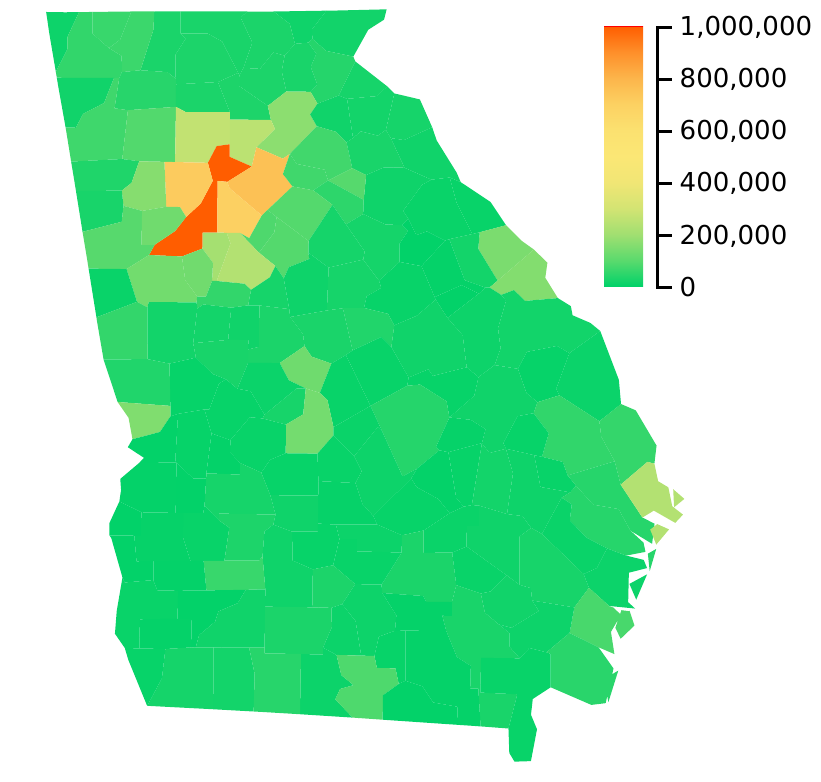}
         \caption{Total Votes in 2020}
         \label{fig:total2020}
     \end{subfigure}
    \hfill
    \begin{subfigure}[b]{0.33\textwidth}
         \centering
         \includegraphics[width=0.72\textwidth]{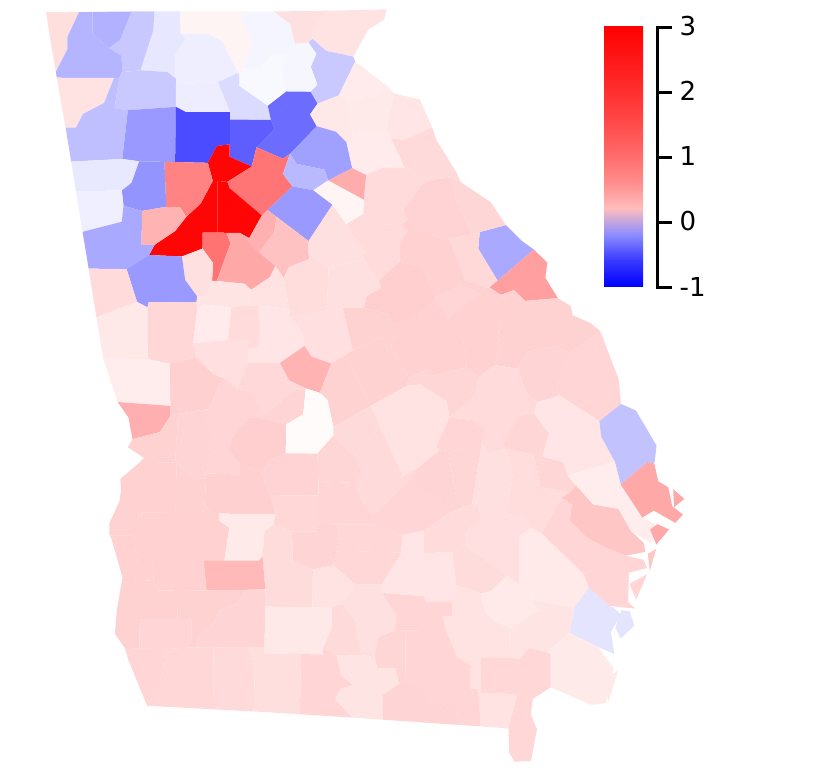}
         \caption{Margin gain after repair by model 1}
         \label{fig:repirnof}
     \end{subfigure}
    \hfill
    \begin{subfigure}[b]{0.33\textwidth}
         \centering
         \includegraphics[width=0.72\textwidth]{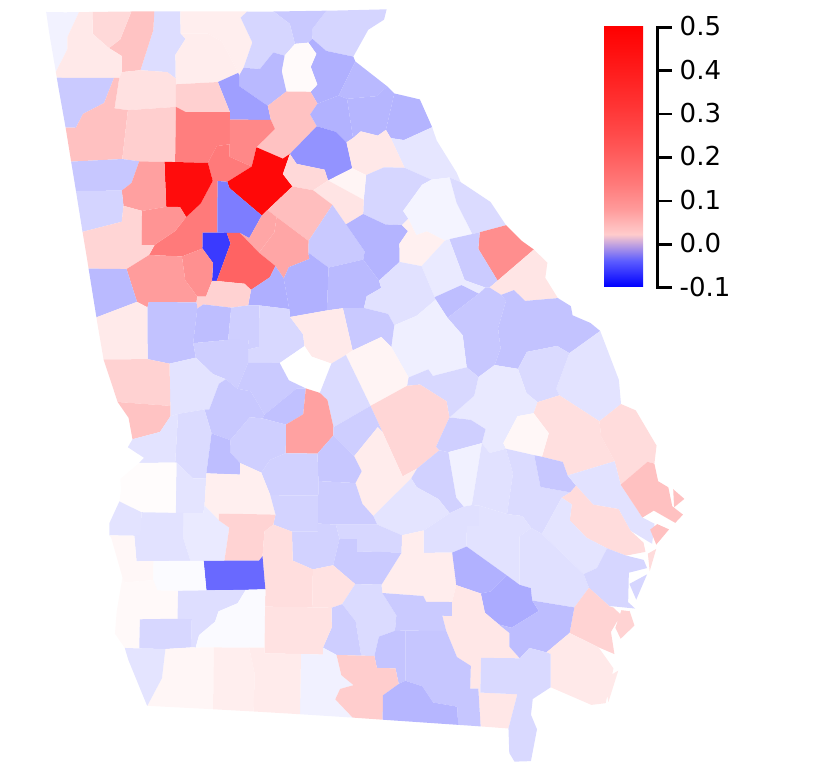}
         \caption{Margin gain after repair by model 2}
         \label{fig:repirwithf}
     \end{subfigure}
    \hfill
    \begin{subfigure}[b]{0.33\textwidth}
         \centering
         \includegraphics[width=0.72\textwidth]{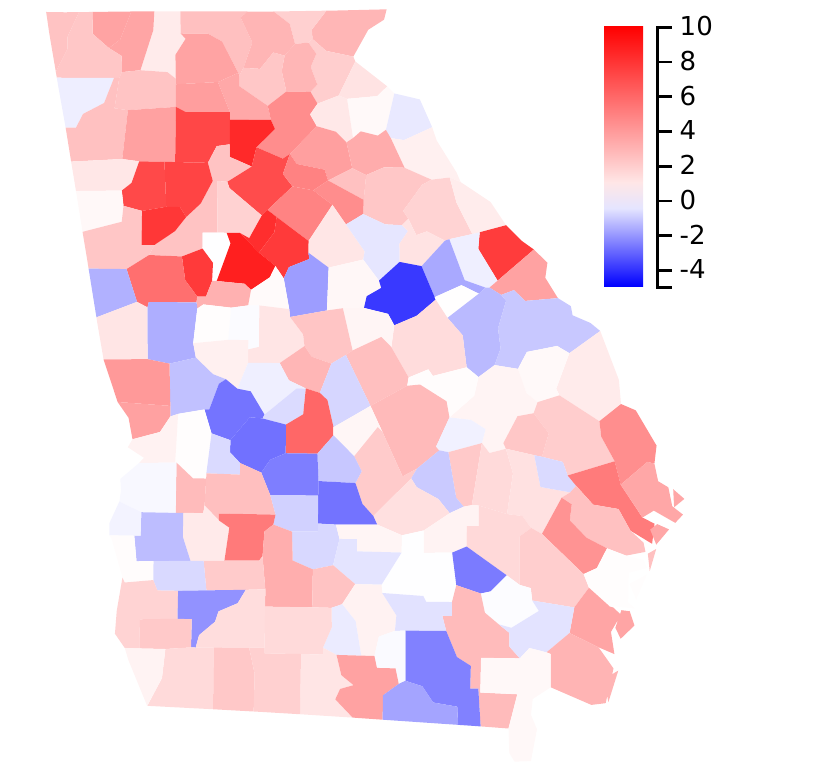}
         \caption{Percentage of Vote change from 2020 to 2016}
         \label{fig:inversetrend}
     \end{subfigure}
    \hfill
    \begin{subfigure}[b]{0.48\textwidth}
         \centering
         \includegraphics[width=0.5\textwidth]{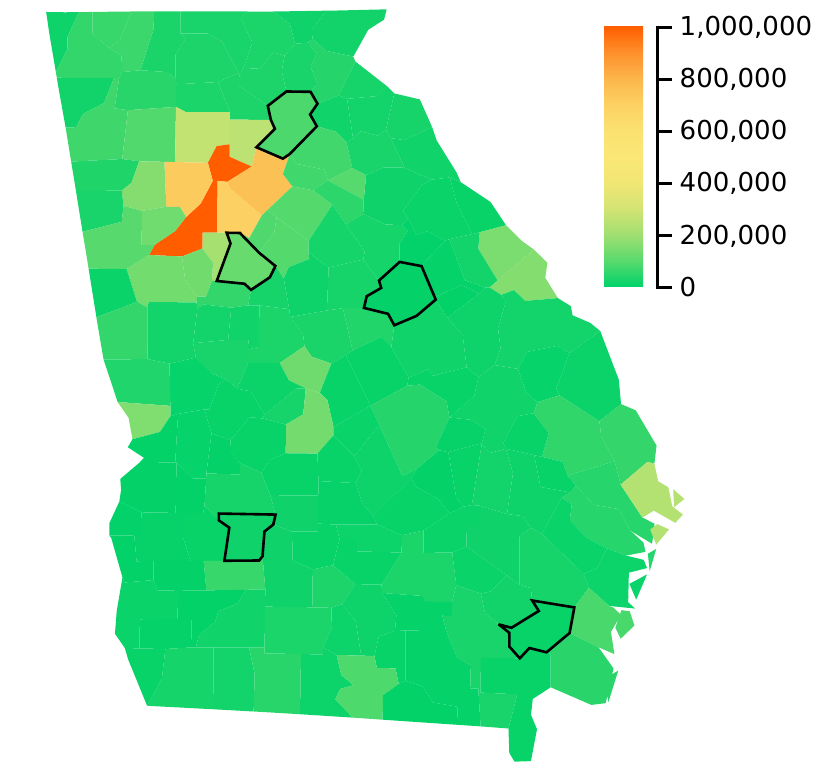}
         \caption{Missing records for Total Votes in 2020}
         \label{fig:newmissing}
     \end{subfigure}
    \begin{subfigure}[b]{0.48\textwidth}
         \centering
         \includegraphics[width=0.5\textwidth]{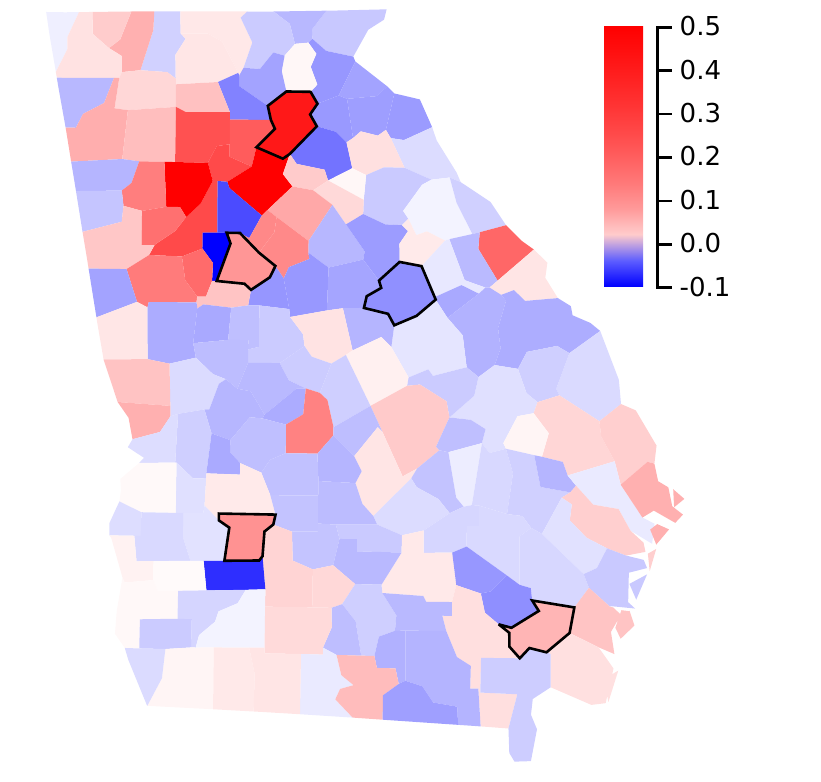}
         \caption{Margin gain of data with missing records}
         \label{fig:missingrepair}
     \end{subfigure}
\caption{Case study of 2020 US presidential election}
        \label{fig:casestudy}
\end{figure*}

Figure \ref{fig:vote2016}, \ref{fig:total2016}, \ref{fig:vote2020} and \ref{fig:total2020} show the Percentage of Votes and total votes of different counties in Georgia in 2016 and 2020. We run \sys using two models with different features. Model 1 is trained by only default feature, and Model 2 is trained by both default feature and external feature. Figure \ref{fig:repirnof} and \ref{fig:repirwithf} show the margin gain of Percentage of Votes after repair by model 1 and model 2. \sys will recommend counties with larger marginal gain as they better resolve the complaint. For model 1, because it only considers default feature, \sys mainly detects outliers in the counties of Georgia. Generally, those counties with low Percentage of Votes are deemed with outliers. Model 2 also considers the Percentage of Votes in 2016, which helps explain counties with low Percentage of Votes in 2020. With model 2, \sys is looking for counties which have abnormal Percentage of Votes or total votes compared to 2016 which after repaired best resolve user complaint. One interpretation of \ref{fig:repirwithf} is that it is calculating the change of Percentage of Vote from 2020 to 2016, which is plotted in \Cref{fig:inversetrend}. While  \Cref{fig:repirwithf} and \Cref{fig:inversetrend} are correlated, there are obvious differences because \sys also takes into account the total votes. To illustrate the effect of total votes, we manually inject missing records to counties highlighted in \Cref{fig:newmissing} by setting total votes to half of its original value. \Cref{fig:missingrepair} shows the margin gain of data with missing records after repair by model 2. The margin gains of counties with missing records changes depending on its original Percentage of Votes. \sys combines signals from all aggregation functions and external dataset to make recommendations.

\end{document}